\documentclass[11pt]{prop_axis} 

\usepackage[super,comma]{natbib} 
\usepackage{apjfonts} 
\usepackage{graphicx} 
\usepackage{amssymb,amsmath} 
\usepackage{color} 
\usepackage{enumitem} 
\usepackage{wrapfig} 

\usepackage[table]{xcolor} 
\usepackage{multirow} 
\usepackage{tabularx} 
\usepackage{array} 

\usepackage{hyperref} 
\hypersetup{
    colorlinks,
    linkcolor={blue!80!black},
    citecolor={blue!80!black},
    urlcolor={blue!80!black}
}

\newcolumntype{L}[1]{>{\raggedright\let\newline\\\arraybackslash\hspace{0pt}}p{#1}}
\newcolumntype{C}[1]{>{\centering\let\newline\\\arraybackslash\hspace{0pt}}p{#1}}
\newcolumntype{R}[1]{>{\raggedleft\let\newline\\\arraybackslash\hspace{0pt}}p{#1}}
\renewcommand\arraystretch{1.2}

\definecolor{headcolor}{rgb}{0.65,0.65,0.65}
\usepackage{fancyhdr}
\renewcommand{\topmargin}{-9mm}

\renewcommand{\textfloatsep}{4mm}

\newcommand{\AXIS}{\textit{AXIS}}
\newcommand{\Chandra}{\textit{Chandra}}
\newcommand{\XMM}{\textit{XMM-Newton}}
\newcommand{\Lynx}{\textit{Lynx}}
\newcommand{\Suzaku}{\textit{Suzaku}}
\newcommand{\Herschel}{\textit{Herschel}}
\newcommand{\Swift}{\textit{Swift}}
\newcommand{\Spitzer}{\textit{Spitzer}}
\newcommand{\NuSTAR}{\textit{NuSTAR}}
\newcommand{\NuStar}{\textit{NuSTAR}}
\newcommand{\ATHENA}{\textit{Athena}}
\newcommand{\Athena}{\textit{Athena}}
\newcommand{\eROSITA}{\textit{eROSITA}}
\newcommand{\XRISM}{\textit{XRISM}}
\newcommand{\Euclid}{\textit{Euclid}}

\newcommand{\HST}{\textit{HST}}
\newcommand{\Hubble}{\textit{Hubble}}
\newcommand{\ALMA}{\textit{ALMA}}
\newcommand{\JWST}{\textit{JWST}}
\newcommand{\LSST}{\textit{LSST}}
\newcommand{\WFIRST}{\textit{WFIRST}}
\newcommand{\JVLA}{\textit{JVLA}}
\newcommand{\SKA}{\textit{SKA}}
\newcommand{\ELT}{\textit{ELT}}
\newcommand{\TMT}{\textit{TMT}}
\newcommand{\GMT}{\textit{GMT}}
\newcommand{\LOFAR}{\textit{LOFAR}}
\newcommand{\GMRT}{\textit{GMRT}}
\newcommand{\MWA}{\textit{MWA}}
\newcommand{\LIGO}{\textit{LIGO}}
\newcommand{\LISA}{\textit{LISA}}
\newcommand{\CTA}{\textit{CTA}}
\newcommand{\TESS}{\textit{TESS}}
\newcommand{\ROSAT}{\textit{ROSAT}}
\newcommand{\RXTE}{\textit{RXTE}}
\newcommand{\JUNO}{\textit{JUNO}}
\newcommand{\VLA}{\textit{VLA}}

\newcommand{\arcsec}{$^{\prime\prime}$}

\newcommand{\fluxergs}{~erg~s$^{-1}$~cm$^{-2}$}
\newcommand{\lumergs}{~erg~s$^{-1}$}
\newcommand{\msunperyr}{~$M_{\odot}$~yr$^{-1}$}
\newcommand{\kms}{~km~s$^{-1}$}
\def\lax{\lesssim}

\newcommand{\bs}{\sffamily\bfseries\small}
\newcommand{\sfsm}{\sffamily\small}
\definecolor{callout}{rgb}{0.25,0.45,0.85}
\definecolor{tablealt}{rgb}{0.77,0.85,1.0}

\begin{document}

\baselineskip=13.2pt
\sloppy
\pagenumbering{roman}

\begin{figure}[t]
\includegraphics[width=0.98\textwidth,viewport=82 125 672 945]{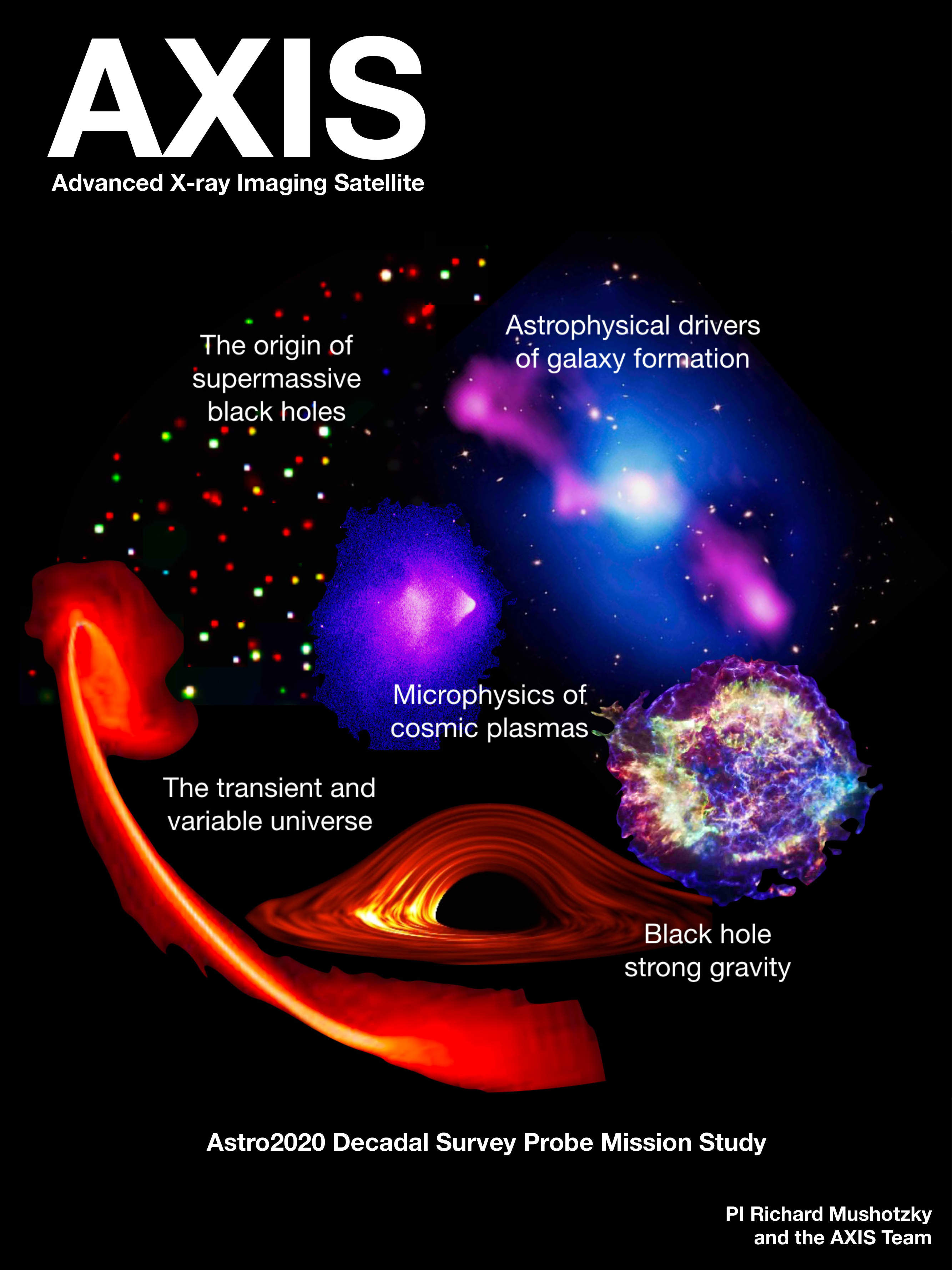}
\end{figure}

\title{\textcolor{black}{\sf\LARGE THE ADVANCED X-RAY IMAGING SATELLITE}}
\maketitle
\thispagestyle{empty}

\vspace*{-10mm}
\begin{center}
\begin{minipage}{16cm}
\hspace{-5mm}
\centering {\em Authors}: Richard~F.~Mushotzky$^{1}$, James~Aird$^{2}$, Amy~J.~Barger$^{3}$,
Nico~Cappelluti$^{4}$, George~Chartas$^{5}$, L\'{i}a~Corrales$^{6}$, Rafael~Eufrasio$^{7,8}$,
Andrew~C.~Fabian$^{9}$, Abraham~D.~Falcone$^{10}$, Elena~Gallo$^{6}$, Roberto~Gilli$^{11}$,
Catherine~E.~Grant$^{12}$, Martin~Hardcastle$^{13}$, Edmund~Hodges-Kluck$^{1,8}$, Erin~Kara$^{1,8,12}$,
Michael~Koss$^{14}$, Hui~Li$^{15}$, Carey~M.~Lisse$^{16}$, Michael~Loewenstein$^{1,8}$, Maxim~Markevitch$^{8}$,
Eileen~T.~Meyer$^{17}$, Eric~D.~Miller$^{12}$, John~Mulchaey$^{18}$, Robert~Petre$^{8}$,
Andrew~J.~Ptak$^{8}$, Christopher~S.~Reynolds$^{9}$, Helen~R.~Russell$^{9}$, Samar~Safi-Harb$^{19}$,
Randall~K.~Smith$^{20}$, Bradford~Snios$^{20}$, Francesco~Tombesi$^{1,9,21,22}$, Lynne~Valencic$^{8,23}$,
Stephen~A.~Walker$^{8}$, Brian~J.~Williams$^{8}$, Lisa~M.~Winter$^{15,24}$, Hiroya~Yamaguchi$^{25}$,
William~W.~Zhang$^{8}$ \\
\vspace{5mm}
{\em Contributors}: Jon Arenberg$^{26}$, Niel Brandt$^{10}$, David~N.~Burrows$^{10}$, Markos~Georganopoulos$^{17}$, Jon~M.~Miller$^{6}$, Colin~A.~Norman$^{17}$, Piero~Rosati$^{27}$
\end{minipage}
\end{center}

\begin{figure}[h]
\vspace*{15mm}
\centering
 \includegraphics[width=0.5\textwidth,angle=-90,viewport=25 78 534 790,clip]{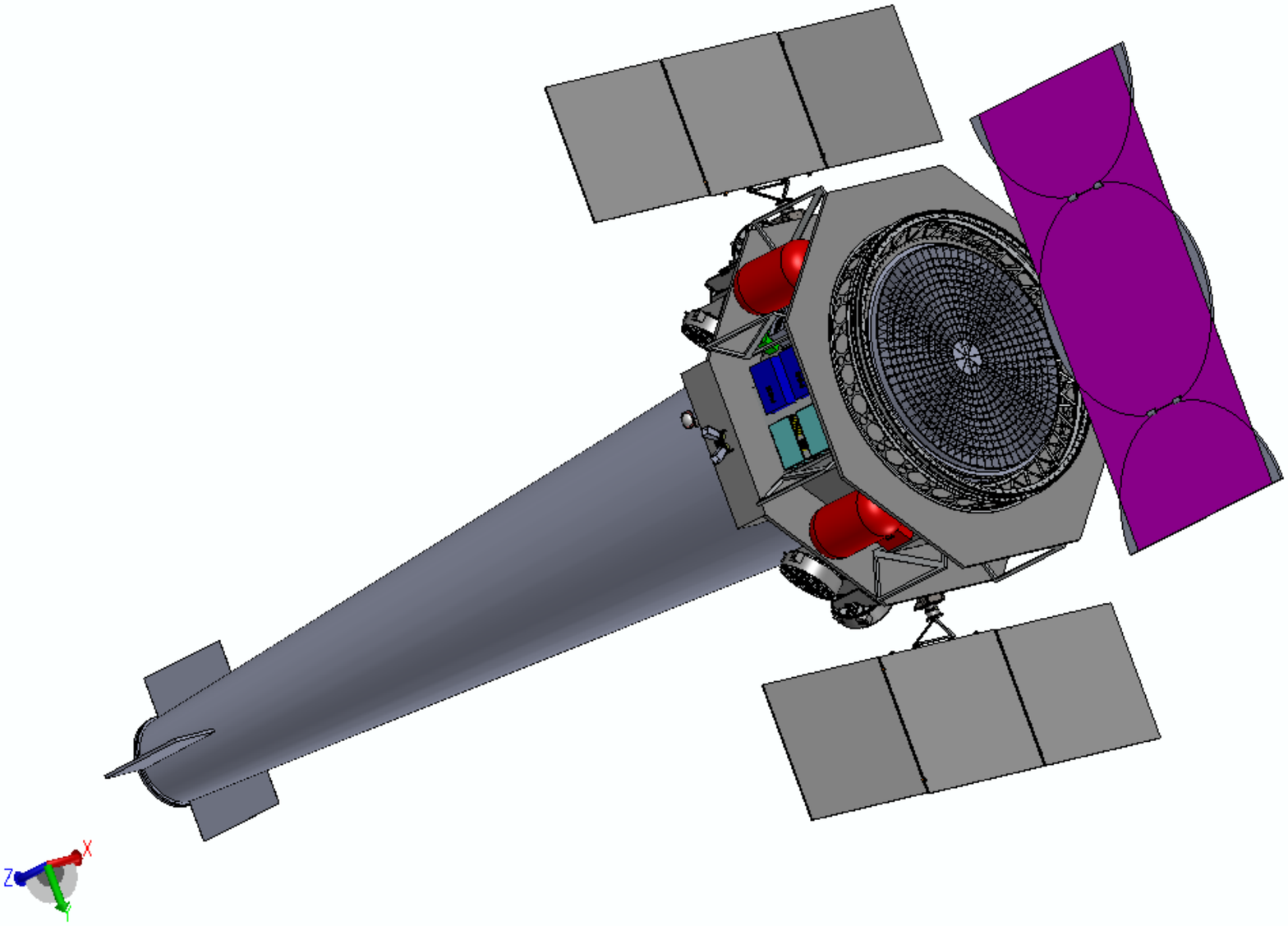}
\end{figure}

\vspace*{25mm}
\centerline{A Probe-class mission study commissioned by NASA for the NAS Astro2020 Decadal Survey} 

\vspace*{5mm}
\centerline{\today}
\clearpage

\vspace*{0mm}
\centerline{\sf\large AUTHOR AFFILIATIONS}
\vspace*{5mm}

\begin{minipage}{\linewidth}
\small
\hspace{-10mm}
\begin{tabularx}{\linewidth}{L{0.98\linewidth}}
$^{1}$~Department of Astronomy, University of Maryland, College Park, MD 20742\\
$^{2}$~Department of Physics and Astronomy, The University of Leicester, Leicester LE1 7RH, UK\\
$^{3}$~Department of Astronomy, University of Wisconsin-Madison, Madison, WI 53706\\
$^{4}$~Physics Department, University of Miami, Coral Gables, FL 33124\\
$^{5}$~Department of Physics and Astronomy, College of Charleston, Charleston, SC 29424\\ 
$^{6}$~Department of Astronomy, University of Michigan, Ann Arbor, MI 48109\\
$^{7}$~The Catholic University of America, Washington, DC 2006\\
$^{8}$~NASA Goddard Space Flight Center, Greenbelt, MD 20771\\
$^{9}$~Institute of Astronomy, Cambridge CB3 0HA, UK\\
$^{10}$~Department of Astronomy and Astrophysics, Pennsylvania State University, University Park, PA 16802\\
$^{11}$~INAF-Osservatorio Astronomico di Bologna, 40129, Bologna, Italy\\
$^{12}$~Kavli Institute for Space Research, Massachusetts Institute of Technology, Cambridge, MA 02139\\
$^{13}$~Centre for Astrophysics Research, University of Hertfordshire, Hatfield, Hertfordshire, QQ62+JJ, UK\\
$^{14}$~Eureka Scientific, Oakland, CA 94602\\
$^{15}$~Center for Theoretical Astrophysics,  Los Alamos National Laboratory, Los Alamos, NM 87545\\
$^{16}$~Johns Hopkins University Applied Physics Laboratory, Laurel, MD 20723\\
$^{17}$~Department of Physics, University of Maryland Baltimore County, Baltimore, MD 21250\\
$^{18}$~Carnegie Observatories, Pasadena, CA 91101\\
$^{19}$~Department of Physics and Astronomy, University of Manitoba, Winnipeg, MB R3T 2N2, Canada\\
$^{20}$~Harvard-Smithsonian  Center for Astrophysics, Cambridge, MA 02138\\
$^{21}$~Department of Physics, University of Rome Tor Vergata, 00133, Rome, Italy \\
$^{22}$~INAF Astronomical Observatory of Rome, 00078, Monteporzio Catone, Italy\\
$^{23}$~Department of Physics \& Astronomy, Johns Hopkins University, Baltimore, MD 21218\\
$^{24}$~National Science Foundation, Alexandria, VA 22314\\
$^{25}$~Institute of Space and Astronautical Science, Sagamihara, Kanagawa 252-5210, Japan\\
$^{26}$~Northrop Grumman Aerospace Systems, Redondo Beach, CA 90278\\
$^{27}$~Department of Physics and Earth Science, University of Ferrara, 44122, Ferrara, Italy
\end{tabularx}
\end{minipage}

\clearpage

\renewcommand{\contentsname}{ }
{\sf\small\vspace*{-17mm}
\tableofcontents
}
\clearpage

\setcounter{page}{1}
\pagenumbering{arabic}

\section{SUMMARY}
\label{section:executive_summary}

Over the last 40 years, X-ray observations have proven crucial for advancing
major areas of astrophysics. Indeed, much of the baryonic matter in the
Universe, including the most active and luminous sources, are best studied
in the X-ray band. Key advances in X-ray optics and detector technology have
paved the way for the Advanced X-ray Imaging Satellite (\AXIS), a 
Probe-class mission that is a major improvement over \Chandra\ ---
with higher-resolution imaging over a larger
field of view at much higher sensitivity, and flexible mission 
operations allowing \Swift-like transient science. The design and 
operations allow an extensive guest observer program open to 
all areas of science.
\AXIS\ can be launched in the 2020s and will transform our 
understanding in several areas of astrophysics. Among them are:
\begin{itemize}[noitemsep,topsep=2mm,fullwidth,itemindent=5mm]
\item {\bs The growth and fueling of supermassive black holes (SMBHs)}: 
Deep surveys
will reveal the formation and evolution of early black holes (BHs). Resolving
dual active galactic nuclei (AGN) will quantify the frequency of BH mergers. Observations, at
an unprecedented angular resolution, of gravitationally-lensed quasars and
hot gas within the Bondi radius of nearby galaxies will allow us to 
study the matter in the immediate vicinity of the SMBHs.
\item {\bs Galaxy formation and evolution}: \AXIS\ will detect and
resolve powerful outflows from AGN and supernovae driven winds during
the peak era of star
formation, and separate X-ray binary (XRB) emission from AGNs at even 
higher redshifts. \AXIS\ will study 
the warm and hot gas in and around nearby galaxies and the 
intergalactic medium near galaxy clusters --- the ultimate reservoir 
of energy and metals expelled from the galaxies over their lifetime.
\item {\bs The microphysics of cosmic plasmas}: \AXIS\ will 
find and resolve shocks, instabilities and perturbations in the 
intracluster medium and supernova remnants (SNRs). These are 
sensitive probes of basic plasma properties and processes, such as viscosity,
heat conductivity, equilibration timescales, and acceleration and diffusion of
cosmic rays --- crucial building blocks for understanding and 
modeling a wide range of astrophysical phenomena.
\item {\bs The time variable universe}: Rapid repointing will enable
observations of violent cosmic events out to high redshift,
including the early stages of supernovae, the electromagnetic counterparts
of gravitational waves from SMBH mergers, tidal disruption 
events, neutron star mergers, AGN flares, and stellar flares.
\item {\bs A wide variety of cutting-edge science}: A high angular
resolution X-ray observatory such as \AXIS\ can provide critical 
data for a wide variety of areas such as the origin of the elements,
the habitability of exoplanets around active stars, proper motions in AGN
jets and SNRs, the nature of dust in the interstellar medium 
(ISM), and allow for a detailed mapping of the elements on the 
surface of the Moon and aurorae on Jupiter.
\end{itemize}

\AXIS\ is designed to make these and
other scientific advances within the constraints of a Probe class mission.
Its groundbreaking capability is due to improved imaging over NASA's existing
flagship X-ray observatory, \Chandra{}, and the European Space Agency's
planned \Athena\ mission. Relative to \Chandra{}, the \AXIS\ on-axis Point Spread
Function (PSF) is nearly twice as sharp; its field of view for subarcsecond
imaging 70 times larger by area; its effective area at 1~keV is 10 times
larger (Table~\ref{table:AXIS_mission_parameters}).
The \AXIS\ low-inclination, low-Earth orbit (LEO) ensures a low and stable detector background,
resulting in 50~times greater sensitivity than
\Chandra{} for extended sources.

\begin{figure}[htp]
    \centering
    \framebox{
    \includegraphics[height=0.30\textwidth]{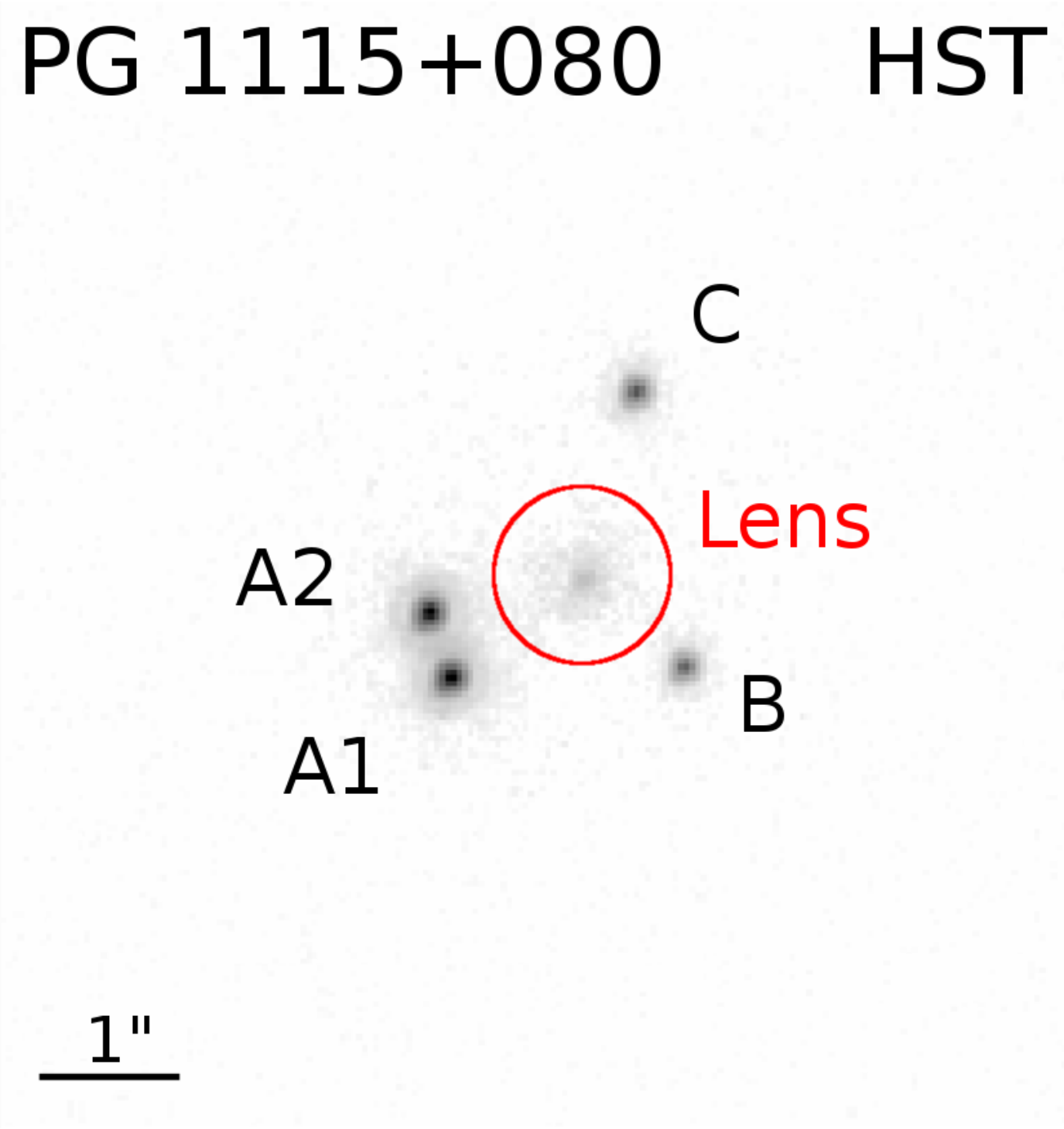}}
    \framebox{
    \includegraphics[height=0.30\textwidth]{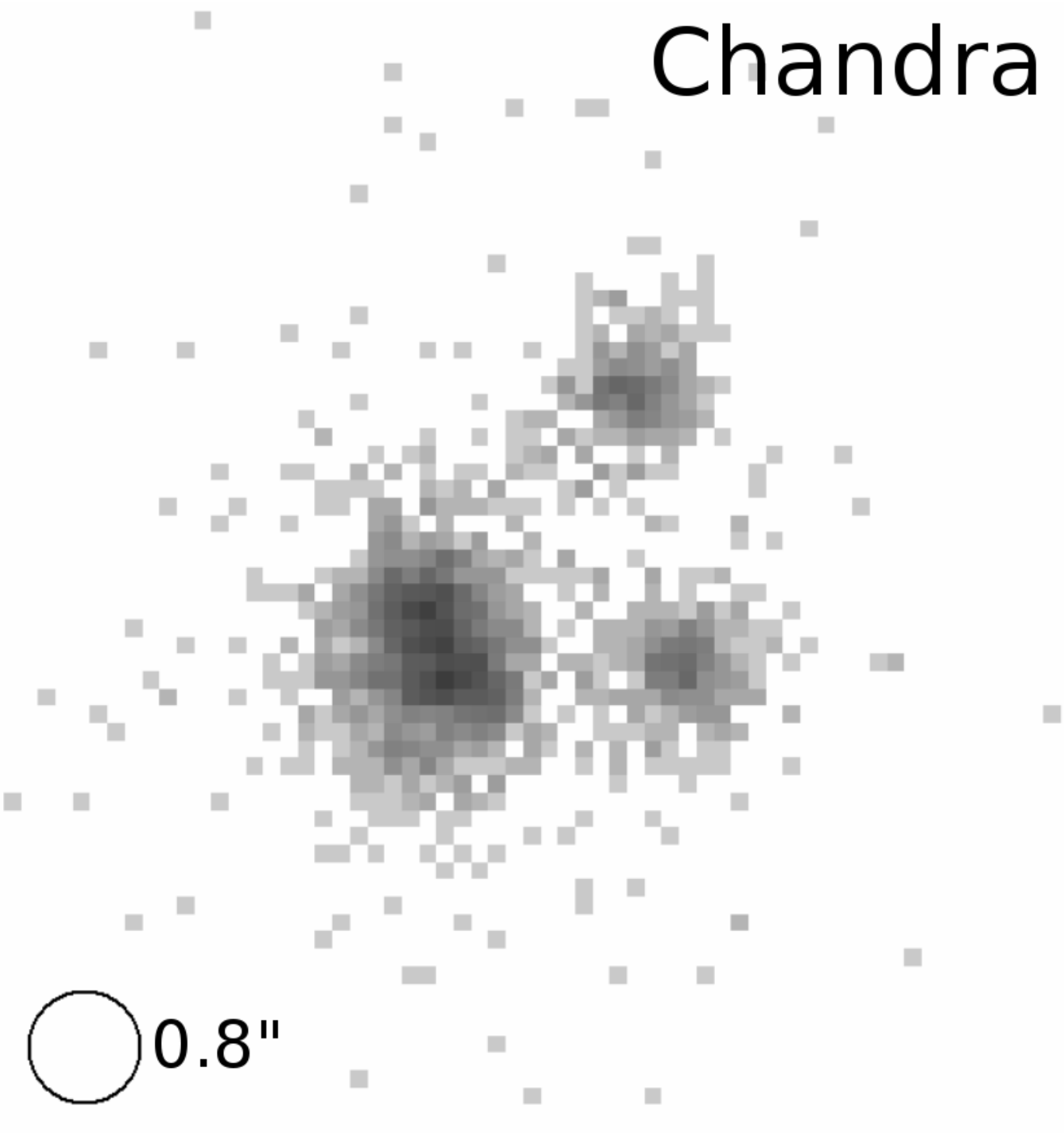}}
    \framebox{
    \includegraphics[height=0.30\textwidth]{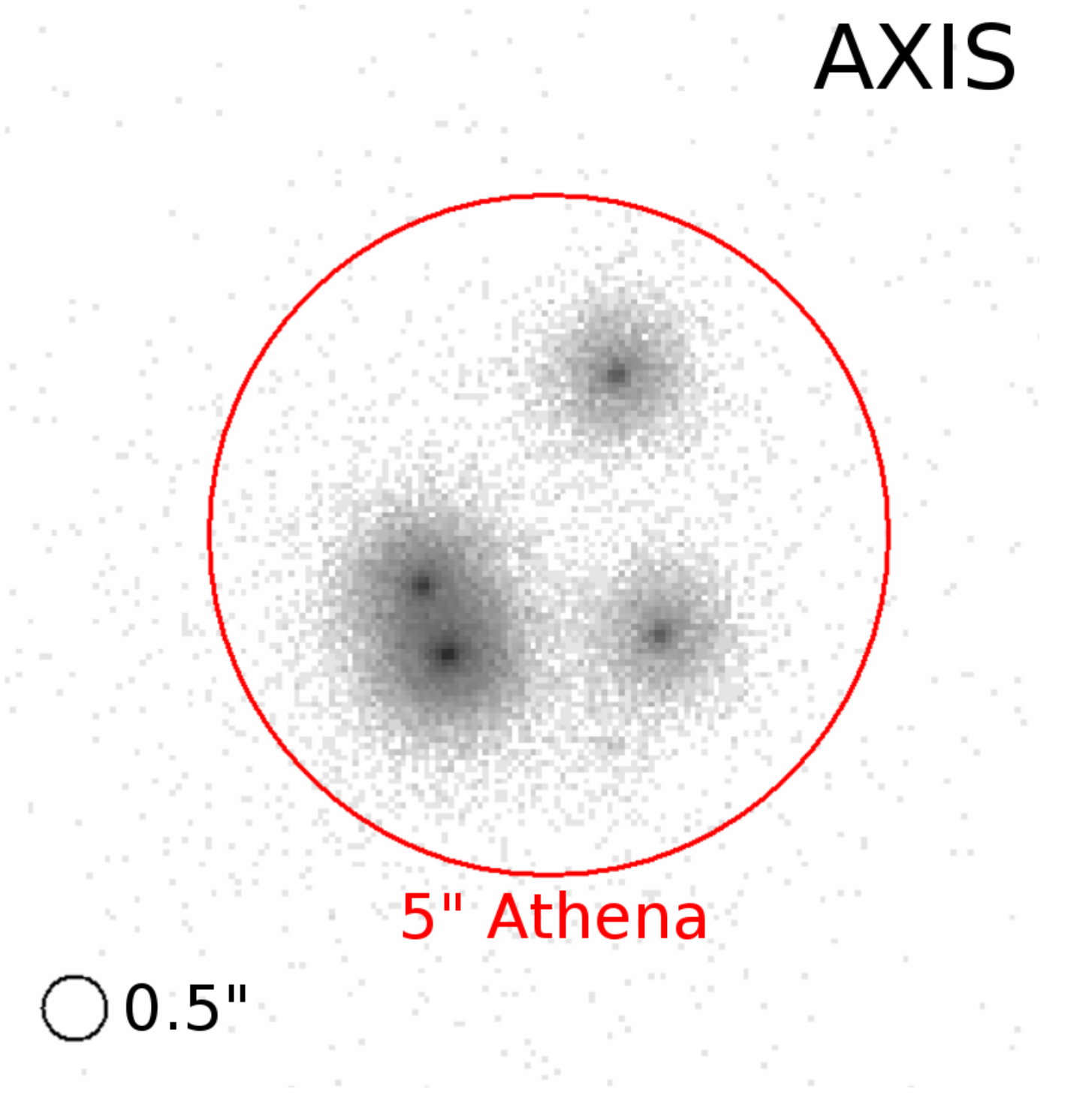}}

    \caption{\AXIS\ will be able to measure the innermost stable
      circular orbit and spin of the SMBH in the
      quadruply-lensed quasar PG1115+080 (HST image at left) through
      monitoring spectral variability of the quasar's multiple lensed images (A1, A2,
      B, C). An archival 30~ks \Chandra{} image (0.8\arcsec\ HPD) has
      limited photon statistics and does not separate the A1 and A2 images,
      whereas a 30~ks \AXIS\ exposure (0.5\arcsec\ HPD) will yield
      high-quality spectra from each quasar image and the variability on
      timescales much shorter than those accessible for \Chandra.}
    \label{fig:INTRO_lens}
\end{figure}

\AXIS\ has a rapid repointing response with operations similar to Swift, but is 100 times more sensitive for time domain science (as
measured by the product of source sensitivity and response time). These
capabilities open up a vast discovery space, and complement the next
generation of astronomical observatories, such as \JWST, \WFIRST, \LSST, \SKA,
\TMT, \ELT, \GMT, or \CTA. As seen
by the strong synergy between \Chandra{} and \XMM{}, a high-throughput, 
high-spectral-resolution mission (\Athena) operating at the same time as 
a high-angular-resolution mission (\AXIS) greatly increases the range 
of scientific discovery.

The simplicity of the \AXIS\ design --- a single mirror and detector, and few
moving parts --- results in a robust, low-cost design. \AXIS\ builds on
developments in X-ray mirror technology over the past decade that 
produce high-angular-resolution, lightweight X-ray optics at reasonable cost,
utilizing precision polishing and thin single-crystal silicon 
mirrors developed at Goddard. An angular resolution better than 
2.2\arcsec\ for a mirror segment pair module was demonstrated in 2018 \cite{Zhang2018}
and {\bs as of early 2019, mirror segments with figure quality of 
0.5\arcsec\ HPD have been regularly fabricated at GSFC.} The baseline
\AXIS\ small-pixel detector array builds on a long legacy of X-ray CCD and
benefits from 25 years of technology development, providing improved
photon localization and thus better effective angular resolution, much 
faster readout time, and broader energy band. Both CCD and CMOS type
detectors with the required properties are already under
development\cite{Bautz2018,Falcone2018}.

\AXIS\ successfully completed NASA/GSFC Instrument (IDL) and Mission (MDL)
Design Lab studies, during which the telescope, detector and spacecraft designs were 
developed using proven components and modern technical approaches compatible with
the \AXIS\ science requirements. The estimated mission costs are consistent with
the \$1B Probe mission cost guideline in 2018 dollars. The single technology area
requiring significant development is the construction of the X-ray mirror.
Successful development of the \AXIS\ optics will serve as the technological
and scientific pathfinder for a major US-led high energy astrophysics
mission in the 2030s.

\section{\AXIS\ IN THE
  FRAMEWORK OF 2020s ASTRONOMY} 
\label{section:2020s}

A new era of astronomical discovery in the imaging, spectral, and time
domains is underway with the next generation of observatories being planned and now in operation
that span the electromagnetic spectrum. \AXIS\ is highly complementary to
these observatories, with its 
ability to respond 
rapidly to transients, observe the highest redshift BHs, identify the
electromagnetic counterparts of gravitational wave sources, and probe
feedback over a wide range of redshift and galaxy mass.

While \Chandra{} and \XMM{} were well-matched to the available sensitivity
at other wavelengths in 1999, current X-ray facilities are insufficient to
address the cutting-edge, high-profile science goals attainable with the latest
space-borne (\textit{Herschel} and, soon, \textit{Euclid} and \JWST) and
ground-based (e.g, \LSST\ and \ALMA) observatories, as well as the planned
\WFIRST\ and 30-m ground based telescopes. \AXIS\ will expand the
frontiers of X-ray astronomy in a manner that complements and enhances these
goals and guides those observatories. \AXIS\ will make major
breakthroughs in the study of the universe by virtue of its high angular
resolution over a wide field of view, high sensitivity to point-like and
diffuse X-ray emission, and its rapid response to transient sources 
(see figures of merit in Figs.\ A.4--6).

\AXIS\ capabilities are critical for identifying unique counterparts and
making morphological comparisons with sources seen at other wavelengths. For
example, at the 1.6~$\mu$m wavelength where \WFIRST\ is near 
peak-sensitivity,%
\footnotemark\footnotetext{https://wfirst.gsfc.nasa.gov/science/WFIRSTScienceSheetFINAL.pdf}
there are 0.16 sources per arcsec$^2$ at an AB~mag of 28\cite{guo2013}, and
\AXIS' resolution is required for unambiguous X-ray counterpart
identification and characterization.

\AXIS\ has the field of view, resolution, and sensitivity to provide 
complementary data to the \WFIRST, \ELT, \TMT, and \GMT\ cameras. In turn these observatories will provide the data to obtain precise redshifts, bolometric luminosities and other physical properties for the sources
\AXIS\ will discover using optical data through to the submillimeter and millimeter (which will include obscured sources). 
The use of wide spectral energy
coverage with matching sensitivity and resolution is key to mapping the
redshift evolution, galaxy masses, and star formation rates of the host
galaxies of the AGN population and partitioning the spectral energy
distribution into AGN and non-AGN components (Section~\ref{section:galaxies}).  
This is critical, since most massive high-redshift galaxies host AGNs.

Three major X-ray missions are under development: \eROSITA\
will launch in 2019, \XRISM\ in 2021, and \Athena\ in the early 
2030s. These missions
form a rich complementary environment for \AXIS. Prior to 
\AXIS, the all-sky \eROSITA\
will discover a large sample of relatively bright X-ray sources that \AXIS\ 
or \ATHENA\ can follow up. Meanwhile, \XRISM\ will take the first high 
spectral resolution observations of moderate samples of bright sources. 
\Athena\ is a planned major ESA X-ray mission with moderate angular
resolution (5\arcsec) and powerful spectroscopic and timing capabilities. 
\AXIS\ could be launched close to the \Athena\ launch date; its high 
angular resolution imaging would complement
\Athena's high spectral resolution. For deep imaging, \Athena\ is
fundamentally limited by confusion
(Fig.~\ref{fig:2020s_ATHENA_confusion}), becoming
confusion-limited\cite{Aird2013} at a 0.5--2~keV flux of $\sim 2\times 10^{-17}$\fluxergs\ for a beam size of
5\arcsec. In contrast, the \AXIS\ confusion limit is
$\sim10^{-19}$\fluxergs.  This faint confusion limit, combined with high
effective area, will allow \AXIS\ to detect sources that are more than an
order of magnitude fainter than the deepest \Chandra{} detections over 
a field of view that is 7 times larger than the higher-resolution (HPD$<$3\arcsec)
portion of the Chandra field used for deep surveys.

\section{MISSION CAPABILITIES
  AND DESIGN DRIVERS} 
\label{section:mission}

{\rowcolors{3}{tablealt}{white!100}
\begin{table}
    \centering
    \begin{tabularx}{\textwidth}{ L{3.3cm} C{3.4cm} C{4.07cm} L{4.055cm} }
        \rowcolor{callout}
        \textcolor{white}{\sfsm Feature}\phantom{\raisebox{-3mm}{\rule{0cm}{8mm}}}&
        \textcolor{white}{\sfsm Value} & \textcolor{white}{\sfsm \AXIS\ vs.\ \Chandra{}} & \textcolor{white}{\sfsm Science Driver} \\
        Angular resolution \newline (HPD, at 1 keV)& 
        0.5\arcsec\ on-axis \newline 
        1\arcsec\ at 15$^{\prime}$ off-axis& 
        1.6$\times$ sharper \newline 
        28$\times$ sharper &
        Point source detection, separation, excision\\ 
        Energy band & 0.2-12~keV & Similar 
        & Soft and hard X-ray sensitivity \\
        Effective area \newline (mirror + detector)& 5800~cm$^2$ @ 0.5 keV \newline 
                                                     7000 cm$^2$ @ 1.0 keV \newline 
                                                     1500~cm$^2$ @ 6.0~keV & 
        15$\times$(launch), $10^3\times$(2018) \newline 
        10$\times$(launch), 40$\times$(2018) \newline 
        6$\times$ & 
        Faint source detection and analysis\\ 
        Energy Resolution & \phantom{1}60~eV @ 1.0 keV \newline 
        150~eV @ 6.0~keV & Similar & Resolving
        emission lines \\ 
        Timing Resolution & $<50$~ms & 64$\times$ faster readout; \newline
        pile-up limit 6$\times$ brighter & Variable sources; \newline observing
        bright sources \\ 
        Field of View & $24^{\prime} \times 24^{\prime}$ & 70$\times$ for
        $<1$\arcsec\ imaging & Extended sources, surveys \\ 
        Detector Background & $2\times 10^{-4}$~ct~s$^{-1}$ keV$^{-1}$
        arcmin$^{-2}$ @ 1~keV & 4$\times$ lower; 50$\times$ better
        sky/background ratio & Sensitivity to low surface brightness objects\\ 
        Slew Rate & 120$^\circ$\,/\,5~min & Comparable to \Swift{} & Observing efficiency, ToO~response\\
        \hline
    \end{tabularx}
    \caption{\AXIS\ mission parameters, compared with \Chandra\ ACIS best-in-class
      values. }
    \label{table:AXIS_mission_parameters}
\end{table}
}

The prime scientific drivers of the mission design are: 
1) The growth of SMBHs and the astrophysical drivers of galaxy evolution for the 0.5\arcsec\ angular resolution; 
2) the evolution of structure over cosmic time and the physics of plasmas in clusters for the $24^{\prime}\times 24^{\prime}$ field of view; 
3) the physics in the immediate vicinity of BHs for the energy resolution;
4) the detection and characterization of the hot baryons in and around galaxies, the early stages of supernovae, and tidal disruptions of stars by SMBHs for the collecting area and low energy range; 
5) the physics of plasma, shocks, and cosmic ray acceleration
for the high energy range and low background, requiring in turn LEO; 
6) time-domain astronomy for the rapid slewing and quick response to  target-of-opportunity (ToO) requests, 
7) the high observing efficiency and the ability to observe bright sources free of photon pile-up and to study rapidly variable sources for the
maximum detector readout rate; 
8) the budget and mass limit for the simple, robust design.

\begin{figure}
\centering
\raisebox{-0.4mm}{\includegraphics[width=0.41\textwidth,viewport=28 27 540 536,clip]{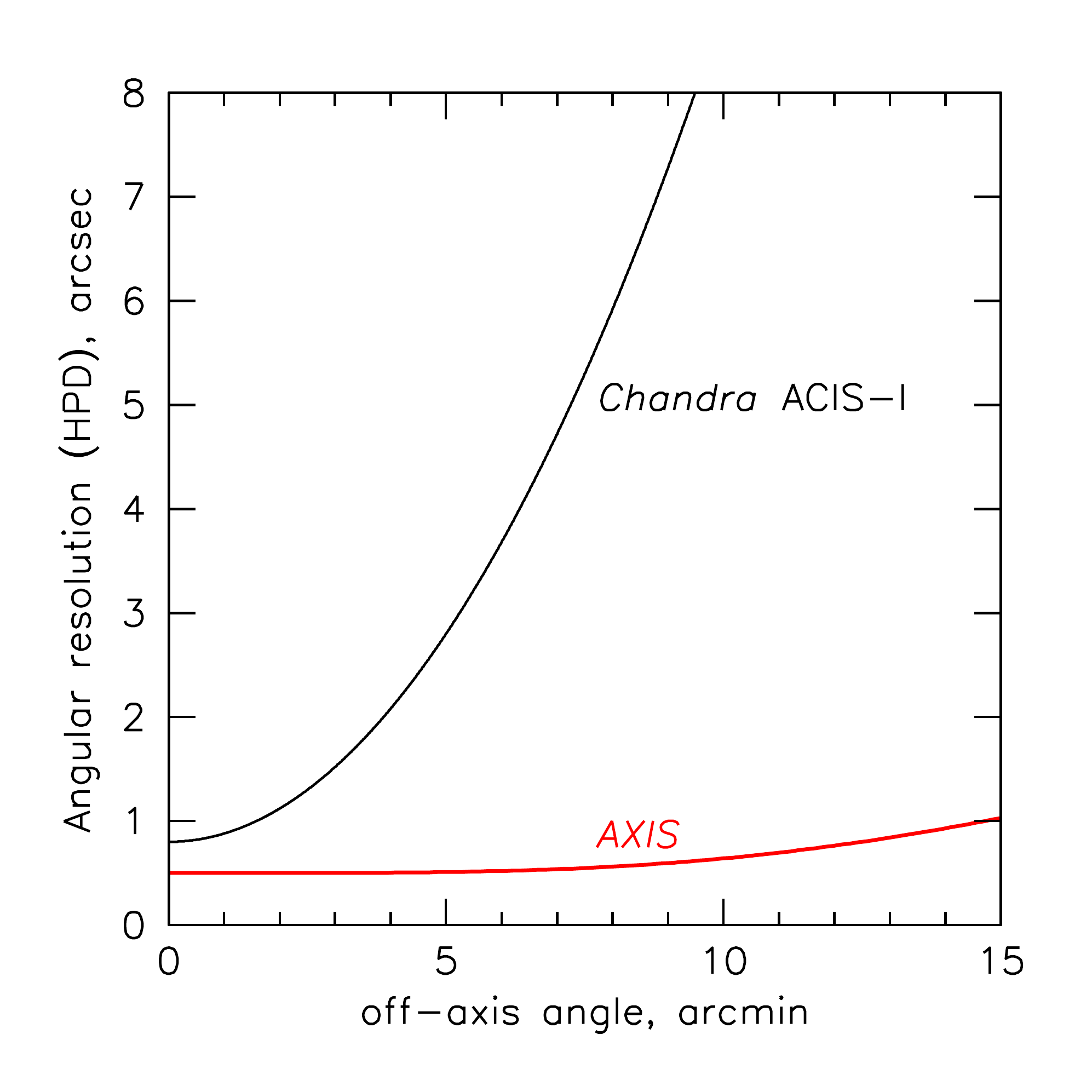}}\hspace*{5mm}
\includegraphics[width=0.54\textwidth,viewport=14 3 426 312,clip]{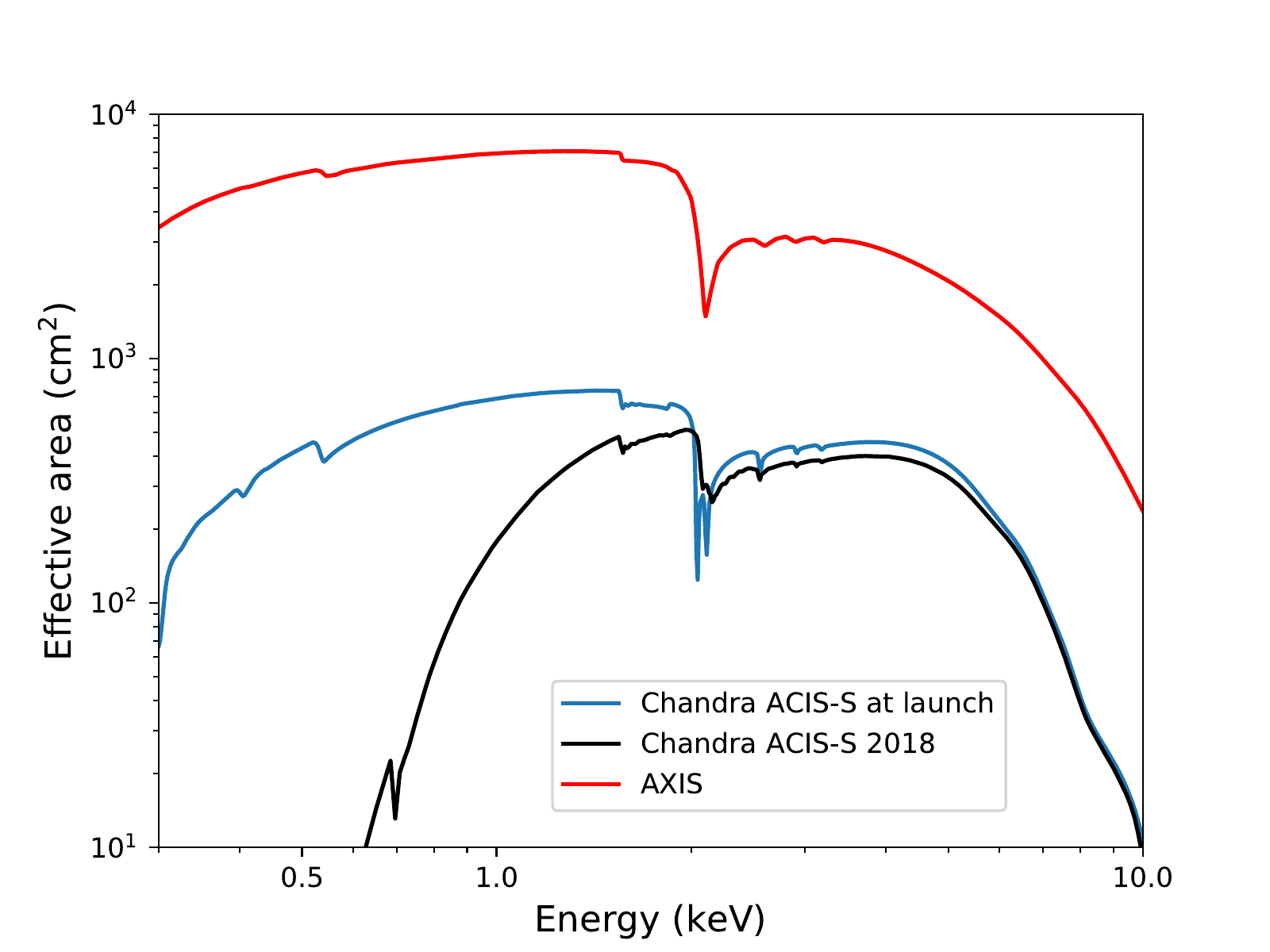}

\caption{\AXIS\ has larger effective area and better imaging
performance than \Chandra. Its on-axis PSF will have a 0.5\arcsec\ 
half-power diameter (HPD) at $E=1$ keV, of which 0.4\arcsec\ is the 
mirror contribution (\S\ref{sec:ops}). The optical design of the 
\AXIS\ mirror is optimized for a sub-1\arcsec\ PSF over the entire 
\AXIS\ field of view. \Chandra's on-axis HPD is 0.8\arcsec\ 
(0.6\arcsec\ mirror-only) and increases more steeply with off-axis angle.}
\label{fig:AXIS_effarea}
\end{figure}

The flowdown of science requirements into technical implementation is
summarized in Science Traceability Matrix (page 6--7); 
a summary of the \AXIS\ capabilities and science
drivers is given in Table~\ref{table:AXIS_mission_parameters}; and a
comparison with \Chandra{} with regard to the PSF and effective area is
shown in Fig.~\ref{fig:AXIS_effarea}. Figures of Merit for imaging, 
low surface brightness, and timing science are shown in Figs.\ A.4--6 
(page 5) in comparison with other current and future missions. The imaging capabilities of 
\AXIS, \Chandra{} and \Athena\ are
compared in  Fig.~\ref{fig:INTRO_lens} using an example of the 
quadruple  gravitational lens PG1115+080.

\section{TECHNICAL IMPLEMENTATION}

\AXIS\ achieves major scientific breakthroughs as a Probe class mission
through the combination of a major advance in X-ray optics, substantial
improvement in detector performance and X-ray filters, a straightforward and
cost-efficient operations philosophy in a LEO that minimizes
background and detector degradation while optimizing response time and
observing efficiency, and low-cost launch capability.

\begin{figure}[t]
\includegraphics[width=0.99\textwidth,viewport=85 90 675 910]{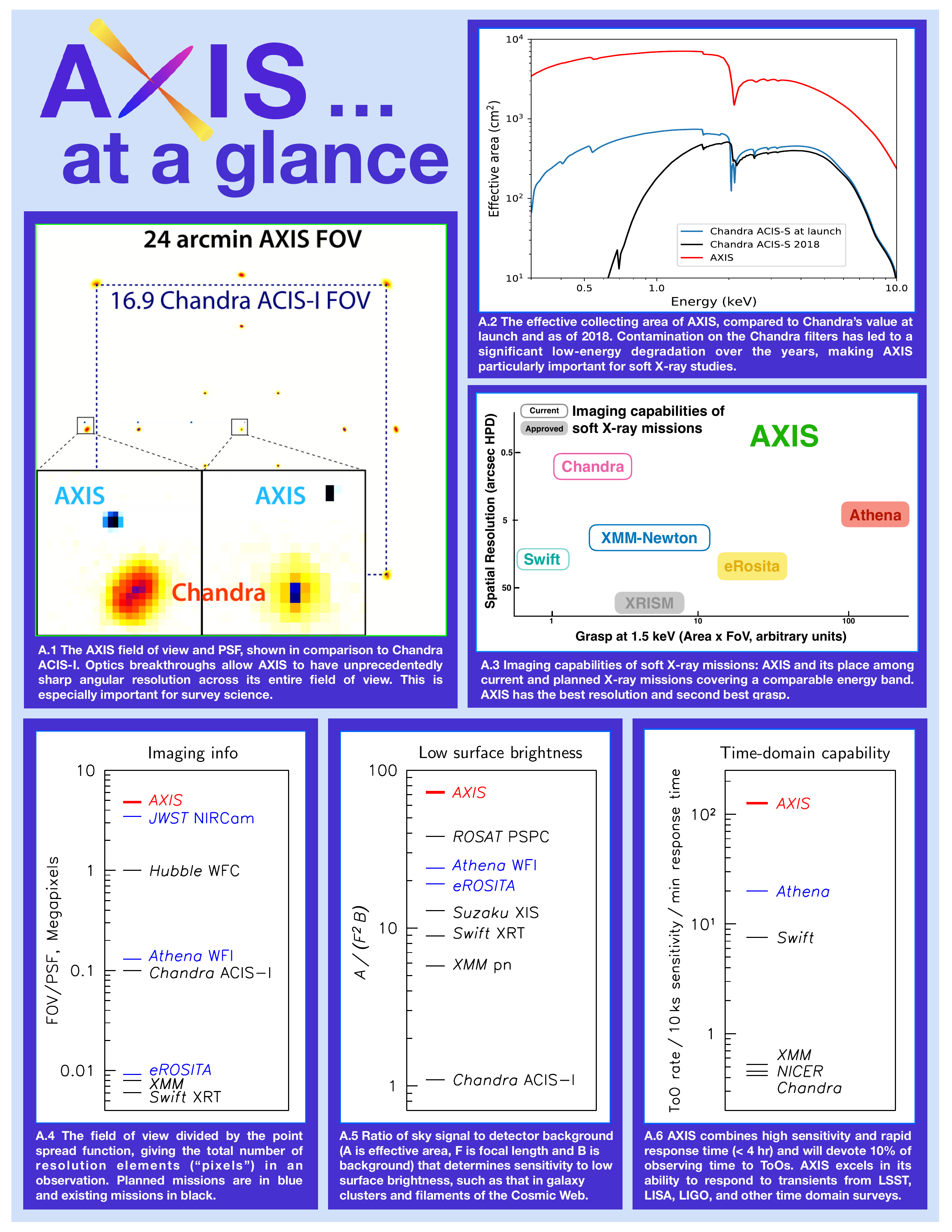}
\end{figure}
\begin{figure}[t]
\includegraphics[width=1.0\textwidth,viewport=71 81 541 723]{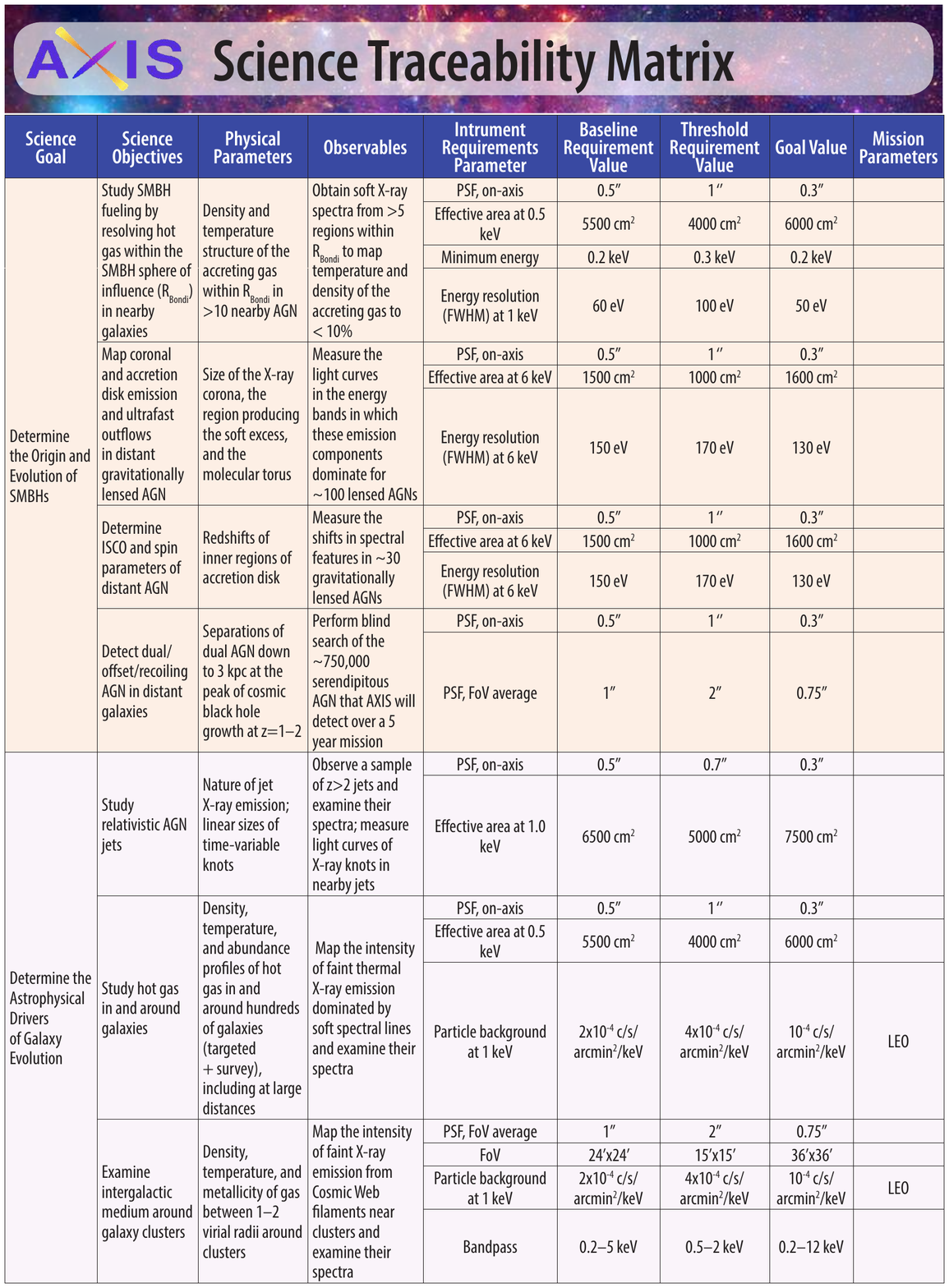}
\end{figure}
\begin{figure}
\includegraphics[width=1.0\textwidth,viewport=71 81 541 723]{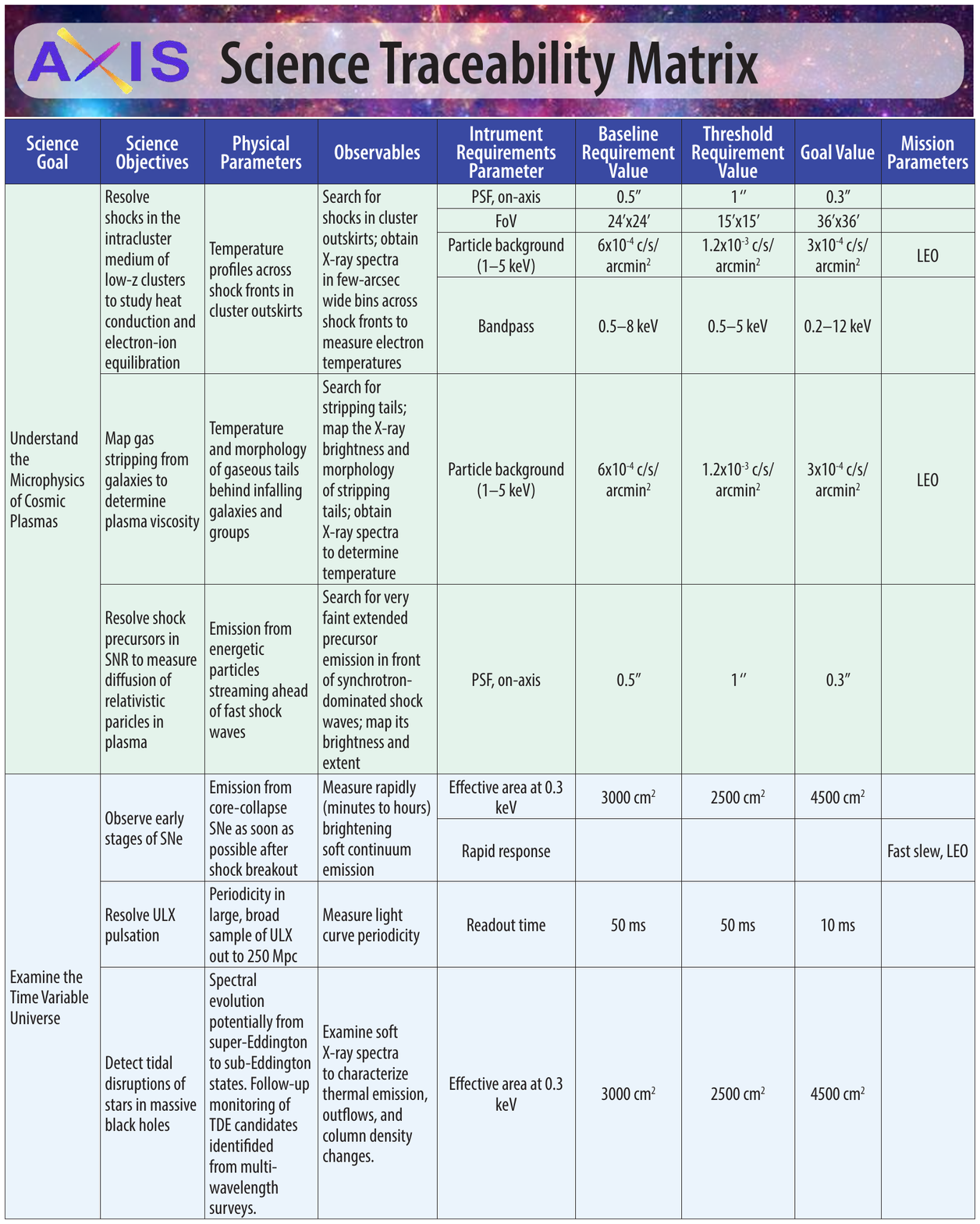}
\label{table:AXIS_STM}
\end{figure}

%

\subsection{Optics}

The \AXIS\ mirror assembly provides a major leap forward from the
state of the art represented by \Chandra{}. This dramatic improvement is
enabled by the silicon metashell optics (SMO) concept, and recent
progress that is the culmination of a $\sim$20-year technology development
effort.

The SMO technology is the same as baselined for the
\Lynx{} Flagship study. Mirrors of 0.5~mm thickness with extremely
accurate surface figure are fabricated from stress-free, mono-crystalline
silicon substrate using rapid, deterministic, precision 
polishing. Taking advantage of the availability of abundant, inexpensive
large blocks of mono-crystalline silicon and the equipment and knowledge
accumulated by the semiconductor industry to process it, this technology
provides a three-fold set of improvements over \Chandra{}: (1) better
angular resolution, on and off axis; (2) an order of magnitude reduction of
mass per unit effective area; and (3) an order of magnitude cost reduction
per unit effective area. \AXIS\ introduces aspects of Type-I
Wolter-Schwarzschild design to improve off-axis PSF performance over that of
the \Chandra{} Type-I Wolter design. Additional improvement of the off-axis
performance is obtained by introducing mirror elements with a smaller 
axial-length-to-focal-length ratio than \Chandra's.

The \AXIS\ mirror is entirely made of silicon, resulting in much 
larger tolerances on mirror on-orbit operating temperature and thermal
control and significant reduction in thermal environmental control during
construction, testing, and transportation on the ground. The highly 
modular approach to mirror construction is amenable to parallel 
production and straightforward de-scoping if needed. The mirror 
segment dimensions are similar to semiconductor industry silicon 
wafer dimensions, allowing the use of commercially available equipment 
and knowledge to minimize cost and production time. The major 
technology developments needed are the coating
of mirror segments to enhance reflectivity without figure distortion 
and their integration into mirror modules (aligning and bonding).

An angular resolution of 2.2\arcsec\ for a segment pair 
(1.3\arcsec\ at 1~keV if effects of gravity distortion and energy 
dependence are subtracted) was achieved during the most recent X-ray 
performance measurement in summer 2018. Mirror development is funded at \$2.4M a year by NASA for 
the foreseeable future, and the schedule anticipates TRL\,5 by the 
end of 2022, making mirror modules meeting the \AXIS\ requirement of 
0.3\arcsec\ HPD (each module). Assembling 188 modules into a mirror 
will yield HPD of 0.4\arcsec\  on-axis (\S\ref{section:mirrors}), 
and detector and aspect errors will result in an in-orbit PSF of 
0.5\arcsec\ HPD (\S\ref{sec:aspect}) --- an improvement over 
\Chandra's 0.8\arcsec.%
\footnote{\Chandra\ Observatory Guide, http://cxc.harvard.edu/proposer/POG}
The \AXIS\ PSF will stay sharper
than 1\arcsec\ over a $r=15'$ field of view, a great improvement over 
$r<2'$ for \Chandra\ (Fig.\ \ref{fig:AXIS_effarea}).

\subsection{Detectors}

The \AXIS\ focal plane exploits the high angular resolution, broad spectral
coverage, wide field of view, and high throughput provided by the optics, and
capitalizes on the extensive heritage from previous X-ray
observatories that utilized solid-state silicon CCD detectors of similar
size, quantum efficiency (QE), and spectral resolution to that required by
\AXIS.  The key technical challenges are pixel size, readout rate, and fast,
low-noise electronics to process and filter the onboard data stream. The
\AXIS\ LEO minimizes cosmic ray damage to the detectors.  Our
baseline conceptual design demonstrates the feasibility of using 
fast parallel-readout CCDs, the workhorse
soft X-ray imaging detector of the last 25 years, and CMOS active pixel
sensors, which are fast, low-power, radiation-hard devices with less
heritage, and our current design incorporates both. 
Considering both technologies at this stage minimizes
technical risk; we plan to closely follow their advances 
and select one type of the detector for the flight design.

The baseline \AXIS\ camera design uses four 1.5k$\times$2.5k CCDs to tile the majority
of the focal plane outside of the center, and a single, smaller 1k$\times$1k
CMOS in the center to minimize photon pile-up (when more than one photon hits the
same pixel between read-outs, which results in undercounting of photons and skews
detected photons to higher energy)
for observations
of bright targets. The CCDs are tilted to approximate the curved focal plane.  Both
detectors have 16\,$\mu$m (0.37\arcsec) pixels and take advantage of
sub-pixel positioning of the electron cloud that results from a photon hit,
thus adequately sampling the exquisite mirror PSF. Both types of detectors
are baselined to read out at 20~frames/s (64$\times$ faster than \Chandra\ ACIS
full-frame) to prevent pile-up and to allow timing of XRBs and 
transients at the 50~ms level. Both detector types have excellent QE and provide moderate,
near-theoretical spectral resolution across the 0.2--12~keV energy band.
With an $i \leq 8^{\circ}$ LEO, \AXIS\ will have $\sim$100 times less
non-ionizing radiation damage than \Suzaku{}. Radiation damage is further
ameliorated through the use of charge injection on the CCDs.

\subsection{Spacecraft and Operations}

\begin{wrapfigure}{r}{0.67\textwidth}
\centering
\vspace*{-5mm}
\includegraphics[width=0.67\textwidth]{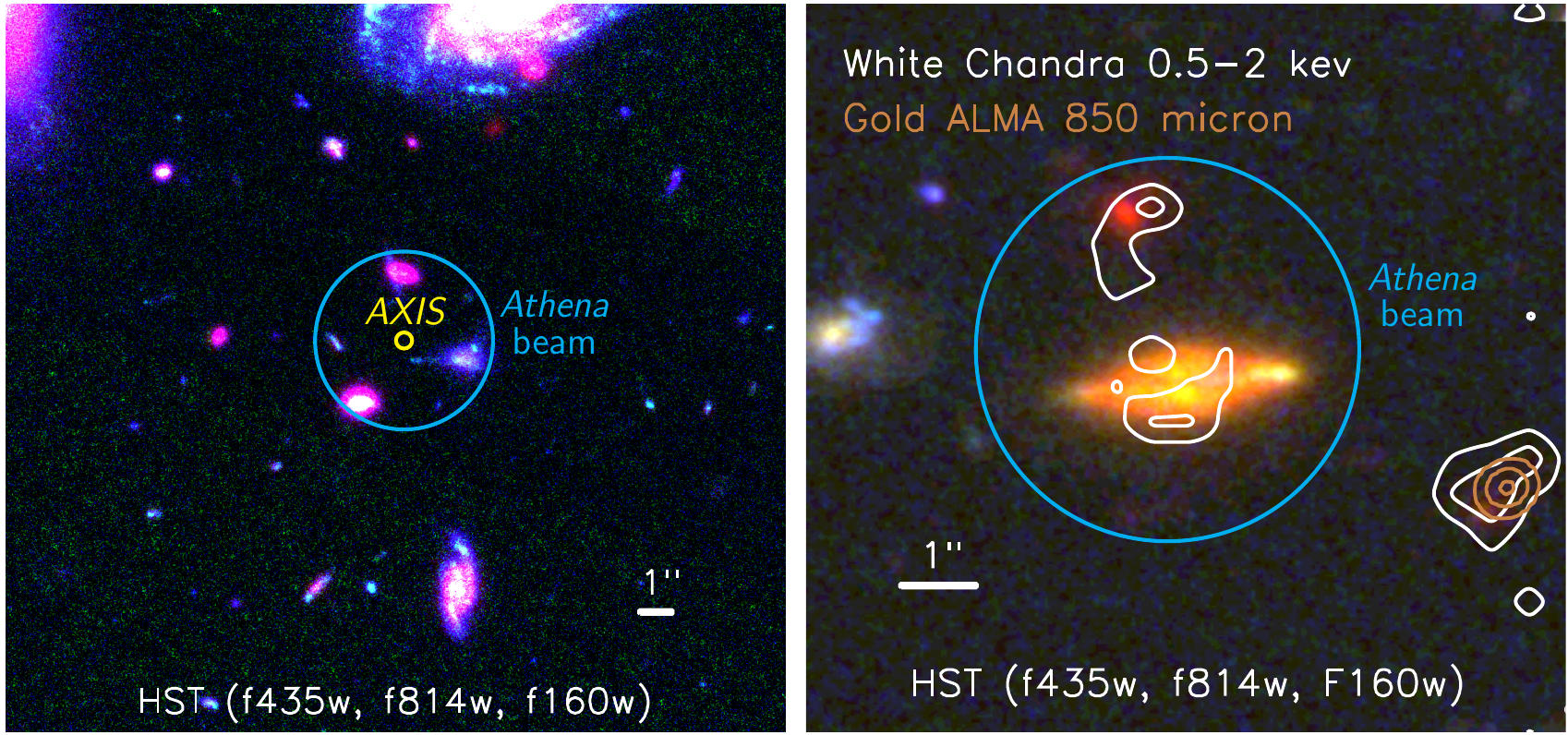}

\caption{\AXIS\ will pinpoint the ultra-faint X-ray sources to their 
high-redshift galaxy hosts, and to the exact location within the host galaxy.
The two panels show \HST\ deep survey images. {\em Left}: \AXIS\ and \Athena\ beam sizes. Unambiguous identification of a detected X-ray source with a 
distant galaxy requires high angular resolution. {\em Right}: \Chandra\/ 
and \ALMA\ contours\cite{Luo2017,Cowie2018} overlaid onto an
image of several closely projected galaxies. The central X-ray sources (white contours) require \Chandra's (and \AXIS') arcsecond resolution to separate
them and match to their very different galaxy counterparts. The \AXIS\ 
resolution is also well-matched to that of
\ALMA, a workhorse for studying high-$z$ galaxies in the sub\-mil\-li\-me\-ter.}
\vspace*{-2mm}
    \label{fig:2020s_ATHENA_confusion}
\end{wrapfigure}

A self-consistent point design for the \AXIS\ instrument system (X-ray
telescope and detector) was produced by the GSFC IDL study in February 2018, and
incorporated into the GSFC MDL mission point design.  
The design meets the \AXIS\ science requirements in a class~B 
mission with a five-year nominal lifetime (with consumables sized for
$>$10 years). The spacecraft uses entirely high-heritage components and
(pre-optimization) meets the mass (2300~kg including 20\% margin),
length, and diameter specifications for launch on a Falcon~9 into a
low-inclination ($\leq$8$^{\circ}$), LEO. The orbit minimizes 
the particle background and allows for rapid
communication and response times for transient events and a 
high observing efficiency of at least 70\% in a low inclination orbit. 

The design
requires six reaction wheels to accommodate the rapid slew rate
($120^{\circ}$ in $<$6~minutes) that optimizes efficiency. Three
magnetic torquers dissipate angular momentum using the Earth's
magnetic field. The only instrument mechanism is a focus adjuster that will be 
used only once, during commissioning.

\AXIS\ has a straightforward, high-heritage concept of operations based on 
over 30 years experience from \Chandra, \Swift, \NuStar\ and  \RXTE. The 
relatively low bit rate permits  an onboard storage and telemetry 
system  requiring only 4 Gbit/day downlinks using two 10-minute S-band ground 
station passes. The requirement for ToO response time is 4 hours,
similar to that performed by Swift; however, a response time as short 
as one hour is possible by taking advantage of TDRSS links.
The mission schedule
assumed for costing estimates a launch $\sim$7 years after the start of
Phase-A, assuming that TRL\,5 is reached before 2023.

\vspace*{5mm}\noindent
\centerline{\textcolor{black}{\sf\LARGE SCIENCE WITH \AXIS}}
\vspace*{-3mm}

\noindent
\AXIS\  will open a large uncharted 
parameter space in many fields of astrophysics. While it is impossible 
to predict what discoveries \AXIS\ will make, below 
we describe how it will address some of the most important problems
in astrophysics. A broader discussion can be found in the
proceedings of the 2018 \AXIS\ workshop.%
\footnote{http://axis.umd.edu/workshop2018.html}
Most of the \AXIS\ observing time will be allocated to the Guest Observer program.

\section{SUPERMASSIVE BLACK HOLES --- ORIGIN, EVOLUTION AND PHYSICS}
\label{section:bh_origin}

\subsection{The Puzzle of Early Massive Black Holes}

The origin of SMBHs in the centers of most big galaxies is a major unsolved problem. The existence of $\sim 10^9 M_{\odot}$ black holes only 700-800 million years after the Big 
Bang requires extremely rapid growth in the early universe. This is in contrast with what is observed in the
local universe, where most SMBHs grow at a sub-Eddington rate. The
nature of the seeds of these SMBHs is still unknown, since X-ray and
optical/near-IR surveys have only sampled the bright end of the AGN
luminosity function at $z>5$.

\begin{wrapfigure}{r}{0.5\textwidth}
    \centering
    \includegraphics[width=0.49\textwidth,viewport=40 16 552 336,clip]{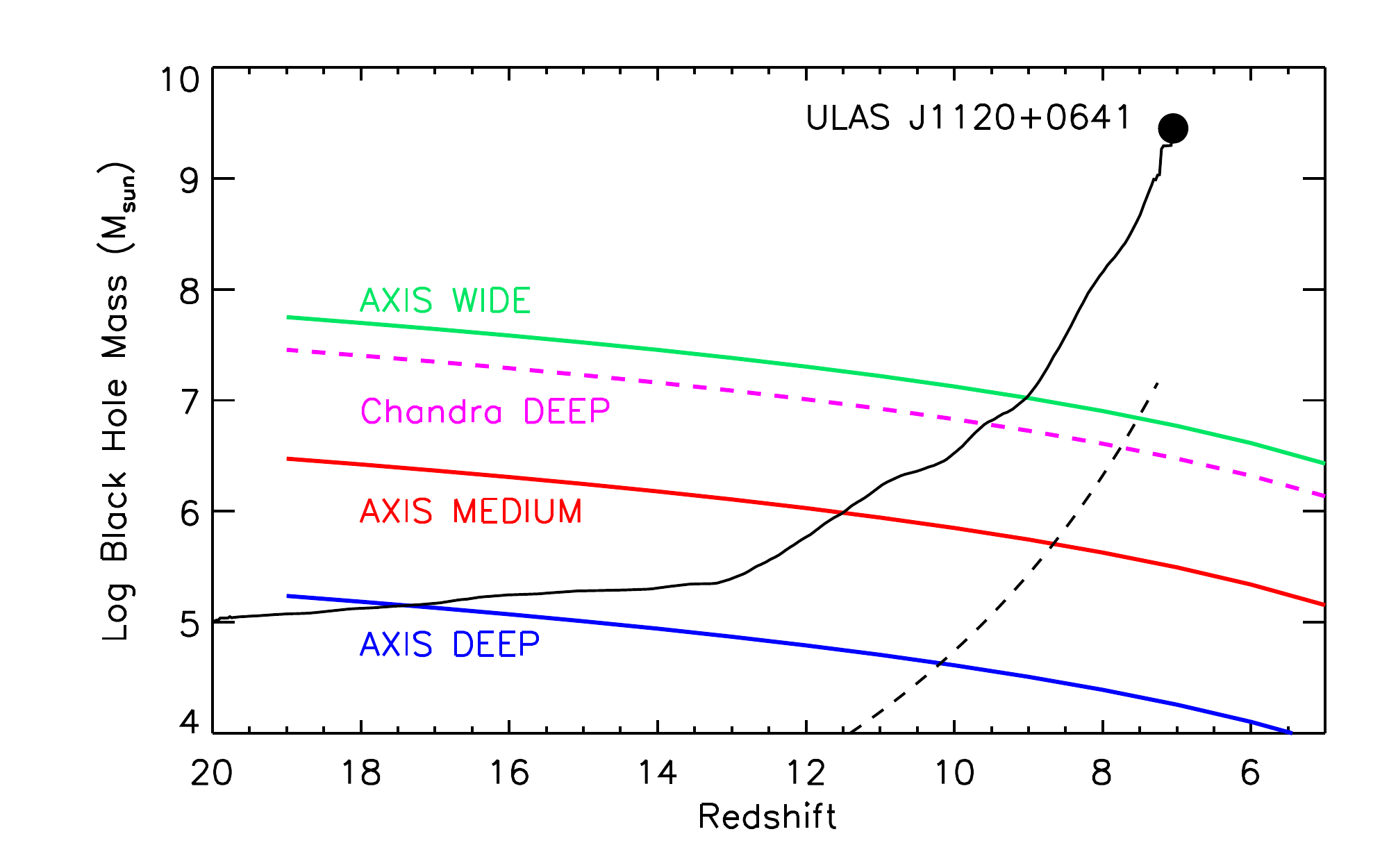}
    \caption{\AXIS\ will resolve the puzzle of high-redshift 
    SMBH. The dot shows the highest-redshift quasar ever discovered\cite{Mortlock2011}.
    Black solid and dashed lines show the mass buildup of a $z=20$ 
    $10^5 M_{\odot}$ and $10^3 M_{\odot}$ seed, respectively\cite{Pacucci2015}.
      The minimum detectable mass curves for the three
      proposed \AXIS\ surveys (wide, medium, and deep) show that \AXIS\ 
      can detect SMBH progenitors to very high $z$\/ for a broad range of masses.}
    \label{fig:SURVEY_BHgrowth}
\vspace*{-2mm}
\end{wrapfigure}

Currently the two main hypotheses for the origin of high-redshift
massive BHs are: (1) massive direct collapse from primordial gas ($10^{4-6}
M_{\odot}$) or (2) from Pop~III stellar remnants ($10^{2-3} M_{\odot}$). These seeds
would imply very different AGN luminosity functions at
high $z$\cite{Ricarte2018}. \AXIS\ can measure those luminosity functions 
by rapidly performing  surveys to very low flux levels. 
\AXIS\ observations will  measure
every stage of SMBH growth from $10^5 M_{\odot}$ at $z\sim 20$ up to
$10^{9-10} M_{\odot}$ at $z \sim 7.5$
(Fig.~\ref{fig:SURVEY_BHgrowth})\cite{Banados2018}.

The main observational challenge is the detection and 
identification of these objects.
While X-rays are the most efficient band for detecting AGNs, star
formation can significantly contribute to the X-ray luminosity of a galaxy,
requiring high angular resolution to locate the accreting
BHs in the objects selected by \AXIS\ and at other wavelengths. \AXIS\ provides a unique 
matching with ultra-distant (and likely irregular and merging) objects that will be discovered by \WFIRST, \JWST, \Euclid, \LSST, 
and 30m class optical telescopes.
\AXIS\ will be fundamental in (a) disentangling high-$z$
AGN progenitors from star forming galaxies\cite{Natarajan2017,Pezzulli2017},
(b) sampling the high-$z$ AGN luminosity function down to $L_X\sim
10^{42}$\lumergs\ at $z\sim 10$ (the luminosity of a $10^5 M_{\odot}$ BH
accreting at 0.1 Eddington) to fully characterize the BH mass assembly, and
(c) detecting low-mass SMBH at $z>10$.

 Since most very high $z$ AGN are expected to be obscured, it
will be difficult for \WFIRST\ or \LSST\ to disentangle the AGN signal from
star formation and to locate the BH within a clumpy, gas-rich,
dusty distant galaxy. Based on recent \ALMA\ data \cite{Cowie2018}, subarcsecond 
resolution is needed to obtain unambiguous X-ray confirmations for optically/NIR
selected SMBH seed candidates.

\subsection{Surveying the Distant Black Holes}

To hunt for high-redshift sources, \AXIS\ will allocate $\sim$10\% of its 
5-year mission (15 Ms) to surveys, overlapping with regions 
surveyed at other wavelengths. We envision deep, 
medium and wide surveys of 5 Ms each. The wide-field survey will select the
rarest objects, while the medium and deep surveys will detect the
faintest remote objects with the advantages of pre-selection of candidates
from either \JWST\ or \WFIRST. \AXIS\ will serendipitously find
$\sim 100$ of $z>6$ SMBHs, which will strongly constrain models of 
the origin of SMBH, and in particular, the  progenitors of the recently discovered 
$10^9 M_{\odot}$ quasar at $z\sim 7$ (Fig.\ \ref{fig:SURVEY_BHgrowth}).

\begin{wrapfigure}{r}{0.55\textwidth}
    \centering
    \includegraphics[width=0.5\textwidth,viewport=18 14 550 528,clip]{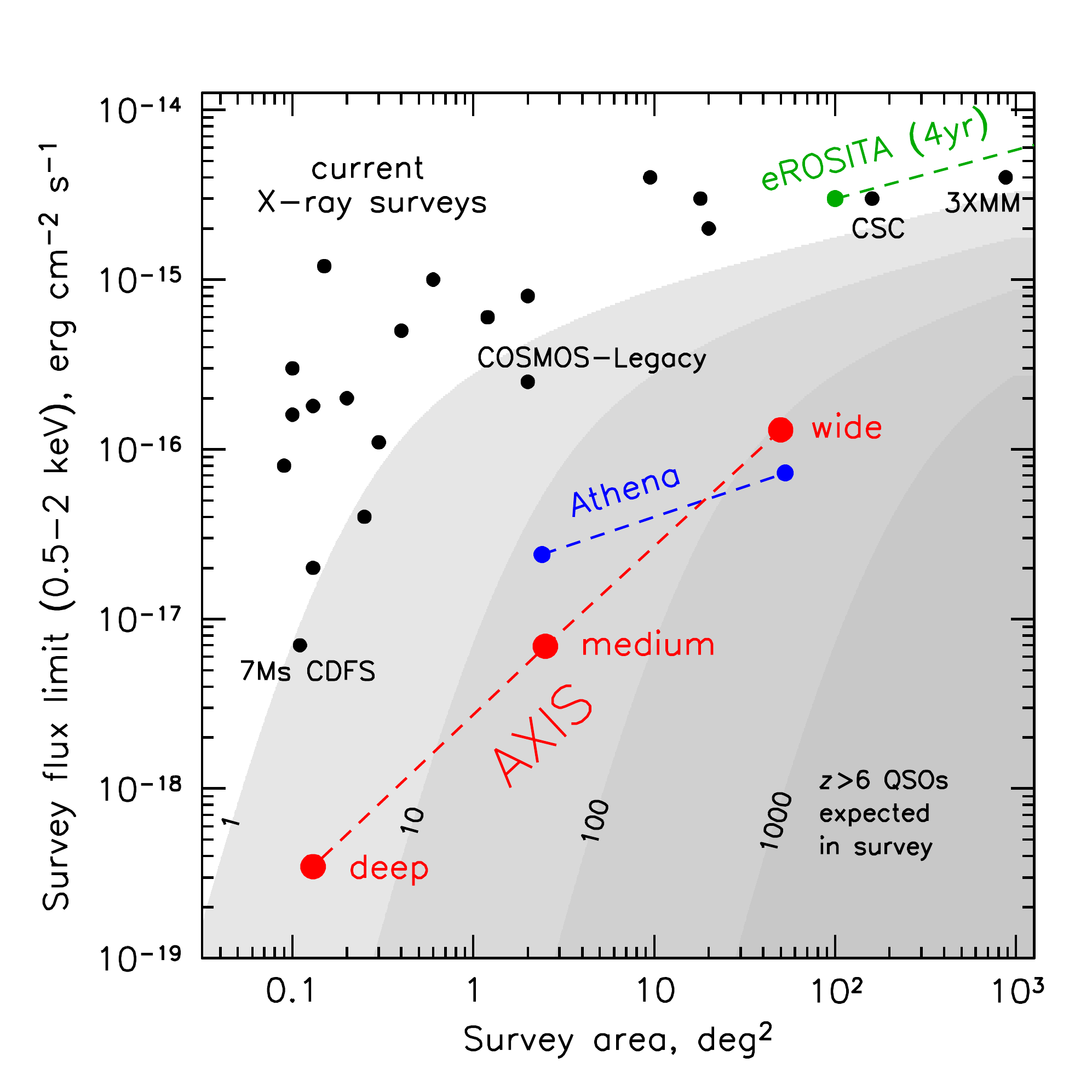}
    \caption{\AXIS\ will be able to
      probe more and fainter high-$z$ sources than any other X-ray instrument, including \ATHENA. The plot shows survey area vs.\ flux limit for  
      planned \AXIS\ surveys compared with \Chandra, \XMM\ and future 
      \Athena\ (1-year) and \eROSITA\ surveys. Grayscale shows the number of expected $z>6$ AGN detectable in the survey\cite{Vito2014,Vito2018}.
      \AXIS' subarcsecond angular resolution will provide unequivocal identifications for its detections (Fig.\ \ref{fig:2020s_ATHENA_confusion}).}
    \label{fig:SURVEY_flux}
\vspace*{-2mm}
\end{wrapfigure}

Figure~\ref{fig:SURVEY_flux} illustrates how \AXIS\ will exceed current and future
surveys in its ability to find the most distant BHs.
\AXIS\ will go lower than \Chandra{} in flux by 1-2 orders of magnitude and cover a larger solid angle, providing secure source identifications. While future large-area X-ray telescopes, such as
\Athena, will detect a similar number of brighter high-$z$\ AGN, their precise 
identification will require \AXIS' angular resolution. Figure
\ref{fig:2020s_ATHENA_confusion} illustrates how much easier it will be for 
\AXIS\ to pinpoint those X-ray sources within their host galaxies, allowing the location of the BH for studies at other wavelengths.

\begin{figure}[b]
\colorbox{callout}{\color{white}\sfsm
\begin{minipage}{0.99\textwidth}\begin{minipage}{0.97\textwidth}
\vspace*{3mm}
\begin{itemize}[itemsep=5pt,labelwidth=0pt,labelindent=0pt]
\item \AXIS\ will provide a complete history of SMBHs up
to {\em z}\,=\,6 with the deepest survey ever performed
\item \AXIS\ will observe the largest sample of ``infant'' X-ray emitting SMBHs right
after seeding at {\em z}$\,>\,$6, and potentially up to {\em
  z}\,=\,15--20
\item \AXIS\ will pinpoint the exact locations of SMBHs in distant irregular, merging host galaxies, and characterize the large-scale environment of early SMBHs
\end{itemize}
\vspace*{1pt}
\end{minipage}\end{minipage}}
\end{figure}

\AXIS\ surveys and serendipitous observations will detect $>10^{5}$ AGN across all redshifts. For a significant fraction of them, 
moderate signal to noise X-ray spectra will be
obtained and basic spectral quantities such as absorption and continuum shape will be measured. These data will determine
the AGN nuclear properties as a function of host and environmental (e.g.,
halo mass) properties, since in the 2030s deep and wide multiband surveys
will provide multiwavelength counterparts and
environmental information for most \AXIS\ sources. At the highest $z$, 
\AXIS\ will probe how
BHs grow within the first large-scale structures by detecting faint
companions of luminous AGN in the overdense regions. Simulations
suggest that large amounts of gas and galaxy mergers in those 
regions may trigger accretion. \AXIS\ will allow unprecedented
studies of how the AGN output affects the physics of their hosts
from the local to the distant Universe.

AGNs are a tracer of the large-scale mass distribution in the early
Universe\cite{Cappelluti2012}. The large number of detected AGNs in \AXIS\
surveys will map the large-scale structure for direct comparison with other
tracers of structure, over a wide range of redshift.

\subsection{Peering at the Vicinity of the Black Hole}
\label{section:bh_vicinity}

\AXIS\ will probe material very close to SMBH --- especially when aided by the Universe's own lenses. In the immediate vicinity of BHs, one of the fundamental energy sources in the Universe is at work ---
accretion of matter onto the BH. For all but the MW Galactic Center, the accretion disk is unresolvable by current 
instruments, but some quasars fortuitously have a massive 
galaxy or cluster on its line of sight, 
which results in gravitational lensing of the quasar. {\em Macrolensing}\/
is the gravitational bending of light produced by the mass
distribution of the foreground lensing galaxy. This lensing often produces 
multiple quasar images with different image flux ratios, lensing magnification, and
time delays. {\em Microlensing}\/ is the gravitational bending of light produced by
stars in the lensing galaxy\cite{Paczynski1986,Wambsganss1990} producing an additional magnification of the affected images. \AXIS\ will
revolutionize X-ray observations of lensed systems and let us peer into the 
immediate environs of massive BHs.

\subsubsection{Sizes of quasar X-ray emitting regions.}


\begin{figure}[b]
\vspace*{-2mm}
\colorbox{callout}{\color{white}\sfsm
\begin{minipage}{0.99\textwidth}\begin{minipage}{0.97\textwidth}
\vspace*{3mm}
\begin{itemize}[itemsep=5pt,labelwidth=0pt,labelindent=0pt]
\item \AXIS\ will constrain the sizes of X-ray emitting regions of quasars
ranging from the hot corona and inner accretion flow to the molecular and
dusty torus
\item \AXIS\ will constrain the evolution of the SMBH spins of quasars, and
provide magnified views of quasar outflows at the epoch of peak AGN activity
\end{itemize}
\vspace*{1pt}
\end{minipage}\end{minipage}}
\vspace*{-3mm}
\end{figure}

Gravitational microlensing can reveal structure within quasars on small
scales that cannot be directly imaged at any wavelength. It produces
caustics that magnify as they sweep over the quasar. The strength of the
magnification depends on the relative size of the emitting region and
caustic, with more compact sources having higher microlensing variability
amplitudes. Comparing the variability amplitudes in the X-ray and optical
bands places tight constraints on the structure of the accretion disk and the
X-ray emitting regions\cite{Kochanek2004}. \Chandra{} observations of
several bright lensed quasars revealed that the X-ray emitting corona is very
compact\cite{Pooley2009,Chartas2016}, $r\sim6-50r_g$ (where $r_g \equiv G
M_{\text{BH}}/c^2$).

These spectacular results beg for extension and enhancement. However, the
imaging and sensitivity limitations of \Chandra{} permits high quality data for only a few lensed
quasars.  \AXIS\ will
enable studies of a much larger sample. \AXIS\ will constrain the
sizes of X-ray emitting regions of quasars ranging from the hot corona and
inner accretion flow to the molecular and dusty torus that surrounds the accretion disk. 
Isolating these emission regions is possible because \AXIS\ has the effective area and angular
resolution to obtain the microlensing light curve as a function of energy as well as the flexible mission operations that allows the observation of these objects during the critical caustic crossings.

These \AXIS\ observations will be made vastly more productive by the discovery in the LSST survey of $>$4,000 additional gravitationally lensed systems with $\sim$300 quadruply lensed and $\sim$1000 double quasars
that are sufficiently X-ray bright to precisely measure with \AXIS. 
\AXIS\ will likely overlap \LSST\ or other optical surveys, allowing
the simultaneous monitoring of caustic crossings of quasars in
multiple optical and X-ray bands. 
\AXIS\ will resolve the vast majority of \LSST\ lensed quasars
over a wide range of quasar redshifts, BH masses, radio loudness values, spin and Eddington
ratios, allowing the  determination of the dependence of the sizes of X-ray emitting regions on these parameters. \eROSITA\ measurements of the X-ray fluxes of \LSST-lensed quasars will provide the sample of that are
bright enough to be {\em monitored}\/ by \AXIS.

\subsubsection{Evolution of SMBH spins in quasars at 0.5$\,<\,$z$\,<\,$5.}

The mechanisms leading to the growth of SMBHs will be
studied by constraining the evolution of the SMBH spins of quasars with $0.5
< z < 5$ with lensing observations.  
X-ray reflection from the inner regions of a quasar accretion disc produces
an iron line that is broadened by gravitational redshift and Doppler shifts\cite{Reynolds1997} 
and is commonly seen in nearby AGN. Quasar
microlensing produces energy shifts of the Fe~K line as microlensing
caustics sweep over the disk\cite{Chartas2017,Krawczynski2017}. Measurement of these shifts, in particular the maximum observed redshift, enables the determination of the innermost stable circular orbit (ISCO), spin, and inclination angle.
The factor of $\sim$10 increase in the collecting area of \AXIS\  and the availability of $\sim$300 X-ray bright, quadruply lensed
quasars from \LSST, will allow \AXIS\ to obtain spin measurements for a large sample of quasars with $0.5 < z < 5$.
 
\subsubsection{Imaging BH accretion disks using caustic crossing method.}

\begin{wrapfigure}{r}{0.67\textwidth}
    \centering
\vspace*{-0mm}
    \includegraphics[height=0.3\textwidth,viewport=1 1 240 224,clip]{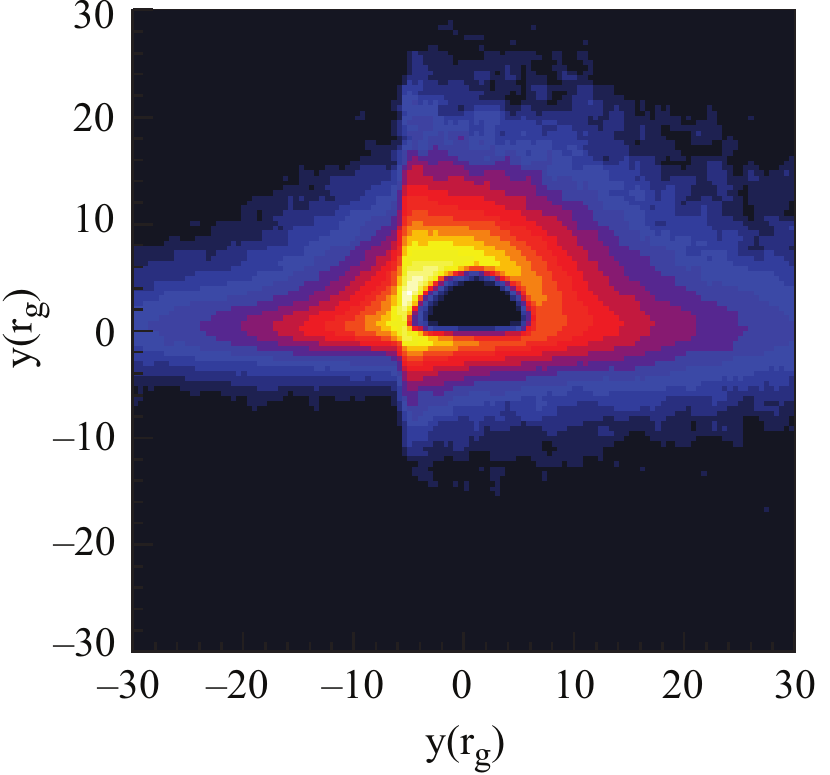}
    \includegraphics[height=0.3\textwidth,viewport=263 1 305 224,clip]{figures/LENS_caustic_cross_chartas_fig3_v2.pdf}
    \includegraphics[height=0.3\textwidth,viewport=320 1 514 224,clip]{figures/LENS_caustic_cross_chartas_fig3_v2.pdf}
    \caption{Gravitational microlensing caustic sweeping over a quasar 
    accretion disk. {\em Left}: Simulated image of the
      Fe-K$\alpha$ line emission as seen by a distant observer. 
      Axes are in units of gravitational radius ($r_g\equiv GM_{\rm BH}/c^2$). 
      {\em Right}: The resulting energy spectrum of the
      Fe-K$\alpha$ emission in the rest frame of the
      source (arbitrary units). Dashed line is the non-microlensed Fe line.}
\vspace*{-2mm}
    \label{fig:LENS_caustic_crossing}
\end{wrapfigure}

The detection of individual caustic crossing events would be spectacular,
revealing the gradual change in the profile and energy of the Fe line as the
caustic sweeps over the accretion disk and corona (see simulation in
Fig.~\ref{fig:LENS_caustic_crossing}). \LSST\ or other surveys will be continuously
monitoring many lensed quasars and thus can provide reliable triggers for
caustic crossing events. Monitoring caustic crossings events in individual
images with \AXIS\ will provide tomographic scans of the accretion disks of
SMBHs.

\subsubsection{Ultrafast outflows in quasars.~}

The detections of ultrafast outflows in distant quasars are rare due to
their X-ray faintness. \AXIS\ observations of a considerable number of the highly
magnified and thus X-ray bright lensed quasars at z$~$1-2 with
magnifications greater than  10 will provide detections of
high signal-to-noise absorption signatures of relativistic outflows at
redshifts during the crucial phase of BH growth at the peak of
cosmic AGN activity. 

\AXIS\ will provide spatially resolved and time-resolved spectra of lensed
images, thus constraining the properties of the outflow in individual
images. Lensed images provide spectra of the quasar at different epochs
separated by image time delays. Detecting the acceleration phase of the
outflowing absorber and constraining its short timescale variability cannot be accomplished with longer exposure times
using current X-ray missions; this requires a mission with
significantly larger collecting area combined with high angular resolution.
The spectral-timing analysis of a sample of high-magnification lensed high-$z$ quasars will thus constrain the energetics of ultrafast outflows near the peak of AGN activity. A comparison between the energetics of these small-scale
ultrafast outflows and the large-scale molecular outflows in the host
galaxies will be used to infer their contribution to regulating the
evolution of their host
galaxies\cite{Tombesi2015,Feruglio2015,Feruglio2017}.

\subsection{Growing a Supermassive Black Hole}
\label{section:SMBH_growth}

BHs grow by accreting matter around them and, rarely, by merging with other BHs. Our understanding of these phenomena will benefit enormously from \AXIS'
high angular resolution and sensitivity.

\begin{figure}[b]
\vspace*{-1mm}
\colorbox{callout}{\color{white}\sfsm
\begin{minipage}{0.99\textwidth}\begin{minipage}{0.97\textwidth}
\vspace*{3mm}
\begin{itemize}[itemsep=5pt,labelwidth=0pt,labelindent=0pt]
\item \AXIS\ will study BH growth in galaxy mergers and perform a blind search for dual AGN in over 750,000 AGN
\item \AXIS\ will study AGN fueling and resolve the accretion flow in the strong gravity region around at least 25 SMBHs in the local universe
\end{itemize}
\vspace*{1pt}
\end{minipage}\end{minipage}}
\vspace*{-5mm}
\end{figure}

\subsubsection{SMBH mergers.}
\label{section:dual_agn}

\begin{wrapfigure}{R}{0.5\textwidth}
    \centering
\vspace*{0mm}
 \includegraphics[width=0.5\textwidth]{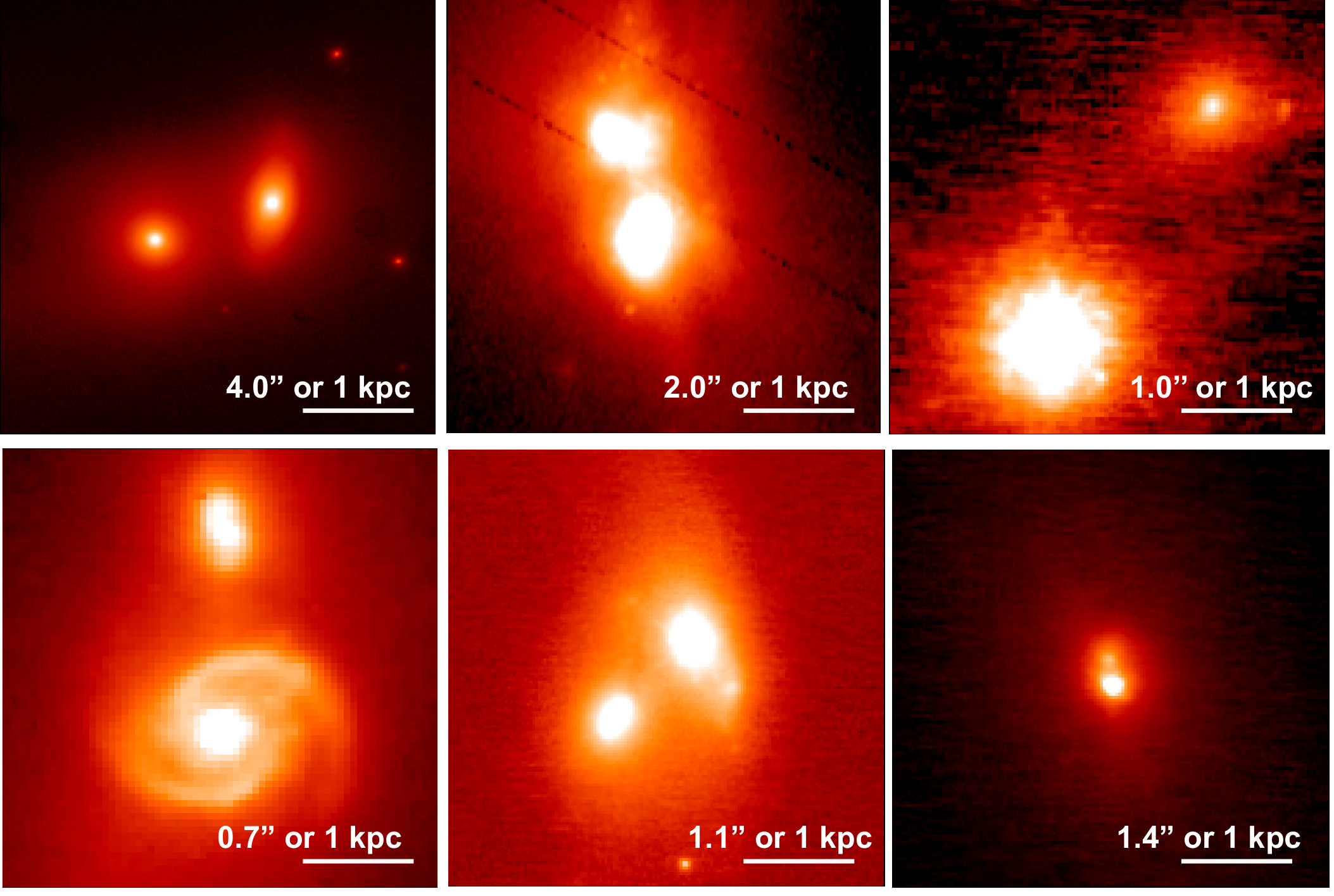}
    \caption{\AXIS\ will resolve potential dual AGN in all of these merger galaxies revealed in adaptive-optics images\cite{Koss2018}, while most are beyond \Chandra's capabilities. Future 30m telescopes and \JWST\ will find 
    many more close merging nuclei.}
    \label{fig:AGN_merging_nuclei}
\vspace*{-2mm}
\end{wrapfigure}
 
\begin{figure}[t]
    \centering
    \includegraphics[width=0.35\textwidth]{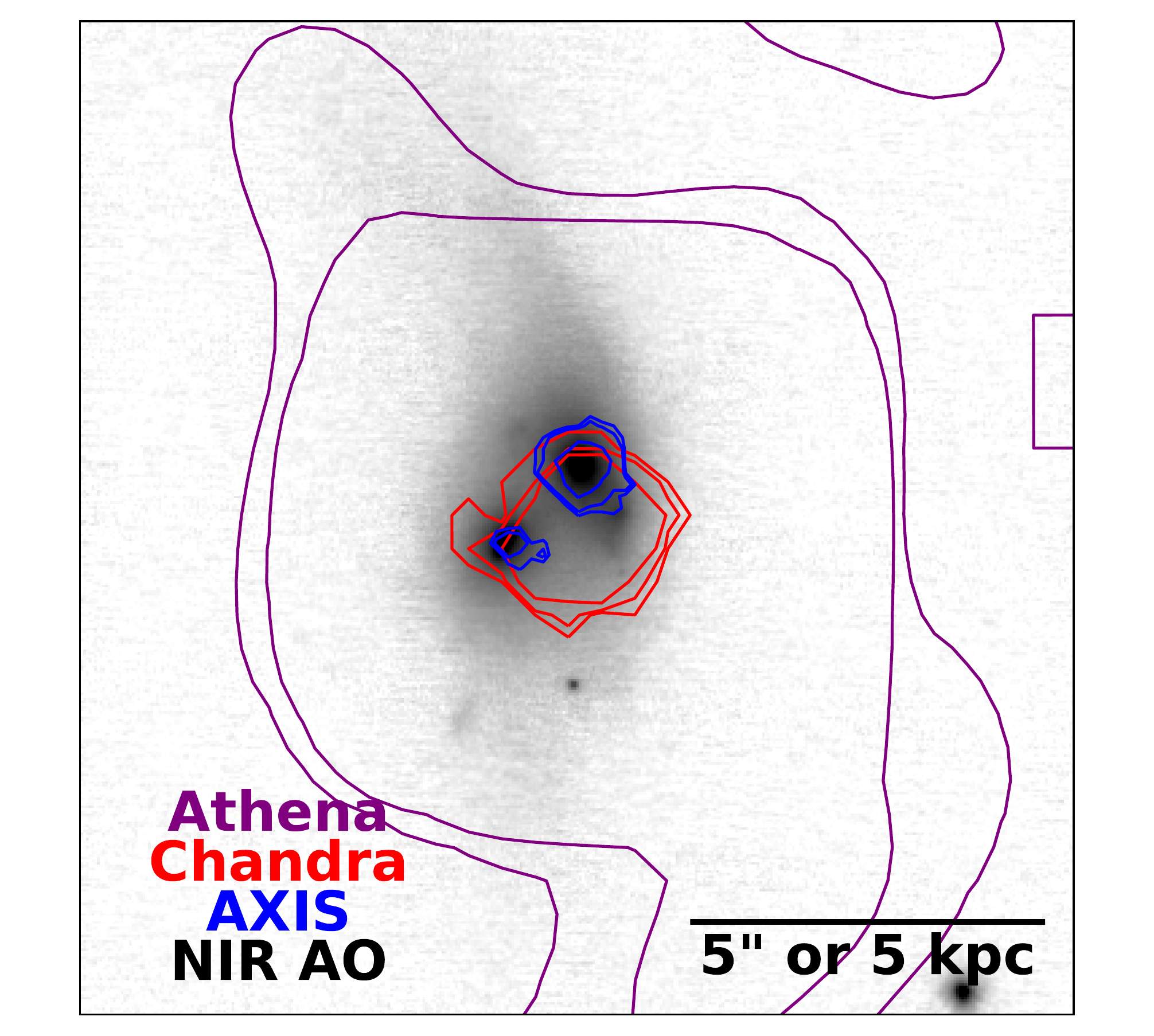}
    \includegraphics[width=0.50\textwidth]{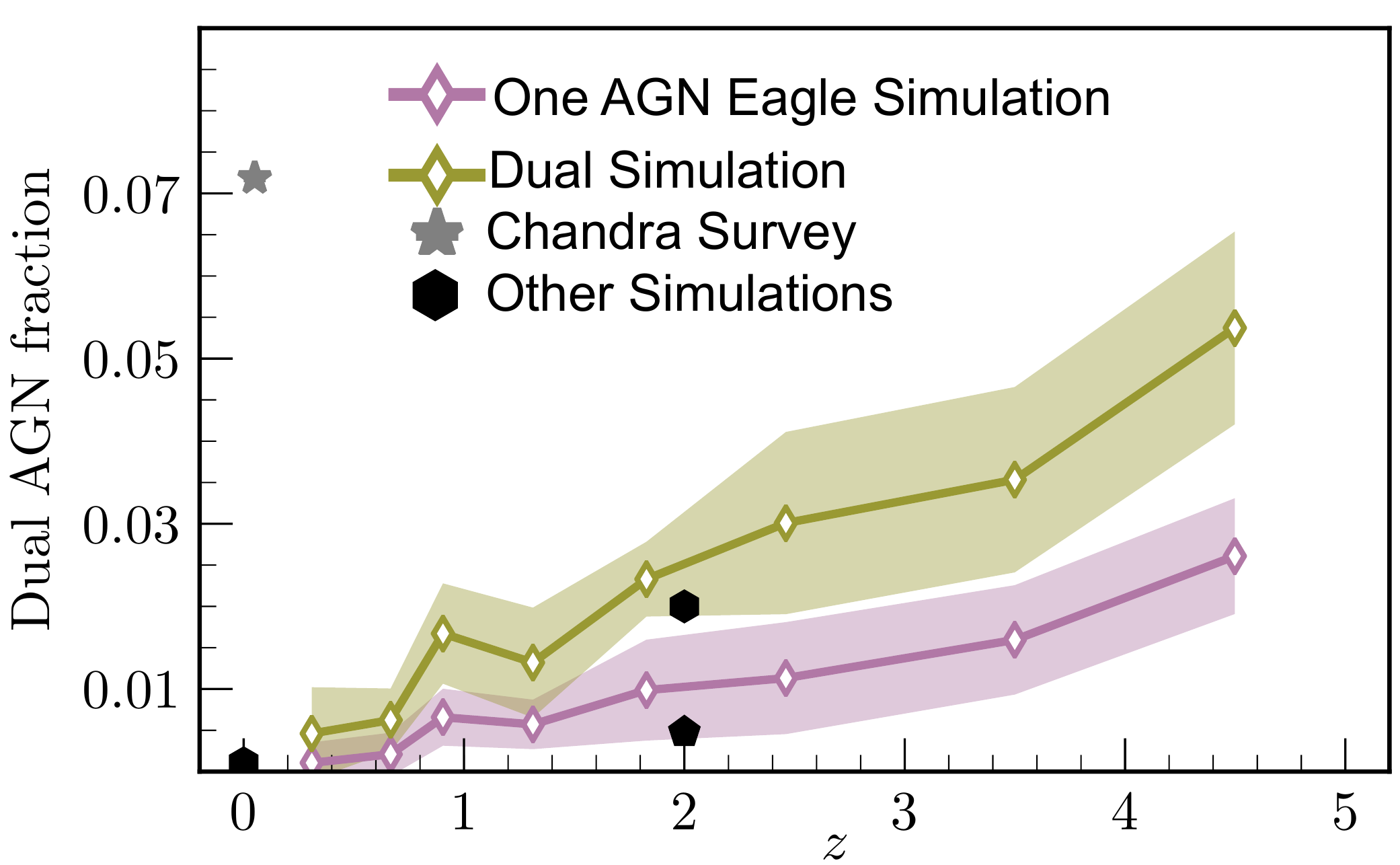}

    \caption{{\em Left}:  The improved PSF and sensitivity of \AXIS\ allows it to easily resolve the two nuclei in the high-resolution NIR image of the merging AGN (CGCG 341-006). The blue contours represent the \AXIS\ image, the purple \Athena, and the red \Chandra. {\em Right}: \AXIS\ will be able to probe the fraction of dual AGN at various redshifts and provide a direct test of the merger history of BHs. Cosmological 
    simulations predict that the fraction of dual AGN 
    increases dramatically with redshift\cite{Rosas-Guevara2019}. A 
    \Chandra\ study of nearby dual AGN\cite{Koss2012} finds a much higher fraction than models predict.}
    \label{fig:AGN_dual_survey}
\end{figure}

The general theory of large-scale structure formation predicts that galaxy
mergers are a major component of galaxy growth. Since almost all massive
galaxies at low redshift contain central SMBHs, it has
long been predicted that when the galaxies merge, their BHs should, too. 
However, SMBH merger timescales are
highly uncertain, and one of the few observational tests of this idea --- other
than gravitational waves at \LISA\ frequencies --- is to search for 
``dual SMBHs'' in nearby
galaxies\cite{Steinborn2016}. Theoretical calculations indicate that a
significant fraction of these sources are actively accreting, which has
stimulated an intensive search for dual AGN\cite{Satyapal2014}. However,
despite these efforts, there is currently little information about
the occurrence rate of dual AGN. Dual AGN are extremely rare in the
radio\cite{Burke-Spolaor2011}, and optical selection techniques for them are
rather inefficient\cite{Nevin2016}, as a large fraction of candidates are ``false
positives.''

In contrast, X-ray observations by \Chandra{} have discovered all three of the
unambiguous, kpc-scale, dual AGN systems known. This result implies a
dual AGN fraction of $>$10\% in the nearby hard X-ray selected
sample\cite{Koss2012}, 100x larger than studies of double peaked optical
sources ($\sim$0.1\%)\cite{Rosario2011}. While \Chandra{} is limited by its
resolution, sensitivity, and field of view in finding more examples, \AXIS\
will perform large surveys of deep fields roughly 300 times faster,
permitting population studies out to high redshift.
(Fig.~\ref{fig:AGN_dual_survey}).

Recent high resolution NIR observations of host galaxies of obscured AGN
have found that very close mergers may exist in a large fraction of obscured
luminous AGN\cite{Koss2018} (Fig.~\ref{fig:AGN_merging_nuclei}). All but one known confirmed dual AGN are separated by greater than 3~kpc
and have luminosities $L_X > 10^{42.5}$\lumergs. At these luminosities and
separations, \AXIS\ can detect dual AGN out to $z = 2$ in 50~ks observations,
allowing a blind search of the $\sim$750,000 serendipitous AGN that \AXIS\
will detect over a 5-year mission. Such data, along with targeted
observations of dual AGN candidates, will permit \AXIS\ to answer
critical issues such as the frequency, environment, and luminosity
dependence of dual AGN, and whether the obscuration level of AGN is
correlated with merger stage. The kpc-scale dual AGN population 
sampled by \AXIS\ provides key constraints for the SMBH mass merger function 
for the expected \LISA\ and pulsar timing array SMBH merger rates\cite{Kelley2016}.

\begin{figure}[tb]
    \centering
    \includegraphics[width=0.85\textwidth]{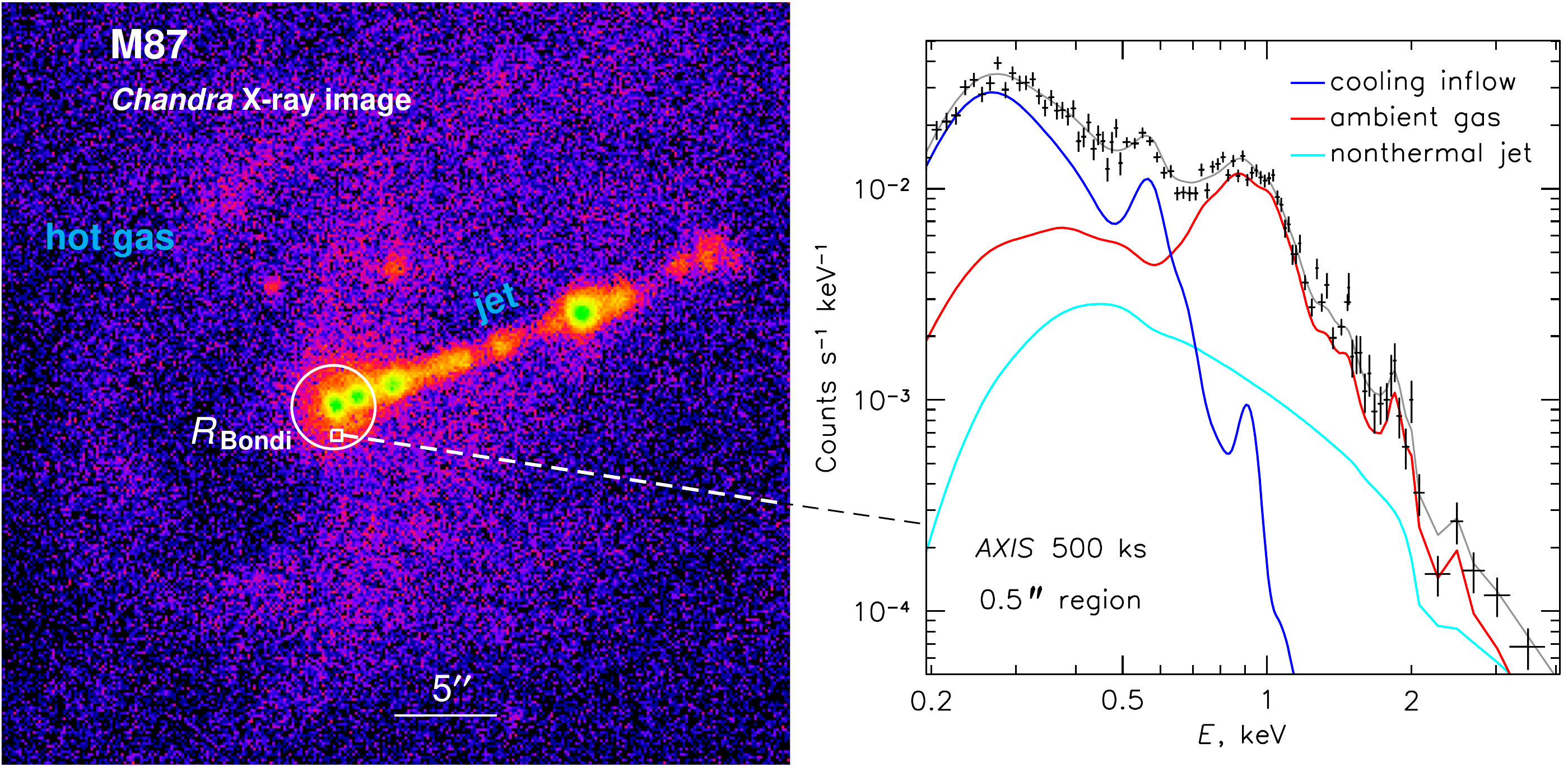}
    \caption{{\em Left}: \AXIS\ will map the accreting gas that fuels SMBH in great detail. The black
      hole's sphere of influence ($R_{\text{Bondi}}=2$\arcsec) is shown by the white circle on this \Chandra\ image of M87. \Chandra{}'s resolution and soft X-ray response limit studies of the gas properties within this region.  {\em Right}: A simulated \AXIS\ spectrum for a 0.5\arcsec$\times$0.5\arcsec\ region (left)
      from a 500~ks observation. \AXIS\ will map the
      rapidly cooling, inflowing gas with $kT<0.5$ keV (blue curve) that
      likely feeds the circumnuclear gas disk and fuels the jet activity.}
    \label{fig:AGN_m87}
\end{figure}

\subsubsection{AGN fueling.}
\label{section:bondi}

According to the simplest model of accretion, the Bondi model, the interstellar hot gas
in massive galaxies will be accreted by and fuel a SMBH if it falls within the Bondi radius, $R_B$, where the BH's
gravitational potential dominates over thermal motions of the gas.
Regardless of whether these assumptions are correct, the hot gas
structure within the SMBH's gravitational sphere of influence is key to
understanding how it is fueled, how it responds to the large scale gas
cooling rate, and how SMBHs evolved from the quasar-era peak to local quiescence.

The virial temperature of gas at $R_B$ for SMBHs is on the
order of $10^7$~K and emits in the X-ray band. This gas can only be studied with \Chandra{} in a few of the
nearest systems. The best sources, NGC\,3115 and M87,%
%
%
\begin{wrapfigure}{r}{0.5\textwidth}
    \centering
\vspace*{4mm}
    \includegraphics[width=0.45\textwidth,viewport=1 31 540 536,clip]{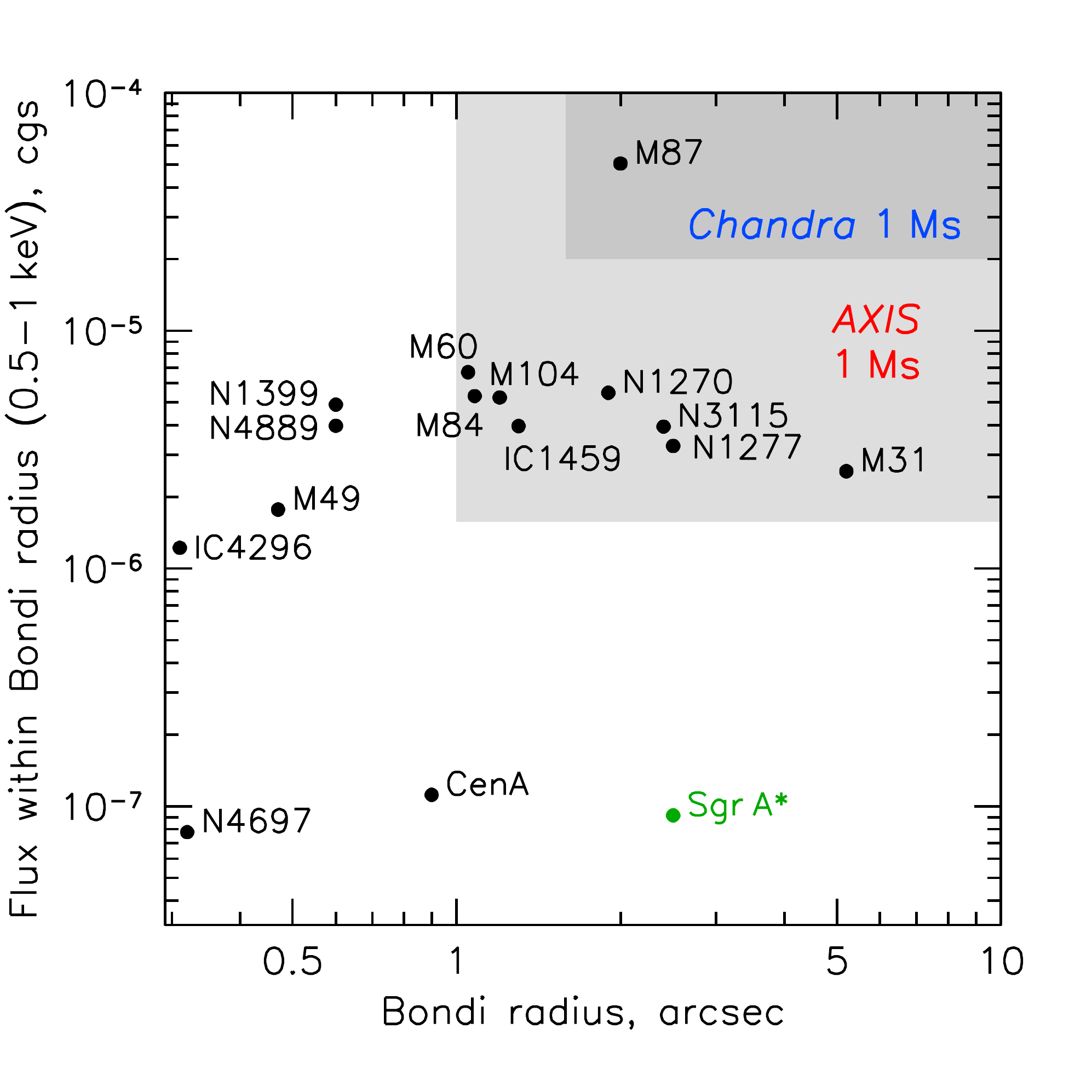}

    \caption{\AXIS\ will dramatically expand the pool of targets for which
      mapping of gas accretion onto the SMBH is possible. Soft-band fluxes
      from cool accreting gas within the gravitational sphere of
      influence of some nearby SMBHs are shown. Shaded regions show targets for which 
      $R_{\text{Bondi}}$ is resolved with \Chandra\ and \AXIS\
      and a 1~Ms observation produces $>10^4$ counts within
      $R_{\text{Bondi}}$, sufficient for detailed mapping of the gas inflow. 
      (Sgr A* is heavily absorbed and needs to be observed at higher energies.)
      }
    \label{fig:AGN_bondi_radius}
\vspace*{-5mm}
\end{wrapfigure}
%
%
were targeted by a
series of Large and Visionary \Chandra{}
projects\cite{Wang_q2013,Wong2014,Russell2018}. Surprisingly, the observations showed a shallow gas profile and declining temperature profile
in M87. This implies the presence of strong outflows that expel the vast majority of the gas
initially captured by the BH, dramatically reducing the accretion
rate.  Instead of $10^7$~K gas, the X-ray atmosphere is dominated by lower
temperature gas at $< 5\times 10^6$~K (Fig.~\ref{fig:AGN_m87}), which appears
spatially coincident with cool circumnuclear gas disks observed with HST.
Unexpectedly, the gas structure on these scales is a complex mixture of a
rapidly cooling inflow fueling the BH and powerful jet-driven
outflows. These can be disentangled with \AXIS\ imaging and spectroscopy.

\AXIS\ will reveal the emergence of the accretion flow
in the strong gravity region around at least 25 SMBHs, and with 
a high accuracy and level of spatial detail similar to or better 
than \Chandra's ultra-deep M87 study, for the galaxies in 
the shaded region in Fig.\ \ref{fig:AGN_bondi_radius}\cite{Garcia2010,vandenBosch2016}.
A potential few-Megasecond \AXIS\
observing program would target several key nearby SMBHs such as 
M87 with very deep exposures and a larger sample with shallower exposures. 
The ability to observe a reasonable sample of such objects in
100~ks is a driver on \AXIS' low energy collecting area and spectral
resolution.  These observations will map the detailed density and
multi-temperature structure within the BH's sphere of influence
(Fig.~\ref{fig:AGN_m87}) to reveal transitions in the inflow that
ultimately fuel the BH activity, and outflows along the jet-axis that
limit the accretion rate, thereby building up the first detailed picture of
these accretion flows.

\section{ASTROPHYSICAL DRIVERS OF GALAXY FORMATION}
\label{section:feedback}

Galaxy formation is governed by competing forces and processes. Gas accretes into a dark matter halo
under the force of gravity; most of it stays diffuse, but some cools down and forms stars, 
while some collapses into a central massive black hole. Stars explode as supernovae, while
the SMBH turns into an Active Galactic Nucleus (AGN) and  
produces powerful relativistic jets and X-ray radiation. Both processes inject energy into 
the newly arriving gas --- heating it and expelling some from the galaxy, thus preventing it
from forming stars and depriving the SMBH of fuel. The result of this
feedback loop is the multitude of galaxies and galaxy clusters that we observe.
\AXIS\ will greatly advance our understanding of these fundamental processes.

\subsection{AGN Jets}
\label{section:jets}

\begin{wrapfigure}{r}{0.5\textwidth}
    \centering
    \vspace*{-10mm}
    \includegraphics[width=0.47\textwidth]{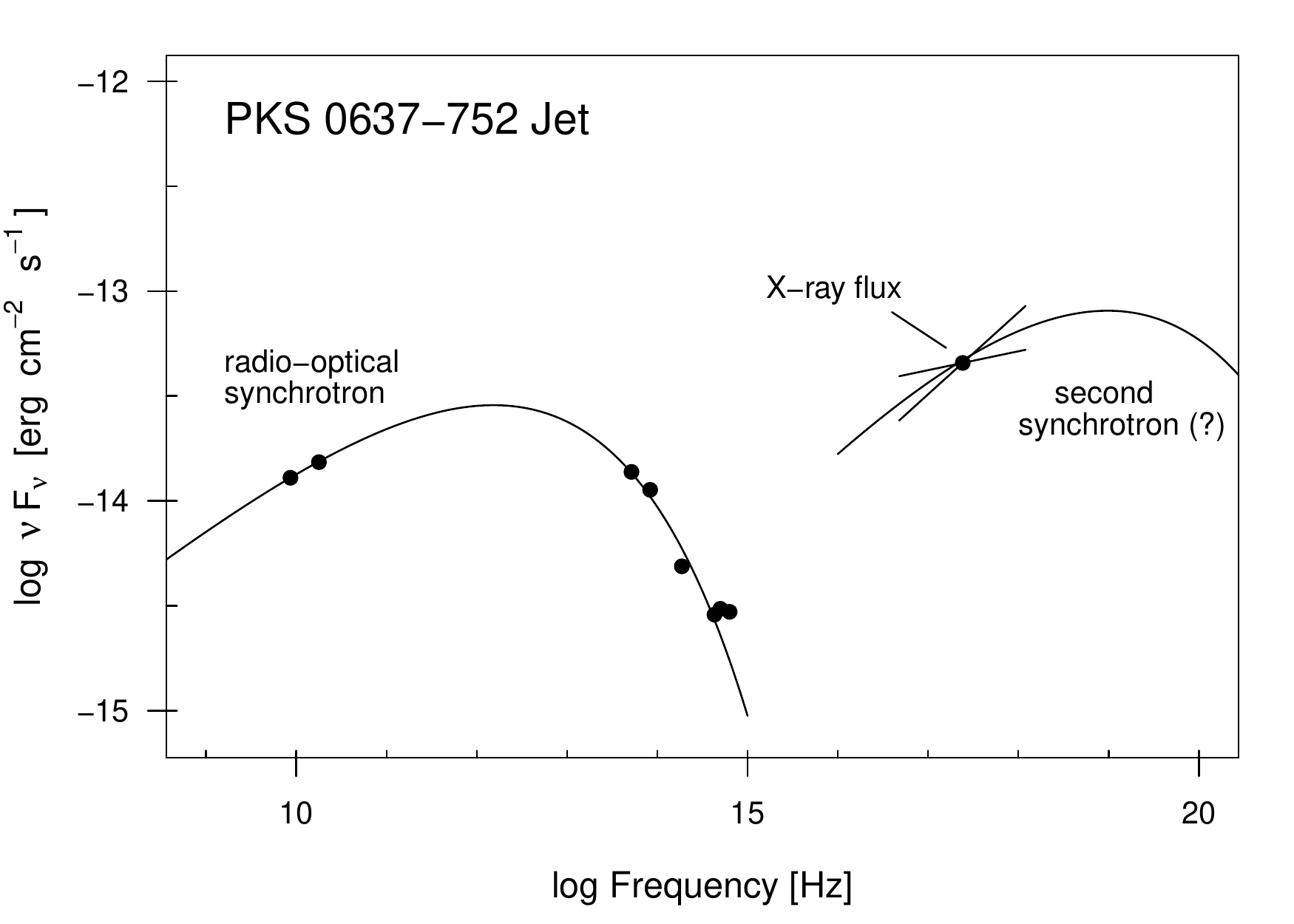}
    \caption{X-ray observations of AGN jets often reveal an unexpected emission component, possibly synchrotron emission from a second, higher-energy population of electrons than produces the radio-optical component, as shown in this SED for the bright jet in PKS 0637-75\cite{Meyer2015}. \AXIS\ will obtain  X-ray data of such quality for objects out to high redshift and over a wide range in AGN and jet parameters.}
    \vspace*{-2mm}
    \label{fig:JETS_SED}
\end{wrapfigure}

\begin{figure}[b]
\vspace*{-1mm}
\colorbox{callout}{\color{white}\sfsm
\begin{minipage}{0.99\textwidth}\begin{minipage}{0.97\textwidth}
\vspace*{3mm}
\begin{itemize}[itemsep=5pt,labelwidth=0pt,labelindent=0pt]
\item \AXIS\ will provide the first sensitive observations of high-redshift jets
\item \AXIS\ will collect the photons necessary to resolve the X-ray spectrum along the jet and conduct variability studies, probing the currently unknown particle acceleration mechanism
\end{itemize}
\vspace*{1pt}
\end{minipage}\end{minipage}}
\end{figure}

Relativistic jets are a crucial link in the feedback loop between 
SMBH and the gas that feeds it. Their physics is extremely complex.
After nearly four decades of study of extragalactic jets,
there are still shockingly basic questions about many of their fundamental
properties, including how the jet properties connect to properties of the
BH (whether spin or mass or some property of the disk), the particle
content of the jets, the exact emission mechanism producing high energy
photons at kpc scales, and how the radiating particles were
accelerated\cite{Blandford2018}. This lack of insight produces a large 
uncertainty in estimates of the total power output of jets, thereby
limiting robust measurement of the impact of jets on their surroundings.

\Chandra{} observations have produced major discoveries, which while based on relatively 
few objects, have significantly changed our understanding of jets such as the detection of an 
``anomalously'' high and hard X-ray flux from a second
spectral component in almost all jets; the discovery in two cases
of time variability in some of the ``knots'' in the jet\cite{Marshall2010,Snios2019};
and from a detailed analysis of Cygnus-A data, the implication that jets are
light, implying that the kinetic power and momentum flux are carried
primarily by the internal energy of the jet plasma rather than by its rest
mass\cite{Snios2018b}.  Extension of these unexpected discoveries to much
larger samples covering a wide range in jet and BH properties, over a
wide range in redshift, requires a significant improvement in both sensitivity
and high-resolution ($<$1\arcsec) imaging.

The bulk of the X-ray emission comes from synch\-ro\-tron emission, as
alternative models such as inverse Compton scattering of the CMB have been
ruled out\cite{Stawarz2004,Hardcastle2006,Meyer2017,Breiding2017}. However,
the synchrotron emission is produced by a separate population than the
electrons which produce the radio through optical
emission\cite{Georganopoulos2016} and makes a significant contribution to
the total jet energetics. The existence of 
this component requires \textit{in situ} particle acceleration (via unknown mechanism) on
kpc scales. The
recent observations of short timescale X-ray variability in resolved
jets\cite{Marshall2010,Hardcastle2016} implies extremely short radiative
loss lifetimes, on the order of years, and thus very small source sizes and
efficient particle acceleration. \AXIS\ can perform a detailed statistical
study of the variability in the jet population which cannot be done by
\Chandra{} because of the very long exposures needed to obtain sufficient
statistics on individual knots in the jets. Only high-resolution,
high-sensitivity X-ray imaging can help us understand the underlying
physical conditions and processes, and probe the currently unknown particle
acceleration mechanisms at play in AGN jets.

\subsubsection{\AXIS\ discovery space for AGN jets.}

\begin{wrapfigure}{r}{0.5\textwidth}
    \centering
    \vspace*{-4mm}
    \includegraphics[width=0.47\textwidth]{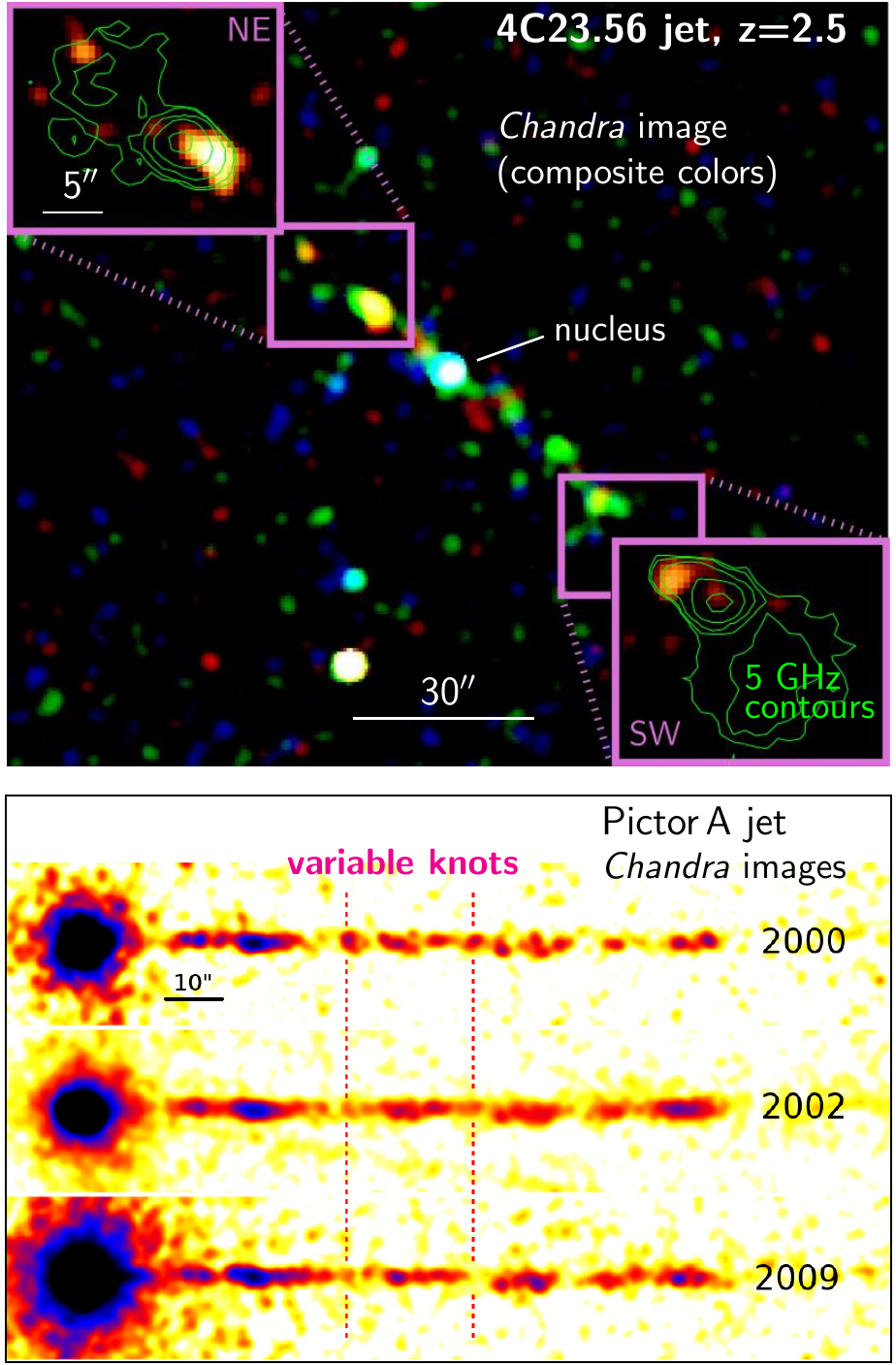}

\vspace*{0.3mm}
    \caption{{\em Top}: \AXIS\ will have the sensitivity and resolution to study high-redshift jets in great detail. Because of the $(1+z)^4$ brightening of the CMB, inverse Compton emission from cosmic-ray electrons in high-$z$\/ jets starts to dominate, which makes these faint objects invaluable for understanding the complex physics of jets. This \Chandra\ image of a $z=2.5$ jet\cite{Blundell2011} has only a handful of photons. {\em Bottom}: \AXIS\ will be able to study jet variability on much shorter timescales than \Chandra. Variable X-ray knots in a well-resolved nearby jet are shown\cite{Marshall2010}; dashed lines indicate knots seen in 2000, which have disappeared in later epochs.}
    \label{fig:JETS_images}
    \vspace*{-6mm}
\end{wrapfigure}


\hspace*{7mm}$\bullet$ \textit{High redshift:} At high redshift ($z>2-3$), X-rays from inverse Compton upscattering of the CMB start to dominate over the synchrotron, and the environment around those jets is vastly different than in the low-$z$ Universe. This provides a unique environment in which to study the jet physics. However, all of the high-$z$\/ jet detections by \Chandra{} have only a handful of counts in the resolved jet. \AXIS, with its angular resolution and large area at low X-ray energies, is an ideal tool for these studies.

$\bullet$ \textit{Spectral evolution along the jet:} So far this has only been
done for a few objects, with contradictory results. The X-ray spectrum is
one of our best clues as to the particle acceleration mechanism, but at
present we have insufficient data to draw strong conclusions.

$\bullet$ \textit{Jet population:} \AXIS\ will sample a large fraction of the  jet population.
The statistics of jet X-ray luminosity and its relation to the radio
emission will constrain the emission mechanism and the bulk speeds of the
X-ray-emitting material.

$\bullet$ \textit{Variability:} \Chandra{} required exposures of more than 50~ks
to find the few X-ray variable jets known.  With the 10x better sensitivity
of \AXIS, variability studies using `snapshot' (5~ks) observations will
allow for studies of populations and measurements of timescales.

%
$\bullet$ \textit{Jet-ISM interaction:} \AXIS\ observations can examine
interactions between the jet and the ISM or IGM that it is traveling
through, allowing measurements of jet-induced star formation as seen in
\Chandra{} observations of Cen-A and NGC4258. They can also examine possible heating mechanisms required in the so-called radio mode feedback models (\S\ref{section:feedback}).

\AXIS' resolution is well-matched to that of existing and upcoming radio facilities (\textit{JVLA}, \textit{SKA}), and complements ground-based optical facilities. It also supports the high-resolution observations of optical synchrotron counterparts from \HST\ and, in the future, \JWST.

\subsection{How Black Holes Heat Galaxies and Clusters}
\label{section:agn_feedback}

\begin{figure}[b]
\vspace*{-1mm}
\colorbox{callout}{\color{white}\sfsm
\begin{minipage}{0.99\textwidth}\begin{minipage}{0.97\textwidth}
\vspace*{3mm}
\begin{itemize}[itemsep=5pt,labelwidth=0pt,labelindent=0pt]
\item \AXIS\ will resolve the length scales over which AGN feedback 
operates in the hot cluster medium

\item \AXIS\ will measure the evolving importance of AGN feedback in
clusters to {\em z}$\sim$3

\item \AXIS\ will map the impact zones where hot AGN winds meet the ISM
\end{itemize}
\vspace*{1pt}
\end{minipage}\end{minipage}}
\vspace*{-3mm}
\end{figure}

\begin{wrapfigure}{R}{0.67\textwidth}
    \centering
    \vspace*{-2mm}
    \includegraphics[width=0.67\textwidth]{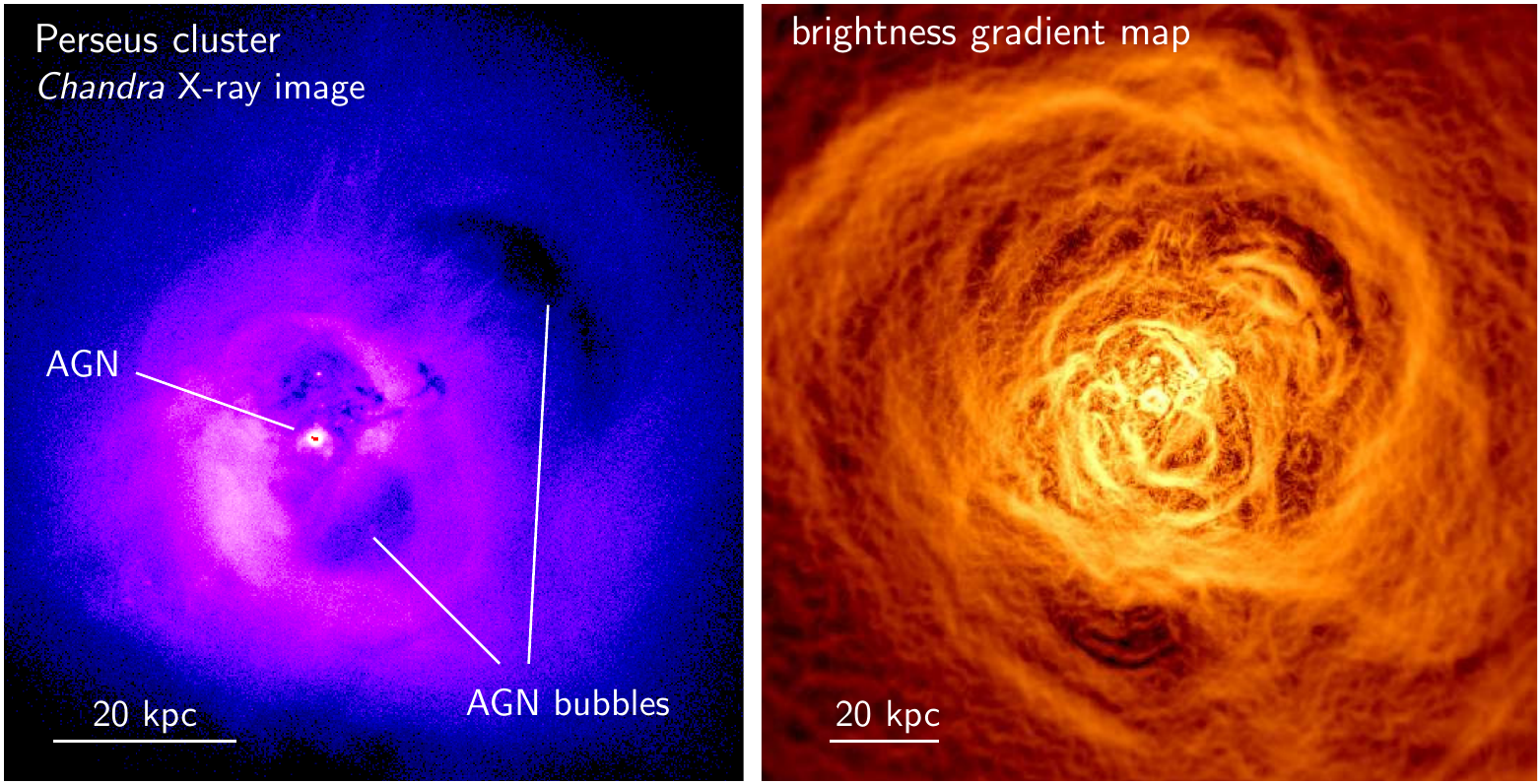}

    \caption{\AXIS\ will obtain maps for many at the level of detail similar to this ultra-deep \Chandra{} image of the nearby Perseus cluster {\em left}. {\em Right}\/ panel shows the surface brightness gradient map\cite{Sanders2016}. Rich structure seen in the hot gas can be due to many different things ---
      acoustic waves, stratified turbulence, or multiple old
      AGN bubbles. \AXIS\ will provide critical information needed to understand the processes that govern 
      feedback between the AGN and the surrounding gas.}
    \label{fig:FEEDBACK_perseus}
\vspace*{-2mm}
\end{wrapfigure}

Even though SMBH's have a mass of only $\sim$0.1\% of their host galaxy, the enormous
energy released as it grows (and shines as an AGN) can heat up and/or expel
star forming gas from the galaxy, thereby slowing or completely truncating
the galaxy's growth\cite{Faucher-giguere2012}. However, the physical processes
by which the energy and momentum are transferred to cooling and star-forming
gas remain unclear. Revealing how this transfer operates, how it scales with galaxy
mass, and how these mechanisms evolve with redshift
is essential to understanding galaxy and cluster evolution. It requires 
resolving complex structures in the hot gas, which contains most of the
feedback energy, on the relevant physical scales (arcseconds for all redshifts). 

The most massive systems in the Universe, clusters of galaxies, allow the
observation of AGN feedback in action. In the vast majority of relaxed,
cool-core clusters, the brightest central galaxy (BCG) hosts a radio-loud
AGN\cite{Burns1990}. \Chandra{} and \XMM{} imaging finds evidence for strong
interactions between the jets from the central AGN and the hot intracluster
medium (ICM) of the cluster---jet blown cavities/bubbles are ubiquitous, and
signs of weak shocks, acoustic waves, and AGN-induced turbulence are also
common. These interactions probably cause the AGN to heat the ICM core,
thereby offsetting radiative cooling and preventing a cooling catastrophe.
The fact that most galaxy clusters possess self-similar ICM temperature and
entropy profiles suggests that a self-regulated feedback loop is established
whereby some residual cooling fuels the AGN which then heats the ICM core
sufficiently to offset 90-99\% of the radiative cooling.

While the energetics of AGN-ICM feedback make sense, current X-ray data have
failed to reveal the actual physical mechanisms by which the jets heat the
ICM or cooling fuels the AGN. High-resolution \Chandra{} images of nearby
clusters such as Perseus and Virgo reveal rich structure
(Fig.~\ref{fig:FEEDBACK_perseus})\cite{Sanders2016}, but, because of the
small sample size and limited statistics, it remains unclear how to
interpret this structure\cite{Zhuravleva2018} --- are we seeing a cluster
core full of acoustic waves, stratified turbulence, or numerous
pancake-layers of old bubbles? Each of these have very different
implications for the physics of ICM heating and how it will depend upon the
density, temperature, and plasma physics of the ICM.  High-resolution
spectroscopy with \XRISM\ is expected to provide important constraints on
core-integrated cluster dynamics in several bright systems. 
\AXIS\ maps of densities, temperatures, and non-thermal emission on 
0.5--10\arcsec\ scales, in combination with high spectral resolution data from \ATHENA for a large sample spanning a wide range of
redshifts, masses and cooling rates, will revolutionize our understanding of
these structures.

\begin{wrapfigure}{R}{0.43\textwidth}
    \centering
    \fcolorbox{black}{black}{
    \includegraphics[width=0.32\textwidth]{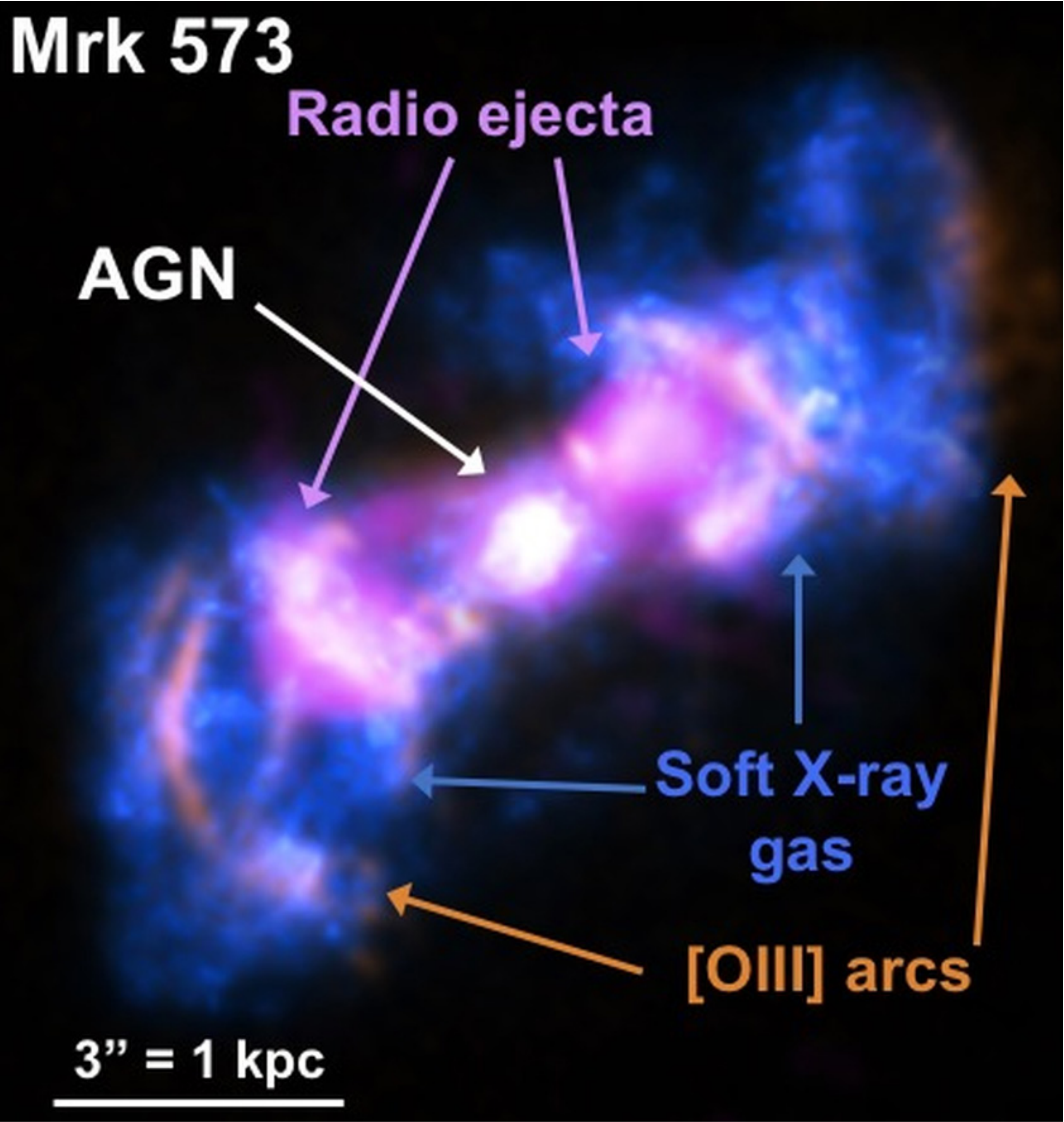}}
    \caption{\AXIS' angular resolution and sensitivity will 
      allow the precise investigation of the multi-phase and
      multi-scale interstellar medium in the innermost regions of Seyfert galaxies and probe SMBH feeding/feedback for a large sample of AGN. Three-color composite image of the central
      10\arcsec$\times$10\arcsec\ region of the Seyfert galaxy
      Mrk~573 is shown\cite{Paggi2012}; soft X-rays from \Chandra{} (blue), radio from \VLA\ (purple), optical from \HST\
      (gold).}
    \label{fig:FEEDBACK_Mrk573}
    \vspace*{-1mm}
\end{wrapfigure}

While observations of a few local clusters have enabled the study of the
physics of this ``kinetic mode'' feedback in great detail, we need to
understand the cosmic evolution of AGN feedback in clusters over a wide
range of cosmic time and mass scale. The best studied case beyond the local
Universe, the Phoenix cluster\cite{Russell2017} ($z=0.6$), possesses
significant quantities of cold gas and AGN-blown X-ray cavities, suggesting
that vigorous AGN-feedback has already been established. However,
observations at these or higher redshifts require extremely long \Chandra{}
exposures, producing a limited very small sample\cite{Hlavacek-larrondo2013}.

There is a strong trend for AGN in BCGs to increase in luminosity with
increasing look-back time, suggesting a possible switch from radiative-mode
to kinetic-mode feedback at some redshift.  If this switch does occur, is it
a function of cluster mass, cosmic time, or some other parameter? To answer
these questions requires large samples of clusters at high redshift with
excellent angular resolution and signal-to-noise ratio. \AXIS\ observations will
build an extensive sample of clusters with sufficiently deep imaging
spectroscopy to see ICM cavities and shocks.

These kinetic-mode feedback processes should also operate in galaxy groups
and individual massive elliptical galaxies. New evidence \cite{Babyk2018} suggests that
they also operate in spiral galaxies, thus implicating them in the formation
of all massive structures. There is a universal scaling of the
entropy profiles of the ICM in clusters, the hot ISM in massive elliptical
galaxies, and the CGM in spiral galaxies at low redshift\cite{Babyk2018}. In
the ICM case, the profile is shaped by self-regulated AGN-feedback,
suggesting that similar processes are at work in the CGM of all galaxies.  A
direct study of these processes is not possible with \Chandra{} --- the CGM
has very low surface brightness and emits primarily below 1~keV where \Chandra{s} sensitivity has been severely reduced.
\AXIS\ will enable the study of feedback in the CGM of individual, $L_*$ 
spiral galaxies within 100-200~Mpc, measuring the distribution of
temperature, metals, entropy, and mass to $R<50$~kpc and relic bubbles 
within $R<20$~kpc. The combination of \Athena's high spectral resolution and 
\AXIS' high angular resolution will reveal how feedback operates over the 
$10^4$ mass range from massive galaxies to rich clusters.

AGN can also influence their galaxies in an impulsive, violent manner.  During
a luminous outburst, an AGN can drive a powerful wind that destroys or
expels molecular gas from the galaxy. This ``radiative-mode'' feedback is
thought to rapidly suppress star formation in a galaxy. Direct evidence for
this phenomenon, however, is elusive. X-ray spectral evidence for high
velocity AGN winds in the form of highly blueshifted ionized iron lines is
commonly seen. At present there are only a very small number of cases in
which the momentum flux of the AGN wind can be related to that of a
large-scale molecular outflow from the galaxy\cite{Tombesi2015}, providing
strong circumstantial evidence that the AGN is indeed responsible for
driving molecular gas out of the galaxy. The combination of \AXIS, \Athena, and \ALMA\
data can increase the sample substantially, allowing tests of theoretical
models of this process\cite{Faucher-giguere2012}: \AXIS\ can spatially
resolve the shock interaction between the ISM and AGN wind in $>20$
nearby systems, while \ALMA\ can provide velocity-resolved molecular
gas maps, and \Athena\  high spectral resolution X-ray
spectroscopy at 5\arcsec\ scales, thereby constraining models of 
radiative-mode feedback. The intrinsic
multi-phase and multi-scale structure of AGN feeding and feedback is
demonstrated by the composite X-ray, optical, and radio image of
Mrk~573\cite{Paggi2012} (Fig.~\ref{fig:FEEDBACK_Mrk573}), one of the few such data sets with \Chandra{}, showing the hot gas phase interacting and
mixing with colder phases.

\subsection{Galaxies Across Cosmic Time}
\label{section:galaxies}

\begin{figure}[t]
\vspace*{2mm}
    \centering
    \includegraphics[width=0.4\textwidth]{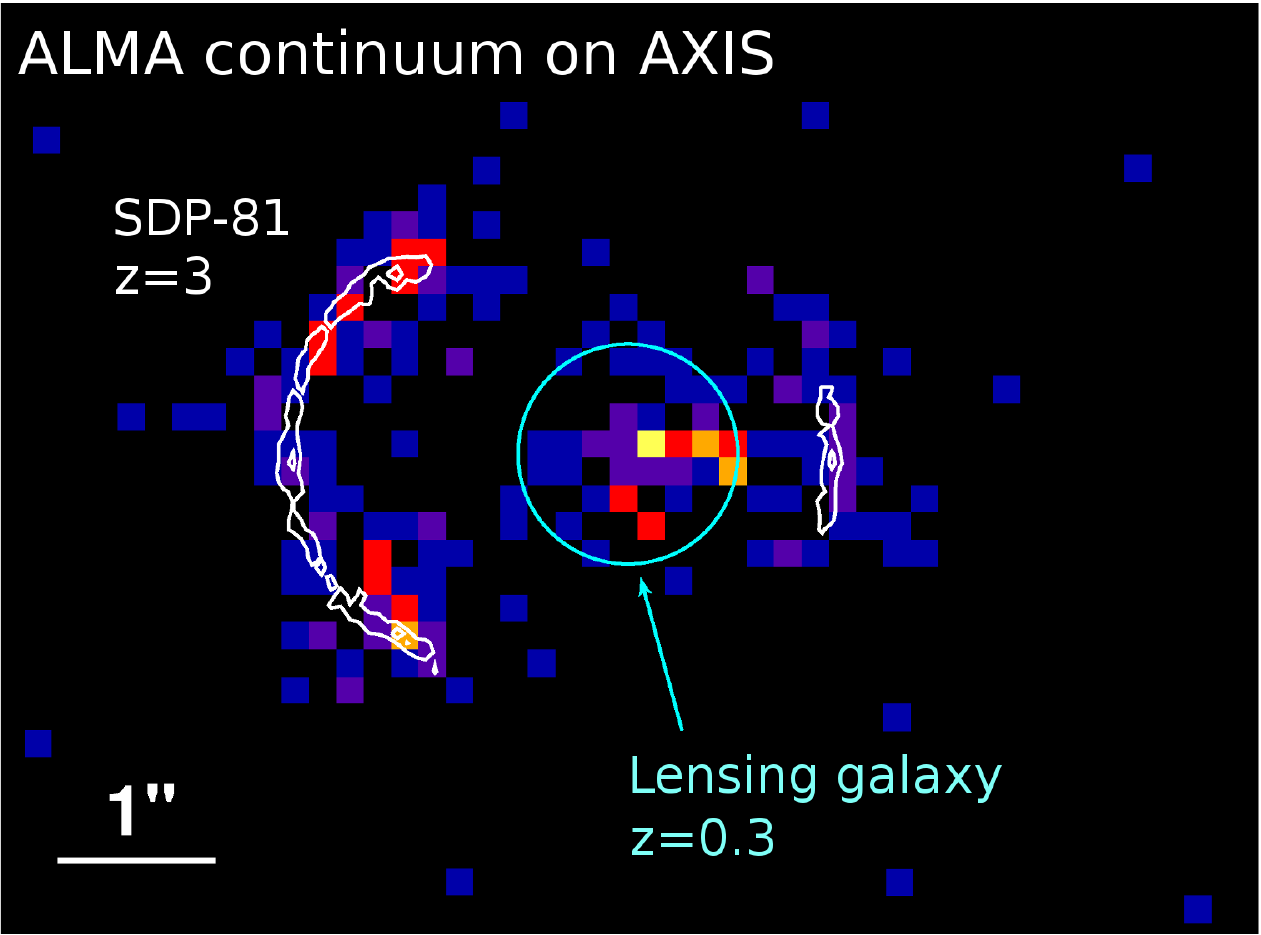}
    \hspace*{5mm}
    \raisebox{-3mm}{
    \includegraphics[width=0.42\textwidth,viewport=13 10 352 261,clip]{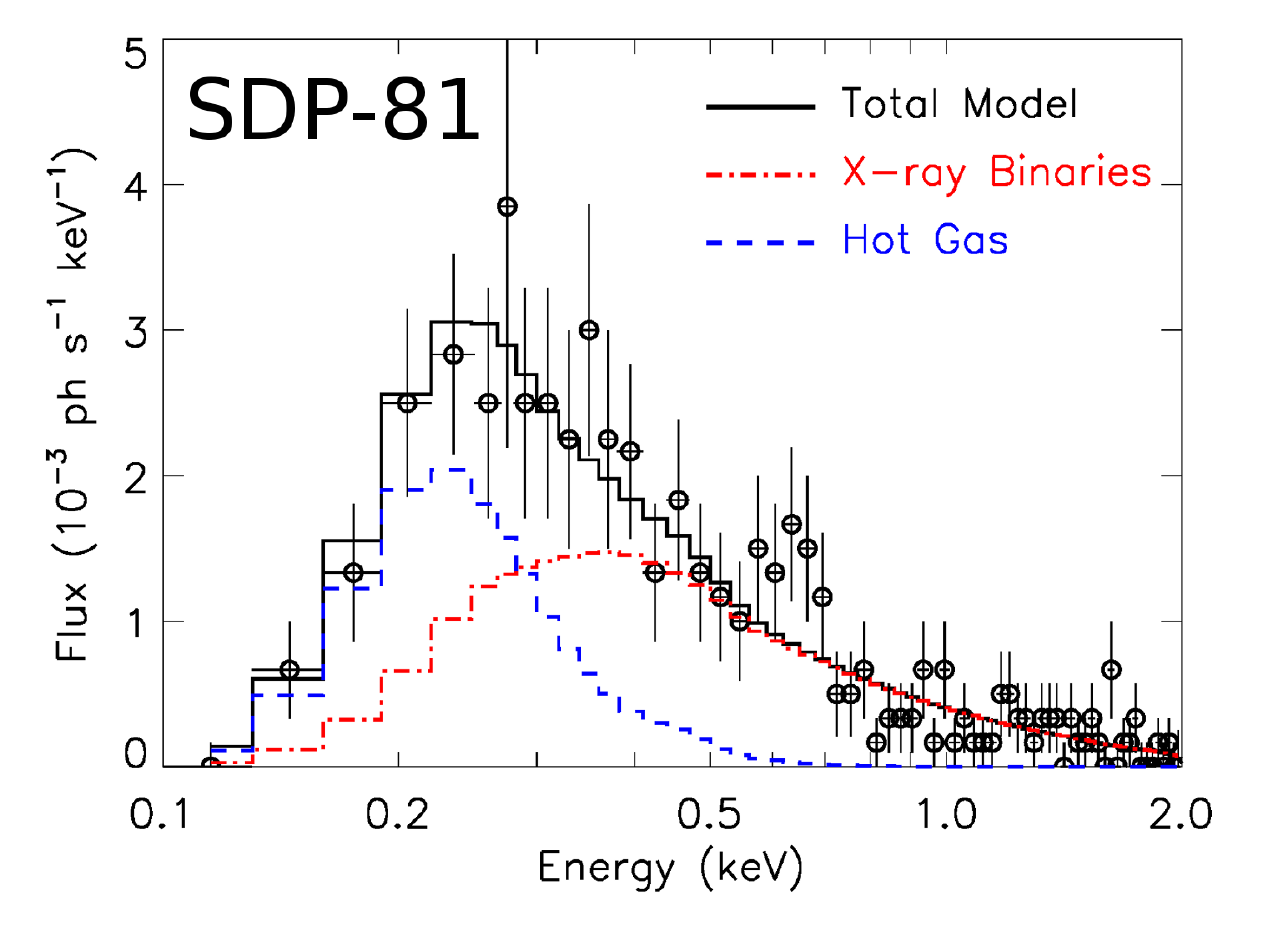}}

    \caption{\AXIS\ can resolve star-forming regions in strongly lensed
      galaxies at $z>2$, such as SDP-81\cite{Dye2015} ($z=3.042$). This 200
      ks exposure shows emission from the lensed galaxy and the lensing
      galaxy (at $z=0.3$). The right panel shows the total spectrum, and
      \AXIS\ is able to detect both XRBs and hot gas.}
    \label{fig:GAL_ALMA_lens}
\end{figure}

\begin{figure}[b]
\vspace*{-1mm}
\colorbox{callout}{\color{white}\sfsm
\begin{minipage}{0.99\textwidth}\begin{minipage}{0.97\textwidth}
\vspace*{3mm}
\begin{itemize}[itemsep=5pt,labelwidth=0pt,labelindent=0pt]
\item \AXIS\ will measure star formation rates in early galaxies through HMXB scaling relations
\item \AXIS\ will identify and measure the impact of galactic winds to {\em
  z}$\,>\,$1
\item \AXIS\ will discover how X-ray binaries contribute to cosmic
reionization
\end{itemize}
\vspace*{1pt}
\end{minipage}\end{minipage}}
\end{figure}

The growth of galaxies is regulated by feedback from supernovae and AGN. 
The balance between the two changes over cosmic time
($0<z<6$) in a way that remains poorly understood because of the diverse
set of processes involved. X-rays probe both the impact of
energetic outflows on galaxies and the star-formation rate (SFR). \AXIS\ will
be able to measure these processes to $z>1$ and will help to identify the
drivers of galaxy evolution over most of cosmic time.  An angular resolution of 
$<1$\arcsec\ is needed to resolve galaxies at $z>0.25$, and a large
collecting area at $E<1$~keV is needed to detect redshifted soft X-ray
emission from the ISM and SNRs.

\subsubsection{Star formation across cosmic time.}

Star formation is regulated both by local ISM processes and by galaxy-scale
gravitational processes. There is a tight correlation (the ``main
sequence'') between SFR and the galaxy stellar mass, $M_*$, for star-forming
galaxies that is independent of redshift\cite{Brinchmann2004}. Galaxy
evolution is also affected by mergers, the density of nearby galaxies, and
the properties of dark-matter halos. Understanding the diversity of galaxies
requires reliable SFR measurements up to and past the peak epoch of star
formation as a function of these parameters.

A major challenge in measuring star formation with far-IR measurements is
the contribution of obscured AGN which are ubiquitous in massive galaxies at
$z>1$. \AXIS\ will unambiguously detect these AGN out past the peak of star
formation, and provide an independent measure of SFR using XRBs
allowing the accurate assessment of the star formation history of most
galaxies.

\AXIS\ measurements of the high-mass X-ray binary (HMXB) population measures
the SFR since it correlates strongly with the total luminosity of
HMXBs\cite{Lehmer2010}. At $z>1$, many galaxies have SFR$\ge 100$\msunperyr,
giving a HMXB contribution that can be detected by \AXIS\ in 100~ks at $z=1$
(or $\sim$1~Ms for $z=2$), with higher SFR galaxies detected to even greater
distances\cite{Lehmer2016}. Stacking analyses can extend these studies
beyond $z=5$. While \Athena\ will also detect these systems, the superior
resolution of \AXIS\ ($<2.5$~kpc at $z>1$) is needed to separate the HMXB and
the (often dominant) AGN contribution.

\AXIS\ can also measure the conditions of the hot ISM around intense
star-forming regions and the resolved SFR (from HMXBs) at $z>3$ by observing
strongly lensed galaxies\cite{Negrello2017}. Fig.~\ref{fig:GAL_ALMA_lens}
shows a simulated 200~ks observation of a $z=3.04$ lensed
galaxy\cite{Dye2015} with SFR$=500$\msunperyr. \Herschel{} observations show
that there are numerous high-$z$ systems bright enough for \AXIS\
observations.

\subsubsection{Stellar feedback.}

\begin{figure}[t]
    \centering
    \includegraphics[width=0.99\textwidth]{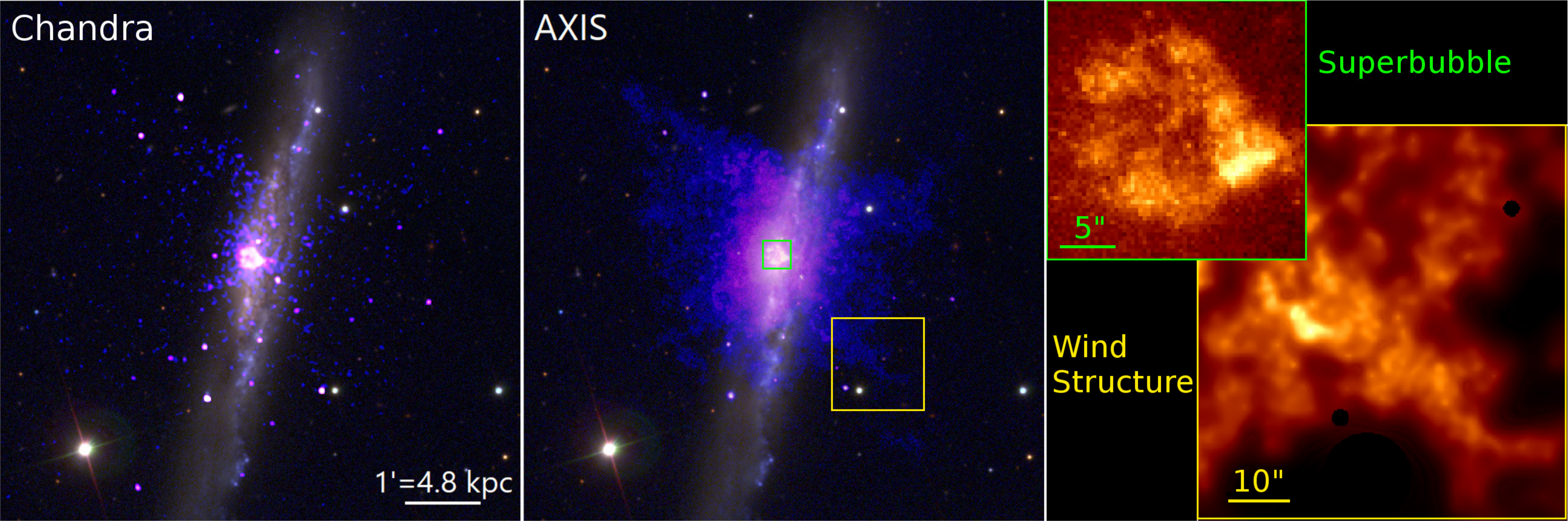}
    \caption{\AXIS\ will measure the temperatures, densities, and abundances of
    filaments in at least 25 galactic superwinds, whereas \Chandra\ can do this for 2-3. 
    For example, we show the 
    soft X-ray wind (0.3--2~keV) around NGC\,3079 as seen with \Chandra\ ({\em left}) 
    and \AXIS\ ({\em center}) for 100~ks each, overlaid on the optical image. \AXIS\ will resolve and accumulate many
    counts in structures 1--3\arcsec\ wide ({\em right}).}
    \label{fig:GAL_NGC3079}
\end{figure}

Star formation regulates itself as stellar winds and supernovae (SNe)
disperse and heat the gas and prevent cooling of the hot, ambient
circumgalactic medium. \AXIS\ will determine the origin of X-rays in local
galactic winds, measure the power and frequency of hot winds up to $z\sim
1$, understand how HMXBs contribute to the Epoch of Reionization, and
characterize the faint, hot ISM in elliptical galaxies.

During periods of intense star formation, multiple SNe combine to form
overpressured hot bubbles that can break out of the disk and drive
winds\cite{Veilleux2005,Heckman2017}. These winds also contain a large mass
in cool gas, which can escape the galaxy, thereby removing fuel for star
formation; they may be the primary reason that star formation has declined
since $z\sim 2-3$, when such winds were common. Determining their impact on
host galaxies and the CGM, as well as the primary driving mechanism (thermal
pressure, radiation pressure, or cosmic-ray momentum) requires precise
measurements of the wind mass, metal, and energy outflow rates which can be
constrained from spatially resolved X-ray spectroscopy. \AXIS\ will obtain spectra
from individual filaments for at least 25 nearby galactic winds in 20-150~ks
exposures (e.g., Fig.~\ref{fig:GAL_NGC3079}). In deeper exposures (survey fields
or targets selected from optical catalogs), \AXIS\ can detect and resolve strong
winds ($L_X > 10^{41}$~erg~s$^{-1}$, as seen in nearby examples such as Arp~220 or NGC~6240)
to $z \sim 1$, where stellar feedback was much more active. The temperatures,
masses, and abundances of these winds will clarify the role of stellar 
vs.\ AGN feedback in quenching galaxies. 

\AXIS\ will also probe how HMXBs contribute to reionizing the Universe at $z>6$.
HMXBs can either directly ionize gas with X-rays or remove gas enshrouding
nearby, very young star clusters that have yet to produce SNe and which emit
ionizing photons. \AXIS\ will provide detailed maps of several local analogs of
Lyman-break galaxies\cite{Heckman2005,Cardamone2009,Izotov2016}, which show
multiple knots of HMXB emission, along with extended hot
gas\cite{Prestwich2015,Basu-zych2016}. These features indicate the locations
of ``channels'' that allow ionizing radiation to escape\cite{Micheva2017}.
\AXIS\ can survey a sample of about 50 low-$z$ Lyman-break analogs
in a total of only 1~Ms.

\subsection{Intergalactic Medium --- Where Everything Ends Up}
\label{section:IGM}

\begin{figure}[b]
\vspace*{-1mm}
\colorbox{callout}{\color{white}\sfsm
\begin{minipage}{0.99\textwidth}\begin{minipage}{0.97\textwidth}
\vspace*{3mm}
\begin{itemize}[itemsep=5pt,labelwidth=0pt,labelindent=0pt]
\item \AXIS\ will allow the first complete census of cosmic baryons and
metals at low redshift, including the majority of the ``missing baryons'' in
the local Universe
\item \AXIS\ will map the unexplored, dynamic regions of the Cosmic Web around
galaxy clusters
\end{itemize}
\vspace*{1pt}
\end{minipage}\end{minipage}}
\end{figure}

The hot intergalactic medium, IGM (a.k.a.\ WHIM, warm-hot intergalactic
medium) is believed to be the main reservoir of the ``missing baryons'' in
the local Universe, and the ultimate depository of metals and entropy
produced in galaxies over cosmological time. UV and X-ray measurements of
the IGM have long been attempted, but so far, only a fraction of it --- the
relatively dense, colder phase visible through the O{\sc VI} absorption in
FUV --- has been unambiguously detected.

\AXIS\ will open a large discovery space by probing the emission from the
theoretically predicted $10^{6-7}$~K IGM that should dominate the baryonic
budget at low redshifts.  In combination with X-ray absorption line studies
from future X-ray missions and the UV absorption line studies of the colder
IGM phase, the \AXIS\ data will allow a complete census of the cosmic baryons
and metals at low redshift.  Owing to a combination of low background 
and high collecting area (Fig.\ A.5), \AXIS\ will be able to measure the flux and
spectrum from very low surface brightness regions, allowing access to
emission from the dynamic regions at the interface between clusters,
galaxies and the Cosmic Web.

\subsubsection{Cluster outskirts and the Cosmic Web.}

\begin{wrapfigure}{r}{0.47\textwidth}
\centering
\vspace*{-5mm}
\includegraphics[width=0.46\textwidth]{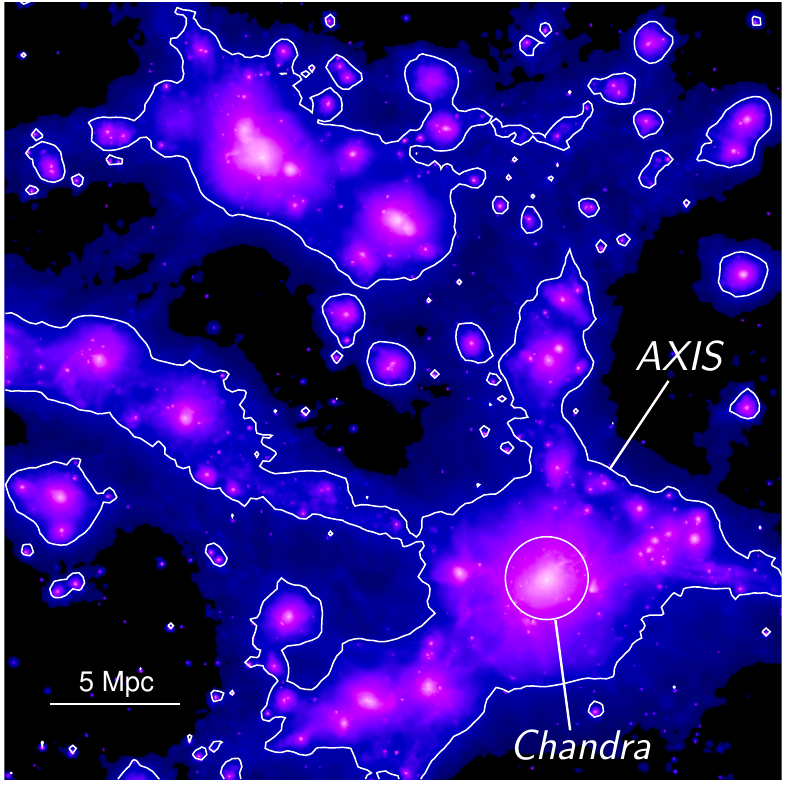}
    \caption{\AXIS\ will see how galaxy clusters connect to the Cosmic Web.
      In this cosmological simulation of Large Scale
      Structure\cite{Dolag2006}, color shows X-ray brightness of the
      intergalactic plasma, revealing a web of giant filaments with galaxy
      clusters as bright nodes. \AXIS\ will reach much farther
      into those dynamic, but very dim regions (out to the white contour)
      than any other X-ray instrument.}
    \label{fig:IGM_LSS}
\vspace*{-5mm}
\end{wrapfigure}

\begin{figure}[t!]
    \centering
    \includegraphics[width=0.85\textwidth]{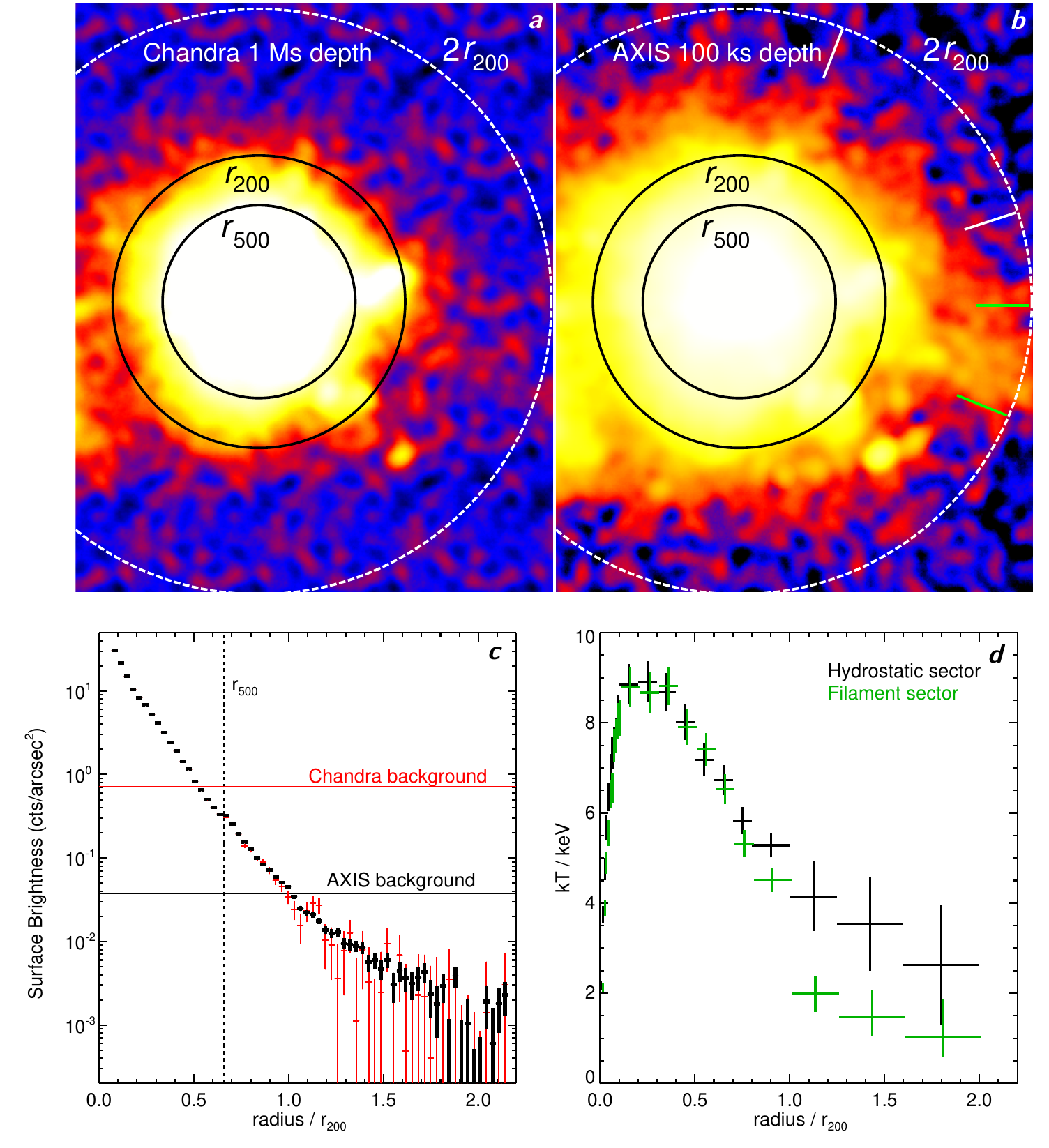}
\caption{\AXIS\ will provide unsurpassed capability for exploring the dynamic outermost regions of galaxy clusters. 
Panels {\em a}\/ and {\em b}\/ show X-ray images of the same cluster 
from cosmological simulations as observed by \Chandra{} (1~Ms) and \AXIS\ (100~ks), respectively. 
They have the same number of cluster counts, but \AXIS\ is able 
to probe the cluster emission to much larger radii, where clusters interface with the Cosmic Web. 
It can determine densities, metallicities and temperatures out to $2\times r_{200}$ in a routine exposure.  
\XMM\ and \Chandra\ can only reach $r_{200}$ in rare ultra-deep exposures. \AXIS\ will be able to study 
the cluster hydrostatic regions (white dashes) and giant filaments connecting the cluster to 
the Cosmic Web (green dashes). (c) X-ray surface brightness  profile in the 0.8-5~keV band extracted from a 
sector shown by white dashes in (b). The background levels (detector + unresolved CXB) are shown by 
horizontal lines. (d) \AXIS\ temperature profiles for the sectors shown in (b) with white and green dashes.}
    \label{fig:IGM_cluster_outskirts}
\end{figure}

Galaxy clusters act as ``IGM traps,'' attracting and compressing the IGM and
making it more readily observable in X-ray emission. Most of this material
lies beyond the virial radius\cite{Walker2019}. The surface brightness in
these outskirts is extremely low, requiring the low detector background of
\AXIS\ (Fig.\ A.5) to characterize. Arcsecond or better angular resolution is
critically important for the removal of the CXB point sources. As shown in
Fig.~\ref{fig:IGM_LSS}, \AXIS\ can trace the IGM out to twice the virial
radius ($2 r_{200}$) of most clusters, where the IGM exists along the giant
filaments of the Cosmic Web. In these regions we expect to find 
infalling galaxy-sized and group-sized objects (with the accompanying shock
fronts and gas stripping), as well as the IGM filaments (see Fig.~\ref{fig:IGM_cluster_outskirts}c).
These observations will test cosmic structure formation theory as it probes
the transition from the weakly- to strongly-nonlinear regimes\cite{Power2018}.

\AXIS\ will measure the plasma temperature and iron abundance of bright
regions of the IGM filaments (Fig.~\ref{fig:IGM_cluster_outskirts}d).
Cosmological simulations indicate that metals in the cluster outskirts
should originate from galaxy formation feedback far in the
past\cite{Biffi2018}, making the IGM metallicity a particularly powerful
probe of the physics of galaxy formation. \AXIS\ mapping of the iron abundance
in the filaments will provide a guide to the \Athena\ calorimeter
observations.

For the ``hydrostatic'' cluster regions
(Fig.~\ref{fig:IGM_cluster_outskirts}d), temperatures provide an
estimate of the total cluster mass. Comparison with estimates from
gravitational lensing from \Euclid\ and \WFIRST\ will allow estimates of the
non-thermal pressure components in the ICM as a function of radius and
constrain the dominant physical processes in the low-density plasma.

\subsubsection{The hot Circumgalactic Medium.}

Regions of the IGM surrounding individual galaxies---the circumgalactic
medium (CGM)---represent another phase of the ``missing baryons.''
$\Lambda$CDM predicts that galaxies with virial temperatures exceeding
$10^6$~K ($\ge L_*$) are surrounded by massive, extended hot halos, as
infalling gas is shock-heated to $T_{\text{vir}}$ and relaxes to
quasi-hydrostatic equilibrium\cite{White1991}. The hot CGM is a reservoir of
fuel for star formation and the dominant repository of mass, energy, and
metals from galactic winds. Measuring these quantities and discovering how
they are linked to cooler CGM components\cite{Tumlinson2017} provides
powerful constraints on galaxy formation models.

Direct detections beyond $R>20$~kpc are presently limited to a very small
sample of massive galaxies with unusually bright and hot
CGM\cite{Bregman2018}. The total extent and mass of the halos of $L_*$
galaxies remain controversial, since measurements are only possible with present capabilities with 
stacking analyses.  \AXIS\ will detect the extended CGM around individual
$L_*$ galaxies within $d<200$~Mpc, and around more massive galaxies to
$z\sim0.1-0.2$. The expected surface brightness falls below that of the soft
X-ray background within $R\sim 20-50$~kpc of the disk (several arcminutes
for a typical target).  With its wide field of view, \AXIS\ will model the
local Galactic foreground and the time-variable solar wind charge exchange
from the same observation allowing robust background subtraction.

\AXIS' high resolution enables precise measurements of the mass, temperature,
metal content, and entropy of the hot gas by removing contaminating point sources
and distinguishing SNe-powered hot fountains (which have much higher surface
brightness) from the extended CGM. This separation is essential to measuring
the enrichment and entropy of the hot CGM. \AXIS\ can obtain secure
measurements of the entropy and metal content around a representative sample
of galaxies with an investment of several Ms.

\section{MICROPHYSICS OF COSMIC PLASMAS}
\label{section:plasma}

\begin{figure}[b]
\vspace*{-1mm}
\colorbox{callout}{\color{white}\sfsm
\begin{minipage}{0.99\textwidth}\begin{minipage}{0.97\textwidth}
\vspace*{3mm}
\begin{itemize}[itemsep=5pt,labelwidth=0pt,labelindent=0pt]
\item \AXIS\ will measure fundamental properties of cosmic plasmas
such as effective viscosity, heat conductivity, equilibration timescales
\item \AXIS\ will advance our understanding of cosmic ray acceleration
by finding and resolving shocks in clusters and supernova remnants
\end{itemize}
\vspace*{1pt}
\end{minipage}\end{minipage}}
\end{figure}

Modern astrophysics relies on computer simulations to understand complex phenomena in the Universe, from solar flares to supernova explosions, black hole accretion, galaxy formation, and the emergence of Large Scale Structure (LSS) in the entire cosmological volume. However, we cannot model all the relevant scales of an astrophysical problem from first principles. For example, turbulence in the cosmological volume is driven by structure formation on the galaxy cluster scales ($10^{24}$ cm), but can cascade down to scales as small as the ion gyroradius ($10^{8-9}$ cm). Such a dynamic range is impossible to implement in codes. The only way to perform realistic simulations of macroscopic phenomena is to measure the relevant microscopic plasma properties experimentally and encode them at the ``subgrid'' (below-resolution) level. However, many key cosmic plasma properties, such as viscosity, heat conductivity, and the energy exchange between thermal and relativistic
particle populations and the magnetic field remain largely unmeasured.
Theoretical estimates span orders of magnitude, resulting in 
qualitatively different numerical predictions for galaxy stripping,
evolution of galaxy clusters and many other processes.

\begin{figure}[t]
    \centering
    \includegraphics[width=\textwidth,viewport=1 13 459 166,clip]{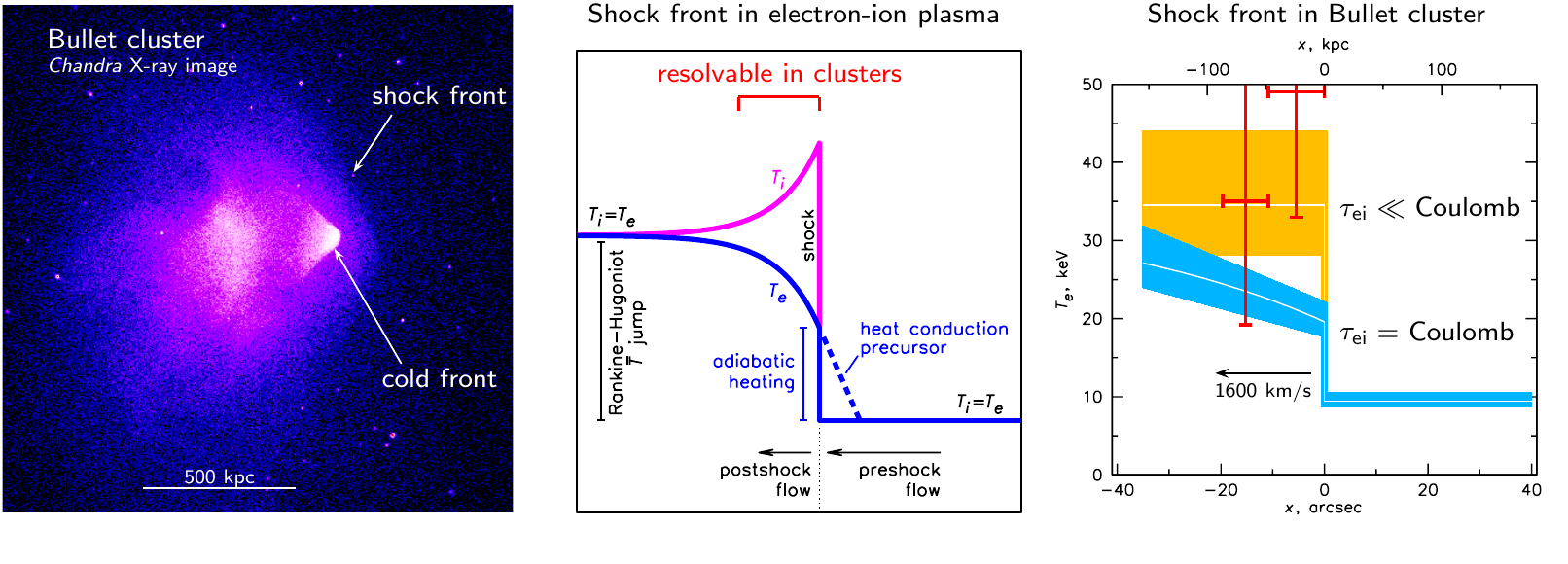}
    \caption{\AXIS\ will advance plasma physics by resolving shock fronts in galaxy clusters.
    {\em Left}: X-ray image of the Bullet cluster, the textbook example of
      a bow shock. The shock is driven by a moving subcluster (the bullet),
      whose front boundary is a ``cold front.'' {\em Middle}: Expected 
      electron and ion temperature profiles across a shock front in
      plasma. Temperatures are unequal immediately after the shock and then
      equalize. If electron heat conduction is not suppressed, a temperature
      precursor is also expected. {\em Right}: \Chandra{} deprojected electron
      temperature profile immediately behind the Bullet shock (crosses; 
      errors are $1\sigma$) with models for Coulomb collisional and instant equipartition\cite{Markevitch2005}. This measurement
      favors fast electron-proton equilibration, but uncertainties are large.
      \AXIS\ will perform a definitive test by using many more shocks.}
    \label{fig:PLASMA_equil}
\end{figure}

Many of these properties can be probed through X-ray studies of galaxy
clusters and SNRs. These objects are filled with a
$10^{6-8}$\,K optically-thin, X-ray emitting plasma permeated by magnetic
fields and ultra-relativistic particles. The plasma is collisionless and
``hot'' in the plasma-physics sense --- the ratio of thermal to 
magnetic pressure is typically $>100$ in clusters and $>10$ in SNR.
This regime is directly relevant to a wide range of astrophysical systems. 

A phenomenon particularly sensitive to plasma physics is shock 
fronts in clusters and SNR. These special locations permit study
of such basic plasma properties as heat conductivity, the electron-ion temperature 
equilibration timescale, and the physics of
cosmic ray acceleration and amplification of magnetic 
fields\cite{Markevitch2007}. As cosmic shock waves occur at 
all scales, these studies can have truly universal applications. 

Another powerful plasma probe is
provided by the ubiquitous, sharp contact discontinuities, or ``cold
fronts,'' found in the intracluster medium\cite{Markevitch2007}. While
\Chandra{} has provided tantalizing new results for shocks and cold fronts,
it has only scratched the surface of what can be learned. Using these
natural laboratories for plasma physics requires
much greater collecting area and lower background, while maintaining high
angular resolution to resolve the sharp spatial features and remove
contamination from faint point sources.

\subsection{Plasma Equilibration Times}

The common assumption that all particles in a plasma have the same local
temperature may not be true if the electron-ion equilibration
timescale is long compared to heating timescales. This timescale is 
fundamental for such processes as accretion onto black holes and 
X-ray emission from the intergalactic medium. It can be 
directly measured using cluster shocks. 

At a low-Mach shock, ions are
dissipatively heated to a high temperature, $T_i$ (which cannot be 
directly measured), while electrons are adiabatically compressed to a
lower temperature, $T_e$. The two species then equilibrate to the
mean post-shock temperature\cite{Zeldovich1966} (Fig.~\ref{fig:PLASMA_equil}). 
From the X-ray brightness and spectra, we can measure the plasma density
and $T_e$ across the shock. Because sonic Mach numbers of cluster shocks
are low ($M=2-3$), the density jump at the shock far from its asymptotic
value and thus provides an accurate value of $M$, giving the
post-shock temperature. If the equilibration is via Coulomb collisions,
the region over which the electron temperature $T_e$ increases 
is tens of kpc wide --- resolvable with \AXIS\ at distances of $z<2$.  
This direct test is unique to cluster shocks because of the
fortuitous combination of the linear scales and relatively low Mach numbers;
it cannot be done for the solar wind or SNR shocks.

A \Chandra{} $T_e$ profile across the prominent shock in the 
Bullet cluster (Fig.~\ref{fig:PLASMA_equil}) suggests that 
$T_e-T_i$ equilibration is much quicker than 
Coulomb\cite{Markevitch2006}, although with low statistical confidence 
and a considerable systematic uncertainty that requires a sample of shocks. 
This measurement at present is limited to only three shocks and the results
are contradictory\cite{Markevitch2006,Russell2012,Wang_Qian2018}.
With \Chandra, suitable shock fronts are rare. \AXIS\ will be able to
find hundreds more shocks, select a sample of suitable ones, and robustly
determine this basic plasma property.

\subsection{Heat Conductivity}

\begin{wrapfigure}{r}{0.5\textwidth}
\vspace*{-12mm}
    \centering
    \includegraphics[width=0.45\textwidth]{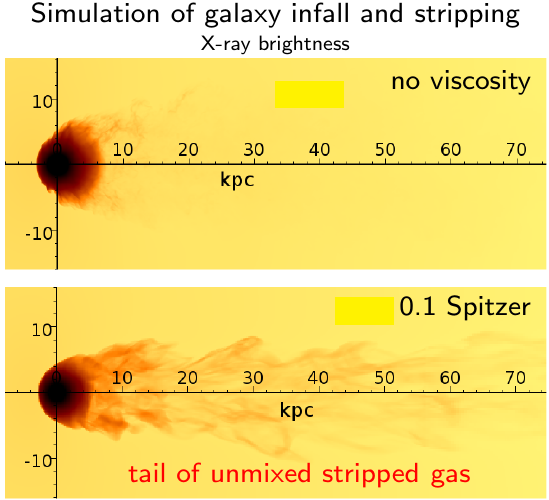}
    \caption{\AXIS\ will be able to probe plasma viscosity by searching for very low-contrast 
    extended X-ray features that should accompany galaxies and groups as they fall into clusters. 
    Viscosity determines how the 
      gas is stripped from the infalling galaxies and how easily it mixes with the ambient
      gas. If viscosity is not completely suppressed, the galaxies should exhibit tails of 
      stripped gas\cite{Roediger2015}.}
    \label{fig:PLASMA_viscosity}
\vspace*{-2mm}
\end{wrapfigure}

Heat conduction erases temperature gradients and competes with radiative cooling, and is of utmost 
importance for galaxy and cluster formation. The effective heat
conductivity in a plasma with tangled magnetic fields can be 
anywhere from zero to one-third of the Spitzer value, with a 
large range of predictions for the conductivity parallel 
to the field lines \cite{Schekochihin2008,Kunz2014}. 
The existence of temperature gradients in clusters confirms that
conduction across magnetic field lines is very
low\cite{Ettori2000,Vikhlinin2001,Wang_Qian2016}, but 
estimates for the average conductivity\cite{Markevitch2003} or the 
parallel component\cite{Wang_Qian2016} are poor. 
Shock fronts are unique locations where the parallel component can
be constrained. Electron-dominated conduction should result in an 
observable $T_e$ precursor to the shock
temperature jump (Fig.~\ref{fig:PLASMA_equil}). \AXIS\ will have the
requisite sensitivity and resolution to measure this effect in
the temperature profiles of cluster shocks.

\subsection{Viscosity}

Plasma viscosity governs the damping of turbulence and sound waves,
suppression of hydrodynamic instabilities, and mixing of different gas
phases, but its value is largely unconstrained. X-ray observations 
of galaxy clusters can measure it in two ways. One is via detailed 
observations of cold fronts --- sharp X-ray brightness edges 
ubiquitous in merging subclusters (e.g.,
Fig.~\ref{fig:PLASMA_equil}) and cluster cores\cite{Markevitch2007}, where 
they are produced by ``sloshing'' of the gas in the cluster potential 
well\cite{Ascasibar2006}. Sloshing produces velocity shear across 
the cold front, which should generate Kelvin-Helmholtz instabilities. 
However, if the ICM is even mildly viscous, these instabilities will be
suppressed\cite{Roediger2013b,Zuhone2015}. These subtle features can 
only be seen with high resolution and lots of photons. \Chandra\
has discovered Kelvin-Helmholtz instabilities in a few cold fronts and put an 
upper limit on the effective isotropic viscosity of ~1/10 the classical 
value\cite{Roediger2013a,Su2017,Ichinohe2017,Wang_Qian2018b}. 
To constrain the viscosity from below requires finding 
instabilities at different growth stages and for a range of 
density contrasts. \AXIS\ will resolve the structure
for many more cold fronts and thus measure the effective viscosity.

Plasma viscosity can also be probed by observing stripping of galaxies and
groups as they fly through the ICM. Figure~\ref{fig:PLASMA_viscosity} shows
a striking difference in the simulated X-ray appearance of the tail of the
cool stripped gas behind an infalling galaxy. In an inviscid plasma, the gas
promptly mixes with the ambient ICM, but a modest viscosity suppresses the
mixing and makes the long tail visible. A deep \Chandra{} image of a 
Virgo elliptical M89 favors efficient mixing and a reduced 
viscosity\cite{Kraft2017}. Other infalling groups in the cluster periphery
do exhibit unmixed tails\cite{Eckert2014}. 
\AXIS\ will have the requisite sensitivity to study these subtle, 
low-contrast extended features, most of which will be found in the 
low-brightness cluster outskirts, to constrain effective viscosity 
--- and directly observe its effect on gas mixing.

\subsection{Cosmic Ray Acceleration at Shocks}

Across the universe, shocks accelerate particles to very high energies
via the first-order Fermi
mechanism. Microscopic details of this fundamental process remain poorly
known for astrophysical plasmas, and particle-in-cell simulations are still
very far from covering realistic plasma parameters.

Many galaxy clusters exhibit striking ``radio relics'' in their 
outskirts\cite{vanWeeren2010}.
These Mpc-long, arc-like structures are synchrotron signatures of
ultrarelativistic ($\gamma \sim 10^4$) electrons. In some clusters, radio
relics coincide with X-ray shock fronts, strongly suggesting that ICM shocks
are responsible for those high energy electrons\cite{Giacintucci2008,Shimwell2015}. 
However, the shock Mach numbers are low
($M=1.5-2.5$), and it is puzzling how they can have the acceleration
efficiency needed to produce the relics. To gain insight, we need a
systematic comparison of shocks in the X-ray and radio. A large number of
relics have been found recently by \LOFAR, \GMRT\ and \MWA, but most are located
far in the cluster outskirts, where the X-ray emission is too dim for
\Chandra\ but accessible to \AXIS\ with its superb sensitivity to low surface brightness objects.

\subsection{Magnetic Field Amplification and Damping at Shocks}

\begin{wrapfigure}{R}{0.65\textwidth}
\vspace*{-2mm}
    \centering
    \includegraphics[width=0.65\textwidth,viewport=231 12 1015 396,clip]{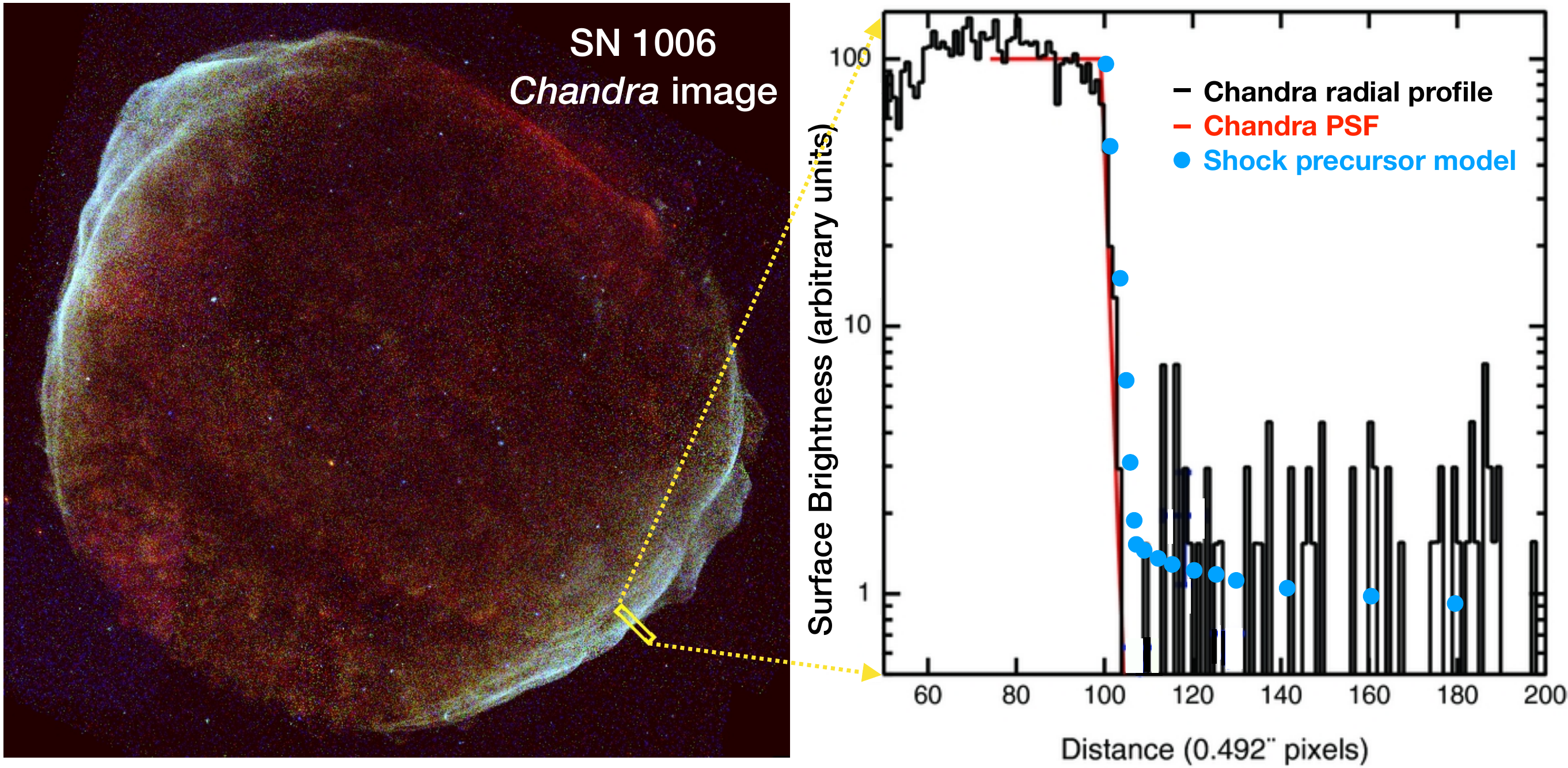}
    \caption{\AXIS\ will measure the X-ray precursors of fast shocks as particles diffuse upstream. {\em Left}: the \Chandra{} image of SN~1006\cite{Winkler2014}, with a
      radial profile extracted from the yellow box shown in right panel. The
      red curve shows the \Chandra{} PSF, while the blue dots show a
      potential shock precursor model\cite{Morlino2010}. Even with
      \Chandra{}'s resolution, the background level is too high to detect
      whether this precursor is present. \AXIS' much lower background and increased sensitivity will reveal if the predicted precursor exists.}
    \label{fig:PLASMA_shock}
\vspace*{-2mm}
\end{wrapfigure}

Shocks should produce large variations of the magnetic
field on small linear scales, but the exact mechanism for field amplification 
is unclear. SNRs
offer the chance to study shocks on the relevant scales --- much smaller than
those in clusters but much greater than those accessible to \textit{in situ}\/
measurements in the solar wind. Several SNRs show thin filaments of X-ray
synchrotron emission in areas where shock velocities exceed several
thousand~\kms. In SN\,1006, the width of these rims varies with energy,
implying that the relativistic electrons rapidly age in a field as strong as
$100\,\mu$G\cite{Ressler2014} --- inconsistent with the field damping quickly
behind the shock. However, in another well-studied SNR, Tycho, the thin
synchrotron rims suggest strong amplification at the shock followed by quick
damping\cite{Tran2015}. Could there be two different mechanisms by which
magnetic fields are amplified and subsequently damped in shock waves?
Observing more synchrotron-dominated shocks will answer this question.
SN\,1006 and Tycho observations required very long integration times with
\Chandra{}; all other similar remnants are simply too faint. \AXIS\ properties
are perfectly suited to drastically increase the sample size and accuracy
for solving this problem.

\subsection{Diffusion of Cosmic Rays}

Some accelerated particles must diffuse from behind the shock into
the upstream medium. In SNR shocks dominated by non-thermal synchrotron
emission from accelerated particles, faint X-ray emission should be
present ahead of the shock, yet this emission has never been
detected\cite{Winkler2014} (Fig.~\ref{fig:PLASMA_shock}). Finding and
characterizing this precursor will put tight constraints on the universal
properties of fast shock waves, such as the degree of magnetic field
amplification and the diffusion and scattering length of energetic
particles. Angular resolution is critical here; searching for this in Galactic SNRs will be possible with \AXIS.

\subsection{Feedback in Shocks from Particle Acceleration}

When the energy in the accelerated particles becomes comparable to the
energy in thermal gas, the shock dynamics change. In Tycho's SNR, the
locations of the forward and reverse shocks and the contact discontinuity
are inconsistent with theoretical predictions\cite{Warren_J2005}. This implies
that the forward shock was efficiently accelerating cosmic rays, robbing the
post-shock gas of energy and lowering the gas temperature. Such direct
observations are invaluable for modeling the acceleration
process\cite{Kosenko2011,Warren_D2013}. However, only a very few SNRs have a
high enough surface brightness at the shock for \Chandra{} observations to
test the theories. By greatly expanding the number of accessible objects 
and covering a broad range of shock Mach numbers, \AXIS\ will revolutionize this 
field .

\section{THE TRANSIENT AND VARIABLE UNIVERSE}
\label{section:trans}

\begin{wrapfigure}{R}{0.55\textwidth}
\vspace*{-5mm}
    \centering
    \includegraphics[width=0.55\textwidth]{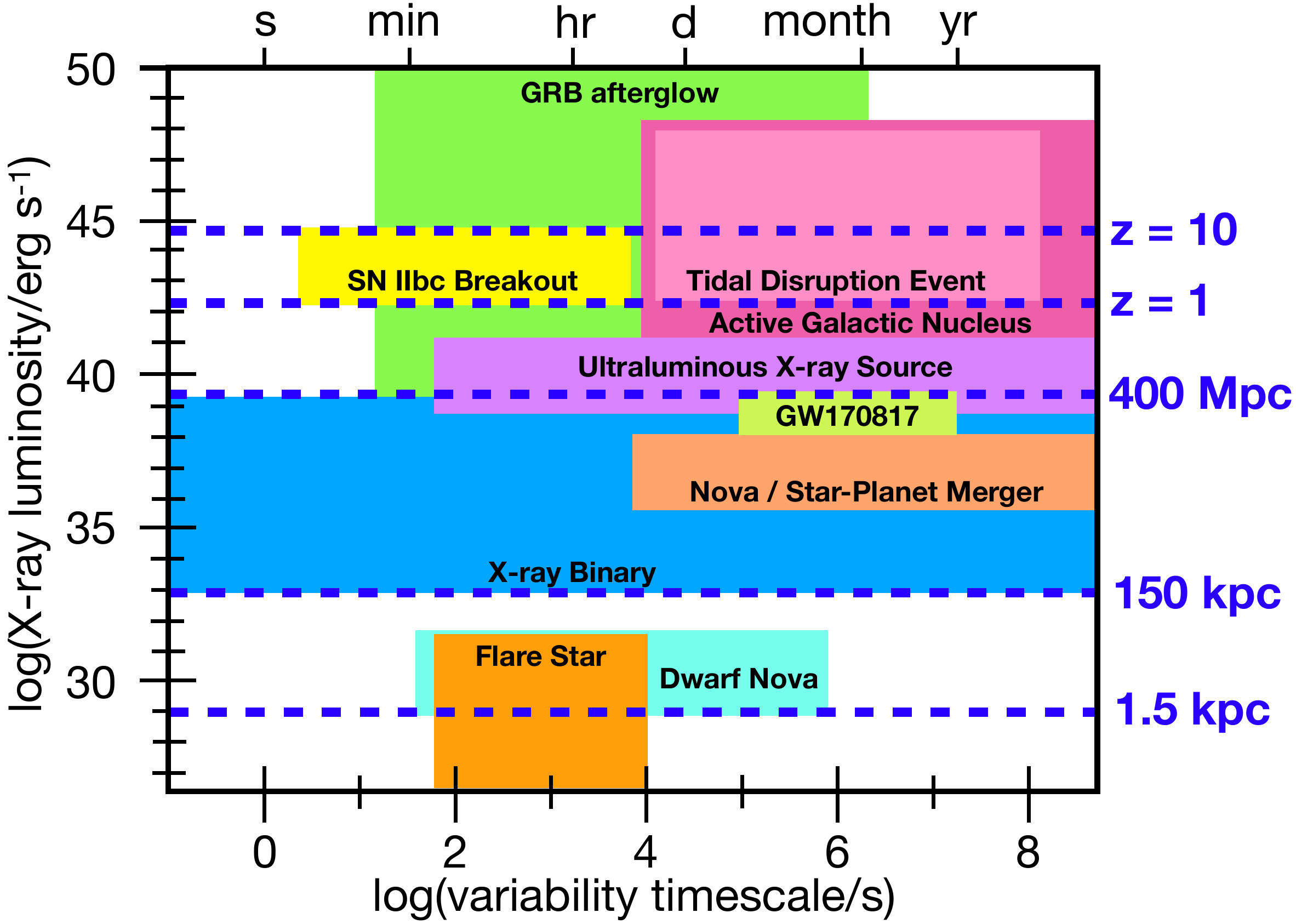}
    \caption{The X-ray luminosity and variability timescale of various
      astrophysical phenomena. On the right axis (in blue), we denote the
      distance out to which \AXIS\ can detect these transients in a 20 ks
      exposure.}
    \label{fig:TRANS_sources}
\vspace*{-1mm}
\end{wrapfigure}

\begin{figure}[b]
\vspace*{-2mm}
\colorbox{callout}{\color{white}\sfsm
\begin{minipage}{0.99\textwidth}\begin{minipage}{0.97\textwidth}
\vspace*{3mm}
\begin{itemize}[itemsep=5pt,labelwidth=0pt,labelindent=0pt]
\item \AXIS\ will have a response time under 4 hours, 100x the effective area
of \Swift{}, and will dedicate 10\% of observing time for follow-up of
transients
\item \AXIS\ will observe 10$^{\sf 6}$ {\em M}$_{\odot}$ SMBH mergers
at {\em z}\,=\,2,
which are too faint for current or proposed wide-field X-ray monitors
\item \AXIS\ will detect and resolve events like the NS-NS merger GW170817 out to
$\sim$400~Mpc
\end{itemize}
\vspace*{1pt}
\end{minipage}\end{minipage}}
\vspace*{-3mm}
\end{figure}

The ASTRO2010 Decadal Survey ranked the \textit{Large Synoptic Survey Telescope}
(\LSST) as the highest priority project for ground-based astronomy,
indicating that time-domain astronomy is an essential component in
understanding our universe. Expectations are that \LSST\ will see
1,000-100,000 transients per night. In the radio, \SKA\ (first light in 2025)
will find tens of transients per night. Next-generation ground-based GW
facilities will observe tens of neutron star-BH and binary neutron
star mergers per year. Finally, by the 2030s, \LISA\ will localize the first
SMBH mergers.  The keys to successfully following up
these multi-messenger transients are (1) high sensitivity (i.e., low
confusion limit and high effective area) to probe high redshifts and faint
targets with short exposures, and (2) rapid and flexible response time to
track transients quickly and monitor their evolution. Guided by the
appropriate decision trees, \AXIS\ will make critical detailed follow-up
observations of a wide range of these new discoveries
(Fig.~\ref{fig:TRANS_sources}).

\subsection{Tidal Disruption Events}

Roughly one star per galaxy every 10 thousand years gets disrupted by the
strong tidal forces of the central SMBH\cite{Rees1988}.
The accretion of the bound stellar material causes a short-lived flare of
emission, known as a Tidal Disruption Event (TDE). Recently, TDEs have
garnered excitement from a range of astronomical communities because, as
opposed to continuously accreting BHs in AGN, they are an impulse of
accretion onto a normally hidden BH. Some of the \AXIS\ TDE science
includes:

\subsubsection{Super-Eddington accretion in TDEs.}

The fallback rate from the tidally disrupted stars is initially highly
super-Eddington for $M_{\text{BH}}<10^7 M_{\odot}$ and drops with time.
Thus, TDEs are laboratories for studying the transition from super-Eddington
to sub-Eddington on timescales of months or less and for testing MHD
simulations of super-Eddington flows\cite{Coughlin2014}. Understanding the
X-ray component of super-Eddington TDEs is key, as this is the emission
coming from the innermost regions, where ultrafast outflows and jets may be
launched. \AXIS' unprecedented sensitivity, angular resolution, and fast
slew capabilities provide fast identification of the X-ray counterpart of
TDEs, allowing the monitoring of their intensity and spectral evolution over
time. Currently, \Swift{} is the key X-ray instrument for monitoring TDEs.
With \AXIS, we have 70x more counts per unit time than Swift, giving a
complete spectrum over a broad energy range.  \AXIS's exquisite sensitivity in the
softest energy bands, where X-ray TDEs emit most of their energy, will
determine the physical origin of the X-ray emission and its connection to
the optical and UV emission.

\subsubsection{Finding the missing intermediate-mass black holes.}

Recently, an off-nuclear X-ray TDE candidate was discovered\cite{Lin2018}
with spectral characteristics indicating it was the disruption of a white
dwarf by a BH of mass $10^{4.5-5} M_{\odot}$.  This result shows
that TDEs are a promising way to find populations of BHs that cannot
be detected dynamically. \AXIS\ has the high angular resolution required to
disentangle these off-nuclear intermediate-mass BHs from the central SMBH. This is especially important for TDEs found from wide-field X-ray
monitors like the Einstein Probe (to launch in 2023) that have much poorer
angular resolution.

TDEs in the next decades will be detected to much larger redshift, requiring
increased X-ray sensitivity to obtain a precise localization and measure the
spectra and timing behavior. At present, due to a small sample size and low
signal to noise, there are very few constraints on when the X-ray emission
is produced, which is crucial to determine the physics of the TDE emission.

\subsection{X-ray Counterparts to \LIGO\ Binary Neutron Star Merger Events}

The recent discovery of gravitational waves (GWs) and EM
detections\cite{Abbott2017b} from a binary neutron star
merger\cite{Abbott2017a} has opened a new field of observational astronomy.
X-ray afterglows of neutron star mergers can constrain the external density,
magnetic fields, and jet structure\cite{Troja2018}. Further, X-rays can
constrain the binary inclination, which, in the GW signature, is highly
degenerate with the distance of the object. This is essential for using GWs
as a cosmological probe.

\begin{wrapfigure}{R}{0.64\textwidth}
\vspace*{-2mm}
    \centering
    \includegraphics[width=0.64\textwidth]{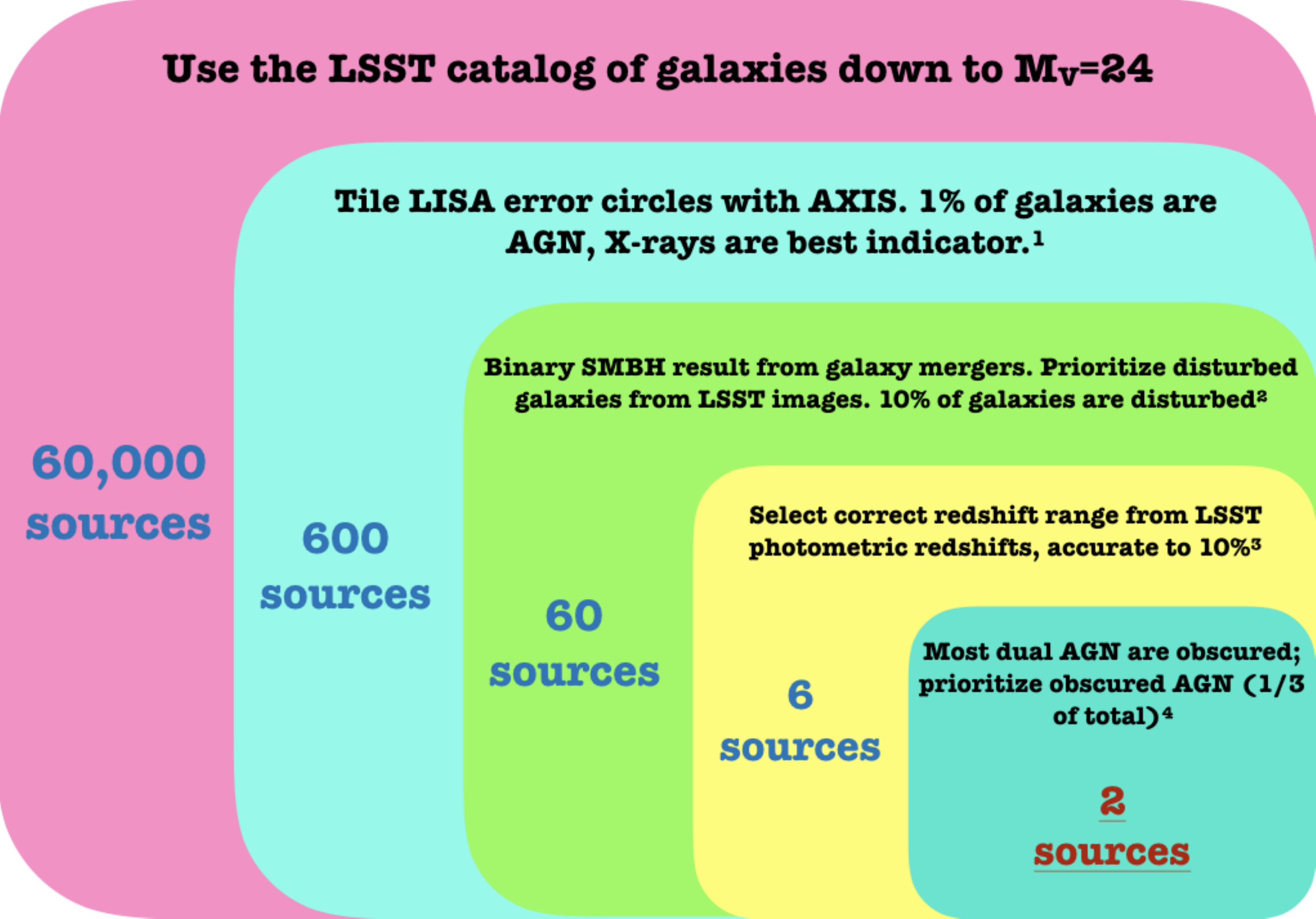}
    \caption{The \AXIS\ team will develop and follow a downselect plan similar to this example 
    for identifying the EM counterparts to GW signals. ($^1$Haggard et al.\ 2010;
      $^2$Lotz et al.\ 2008; $^3$Ivezi\'{c} et al.\ 2011; $^4$Lusso et
      al.\ 2013)}
    \label{fig:TRANS_downselect}
\vspace*{-3mm}
\end{wrapfigure}

By the early 2020s, most GW events will be found at $d \sim 200$~Mpc. If
these events are like GW170817, their X-ray afterglow will have $F_X \sim
8\times 10^{-16}$\fluxergs, below the sensitivity of \Chandra{} for a 100 ks
exposure. \AXIS\ could detect such an afterglow in only $\sim$5 ks,
allowing monitoring of the light curve, crucial for
modeling the nature of the jet. Longer exposures will determine the source
spectrum.

Arcsecond angular resolution was important for the GW170817 X-ray afterglow,
since it was offset from an AGN nucleus by 10\arcsec. At 200~Mpc, a similar
event would be 2\arcsec\ away from the nucleus; \Athena, with its
$\sim$5\arcsec\ resolution, would not be able to distinguish the event from
an AGN. Good angular resolution is key, as AGN, ultraluminous X-ray sources
(ULXs), and magnetars are highly variable and of similar luminosity, and
could be confused with the neutron star merger.

In the late 2020s, the next generation ground-based GW detectors will expand
the detection range to $\sim$400 Mpc. The rate of detections of GW events
will increase by an order of magnitude to $\sim$100 per year. \AXIS' large
area and fast slew capabilities allow for short exposures and good
flexibility, so it can monitor the evolution of several merger events per
year and perform rapid searches for the counterparts.

\subsection{Electromagnetic Counterparts to \LISA\ SMBH Mergers}

ESA's \LISA\ GW mission is scheduled to launch in the early
2030s, during the lifetime of \AXIS. While the exact SED of these events is
not known, it is likely that the candidates will be AGN-like. Some
simulations  predict that most of the
luminosity originates in hot mini-discs around the two SMBHs \cite{Tang2018}, so X-rays may be the best wavelength to find the EM
counterparts of \LISA\ GW events.

\LISA\ can detect BH inspiral and mergers from $10^{3-7} M_{\odot}$
BHs up to a redshift of $z=20$\cite{LISA2017}. \AXIS\ is well-suited
to observe the `average' $10^6 M_{\odot}$ merger, which will occur at the
peak of star formation at $z=2$. We make the conservative assumption that
the source is at 10\% of the Eddington luminosity, where the X-rays are 10\% of the bolometric luminosity. We therefore expect $F_X = 3\times 10^{-16}$\fluxergs, several orders of magnitude below the limiting flux of current or proposed
wide-field X-ray monitors. \AXIS\ can detect this flux with $\sim$15 photons
in 15~ks. However the details of SED are highly
uncertain. Even if the source is moderately obscured ($N_{\text{H}}=5\times
10^{22}$~cm$^{-2}$, as is seen for many dual AGN\cite{Koss2018}), the
estimated flux at $z=2$ is $F_X = 10^{-16}$\fluxergs, and \AXIS\ can
detect $\sim$15 photons in 50~ks.

We show in Figure~\ref{fig:TRANS_downselect} a scenario in which \AXIS\ can identify the EM counterpart. \LISA\
will localize 1~GW event per year with an error circle of $<1$~deg$^2$, with a
distance measure within 10\%, and $\sim$5 events per year with error circles
of 10~deg$^2$ or more\cite{Lang2008}. The ``warning'' time, when a position
can be reasonably estimated, is $\sim$16-35 days\cite{Haiman2017}. \AXIS\ will
tile the 1~deg$^2$ error circle in six 15~ks pointings ($\sim$1~day) and alert
the community, providing roughly 10 high probability
candidate GW events (Fig.~\ref{fig:TRANS_downselect}). \AXIS\ will monitor
those few candidates for the 16-35 days before merger, searching for
periodicities\cite{Graham2015,DOrazio2015} or atypical AGN SEDs, which may
indicate which candidate is the GW source, depending on the actual fluxes.

\subsection{Serendipitous Time Domain Science}

\AXIS\ will serendipitously detect transients, like ULXs, novae, and
core-collapse SN, out to $\sim$ 250~Mpc  (Fig.~\ref{fig:TRANS_sources}). \AXIS\ will detect an average of $\sim$1.5
MW-mass galaxies per field of view; in a 50~ks exposure, it will detect all
of the ULXs in those galaxies, which occur at the rate of $\sim$1 per 3
galaxies. With \AXIS' $<$50~ms time resolution, it can be used to search for ULX
pulsation\cite{Bachetti2014}, since the fastest known ULX has a period of
$0.419$~s. \AXIS\ can detect classical novae ($L_X \sim 10^{35}$\lumergs) out
to a distance of $\sim$1~Mpc, allowing X-ray flux measurements for all
classical novae in the Local Group. Core-collapse SNe have $L_X \sim
10^{38-41}$\lumergs, depending on the type of explosion and how long after
the explosion it is observed. \AXIS\ can detect a SN with $L_X \sim
10^{40}$\lumergs at a distance of 400~Mpc, which allows access to a huge
number of targets: assuming a SN rate of $10^{-2}$~galaxy$^{-1}$~yr$^{-1}$
and a space density of $0.01$~massive galaxies Mpc$^{-3}$, there will be
$2\times 10^4$~SN~yr$^{-1}$, which \AXIS\ can discover serendipitously.

\section{THE MILKY WAY AND NEARBY UNIVERSE}
\label{section:MW}

\begin{figure}[b]
\colorbox{callout}{\color{white}\sfsm
\begin{minipage}{0.99\textwidth}\begin{minipage}{0.97\textwidth}
\vspace*{3mm}
\begin{itemize}[itemsep=5pt,labelwidth=0pt,labelindent=0pt]
\item \AXIS\ will measure the luminosity of Sgr A* for several hundred years
in the past
\item \AXIS\ will map Fe-group elements in supernova remnants to determine how they exploded
\item \AXIS\ will measure interstellar dust grain size and composition from scattering halos
\item \AXIS\ will measure the luminosity functions for XRBs in different
environments
\end{itemize}
\vspace*{1pt}
\end{minipage}\end{minipage}}
\end{figure}

\subsection{The Galactic Center}

Arcsecond X-ray imaging resolution is indispensable for studying the
Galactic Center (GC). \Chandra{} imaging discovered extended X-ray emission
from the accretion flow of Sgr A*\cite{Baganoff2003,Wang_q2013,Roberts2017}
and revealed the population of compact objects in the GC, placing
constraints on the stellar history and dynamical evolution of the Galactic
bulge and GC region and strong constraints on stellar mass BHs near
the GC\cite{Muno2009,Dexter2014,Hailey2018}. X-ray observations have elucidated the
structure of the GC ISM and the accretion history of Sgr A* through iron
fluorescence of GC molecular clouds\cite{Ponti2010,Capelli2012,Ryu2013}.
\AXIS\ will enable a deeper view of the GC population of BHs, Sgr A*
flares and accretion flow and dust echoes. The higher signal-to-noise for
iron fluorescence combined with high spatial resolution will extend by
several hundred years the measurement of Sgr A* X-ray luminosity in the
past, a unique observation for any SMBH.

\subsection{Supernova Remnants}

\begin{wrapfigure}{r}{0.67\textwidth}
\vspace*{-11mm}
    \centering
    \includegraphics[width=0.67\textwidth,viewport=6 0 441 187,clip]{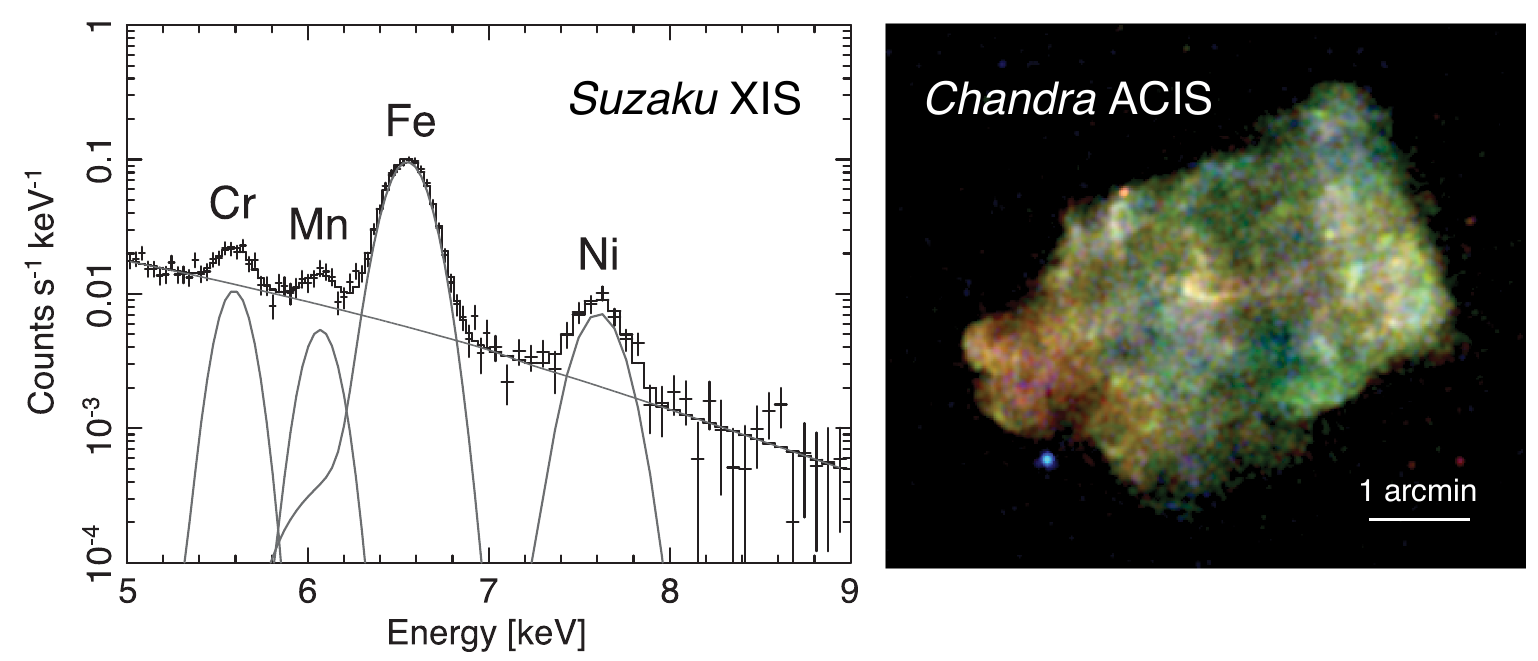}
    \caption{\AXIS\ will map the spatial distribution of Fe-peak elements 
    (Cr, Mn, Fe and Ni) with unprecedented accuracy in many supernova remnants, 
    determining the mechanism of the Type Ia supernova progenitor explosions. 
    {\em Left}: \Suzaku\ spectrum of the Galactic SNR 3C397\cite{Yamaguchi2015},
    integrated over the entire remnant. {\em Right}: \Chandra\ high-resolution 
    false-color image of the remnant\cite{SafiHarb2005} in broad energy 
    bands (red 0.8--1.5 keV, green 1.5--3 keV, blue 3--7 keV bands). The CCD energy
    resolution is sufficient to separate the lines; \AXIS\ will be the first 
    instrument to combine the necessary angular resolution and sensitivity to map
    the emission in individual spectral lines.}
    \label{fig:MW_3C397}
\vspace*{-2mm}
\end{wrapfigure}

Many unsolved problems about SNRs, such as the
progenitors of the various SN types and how they relate to the remnants
produced, and how SNR populations of nearby galaxies correlate with the
local star-formation history, require \AXIS' capabilities to solve. Mapping the
distribution of elements in the SNR is crucial, as the wide range of
explosion models for both Type~Ia and core-collapse SNe predict vastly
different ejecta distributions. It is particularly important to map Cr, Mn, and Ni for SN~Ia remnants, as their locations allow the mass determination of
the Type Ia-progenitor. These lines are faint and difficult
to observe with
current observatories (see Fig.~\ref{fig:MW_3C397}), with the low $S/N$
making element distribution mapping impossible. Only \AXIS\ will have
sufficiently high angular resolution and sensitivity to make such
observations for SNR in the Milky Way and, for the first time, other
nearby galaxies. This capability will allow the comparison of the SNR population with
the local star-formation history in other galaxies, in addition to constraining
SN progenitor models\cite{Margutti2018}. Detailed studies of the  spectral lines
in SNR with the high spectral
resolution of the forthcoming X-ray calorimeter missions (\XRISM, \Athena) will provide complementary dynamical and ionization constraints, which combined with \AXIS\ data provide a major step forward in our understanding.

\subsection{Dust Halos in the Interstellar Medium}

\begin{wrapfigure}{r}{0.6\textwidth}
\vspace*{-1mm}
    \centering

\includegraphics[height=0.35\textwidth,viewport=2 2 254 320,clip]{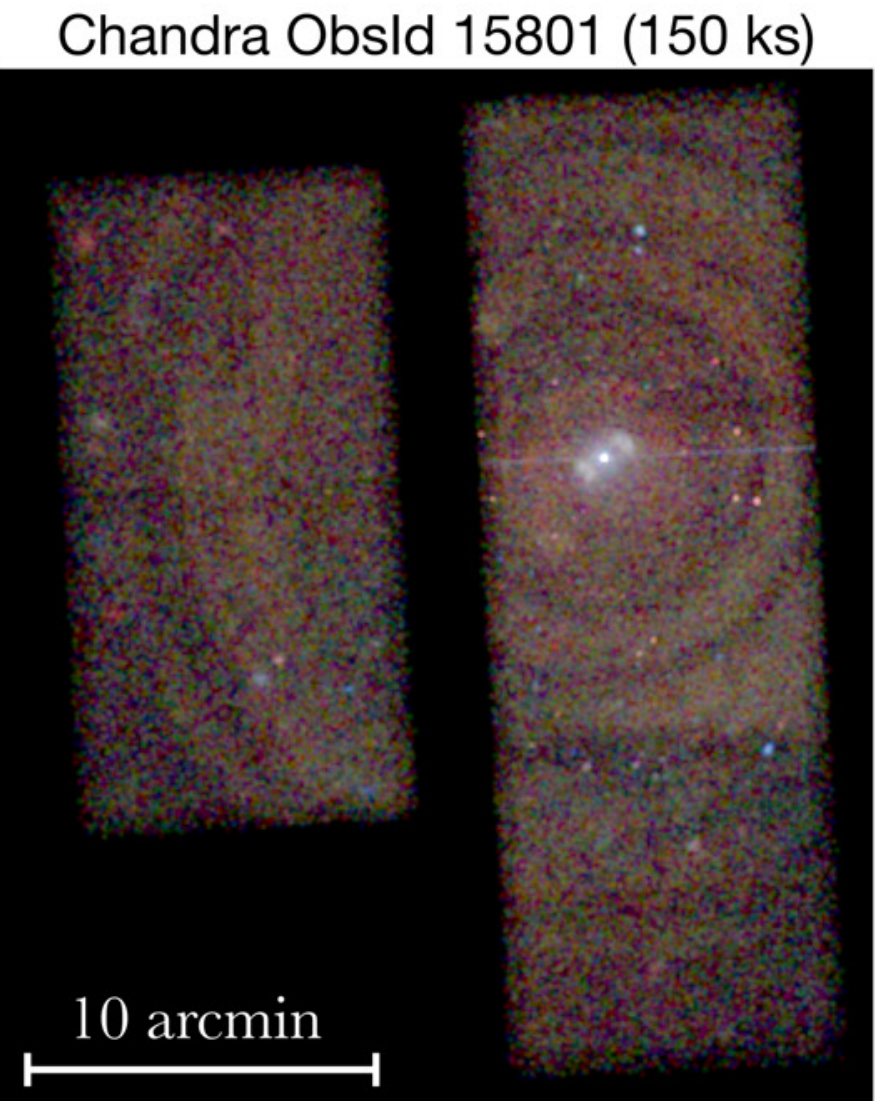}
\includegraphics[height=0.35\textwidth,viewport=2 2 254 322,clip]{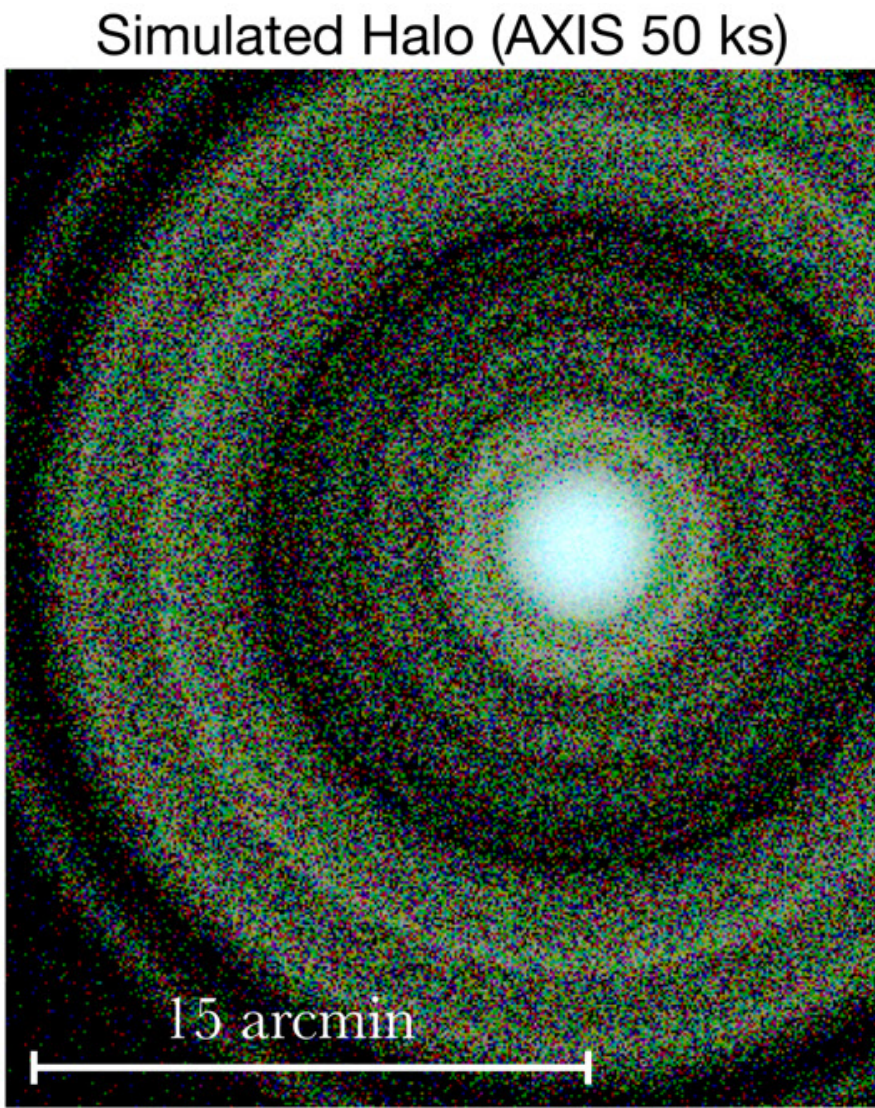}

    \caption{\AXIS\ will produce images of dust scattering halos of unprecedented quality, allowing detailed study of the composition and structure of the scattering medium, as well as a historic record of the source's flares. {\em Left}: \Chandra{} binned composite-color image of Cir X-1. Red: 1-2~keV, green: 2-3~keV, blue: 3-5~keV. {\em Right}: Simulated
      50~ks halo observation with \AXIS\ (no point sources.)}
    \label{fig:MW_dust}
\vspace*{-2mm}
\end{wrapfigure}

Understanding the structure and composition of Milky Way 
dust is vital for studies of the CMB, as dust can greatly impact CMB polarization
measurements\cite{Draine2009,Draine2013}. X-rays also provide otherwise
unobtainable constraints on interstellar dust models, which currently rely
on degenerate results from other
wavelengths\cite{Zubko2004,Jenkins2009,Valencic2015}. \AXIS\ will probe
interstellar dust through observations of scattering halos around
bright flaring sources, produced from small-angle scattering by intervening
dust, which has a scattering cross-section that increases rapidly with grain
size and is sensitive to grain composition\cite{Mathis1991}. \AXIS' rapid
response time and low background allow for X-ray tomography studies of the
ISM for $\sim$50 sightlines over \AXIS' minimum 5-yr lifetime, to greater
precision than current stellar population extinction maps\cite{Heinz2015}
(Fig.~\ref{fig:MW_dust}).

\subsection{Local Volume X-ray Binaries}

XRBs provide insights into star formation histories and stellar dynamics.
The XRB luminosity function (XLF) steepens with age\cite{Lehmer2017} and
depends on metallicity, so measuring the XLF independently constrains the
IMF and metallicity\cite{Coulter2017,Peacock2014,Peacock2017} for direct comparison 
with results derived from
%
%
optical, UV, IR and mm observations. \AXIS\ will
detect XRBs down to a $L_X = 10^{36.5}$\lumergs\ in galaxies within
$d < 20$~Mpc for exposures of $\sim$100~ks, determining the parameters of
the HMXB XLF and providing an independent SFR indicator for
normal galaxies which is insensitive to dust and has totally different
systematics\cite{Gilfanov2004}, with hundreds of XRBs detected per galaxy.
These data will be used in the high-$z$ regime to determine the fraction of
low mass XRBs that originate in dense stellar clusters, identify runaway
HMXBs, and measure the dependence of the HMXB population on metallicity.
Finally, monitoring galaxies within 10~Mpc ($10-30$~ks visits) over the
mission lifetime will quantify the variability among XRBs down to
$10^{37.5}$\lumergs\ and provide spectra and timing for the brightest
sources that will connect the classification of XRBs in the Local Group to
the wider population of galaxies and clarify the origin, nature and lifetime
of ULXs. \AXIS' 50~ms time resolution will allow
studies of the periodicity of $\sim$99\% of all known accreting X-ray
pulsars in the Milky Way, LMC, SMC and other nearby galaxies.

\subsection{Pulsar Wind Nebulae}
\label{section:pwn}

\begin{wrapfigure}{r}{0.53\textwidth}
\vspace*{-11.7mm}
    \centering
    \includegraphics[width=0.45\textwidth,viewport=91 462 318 652,clip]{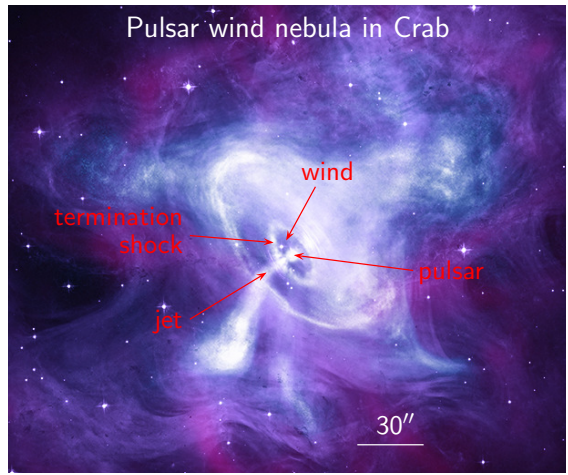}
    \caption{\AXIS\ will produce high-resolution images for many PWN, allowing
      multiband study of these potentially most powerful particle accelerators in the local Universe.
      Figure shows a multiband image of the Crab nebula\cite{Weisskopf2000}
      (white/blue: \Chandra, purple: HST, pink: \Spitzer). With \AXIS, such level of detail will become 
      possible for many PWN.}
    \label{fig:MW_Crab}
\vspace*{-10mm}
\end{wrapfigure}

Pulsar Wind Nebulae (PWNe) are synchrotron bubbles of relativistic plasma
inflated by the rotational energy loss of a fast spinning neutron star.
Confinement of the wind by the SN ejecta leads to the formation of a
`termination shock,' at which particles are accelerated and beyond which the
nebula forms. The termination shock is often at $<0.1$~pc from the pulsar,
thus requiring subarcsecond resolution to resolve it
(Fig.~\ref{fig:MW_Crab}). \Chandra{} has provided insight into some of their
fundamental properties, such as the origin of the characteristic torus-jet
structures and the strength of the nebular magnetic
field\cite{Kargaltsev2008,Bucciantini2011}, but studies have been limited to
a few nearby, bright, young PWNe. \AXIS\ will study a much larger sample of
PWNe in our Galaxy and the Magellanic Clouds, and will
shed light on the physics of pulsar wind magnetization, wind propagation,
and particle acceleration.

\section{SOLAR SYSTEM AND EXOPLANETS}
\label{section:PLANETS}

With its superior sensitivity and angular resolution, \AXIS\ can use variable
X-ray emission to probe the composition of local comets and  the evolution of solar system 
planetary atmospheres. \AXIS\ will also characterize exoplanet habitability and the role
of X-rays in planet formation and evolution through observing exoplanet host
stars and X-ray transits due to exoplanets.

\subsection{Comet Chemistry}

There are two sources of X-rays from comets. First, charge exchange (CX) between
solar wind ions and neutral atoms in cometary atmospheres produces low energy 0.1--1.0 keV
X-rays\cite{Wegmann2004} (first detected in Comet Hyakutake
1996/B2\cite{Lisse1996}). Second, nano-sized dust/ice particles scatter
solar X-rays\cite{Snios2014,Snios2018a} at $E>1$~keV. These mechanisms can
probe the comet surface and subsurface chemistry for comets crossing the
snow line, where the surface layers are sublimated.  \AXIS\ will resolve CX-driven X-ray emission from comets as far out as 2.5 au from the Sun, raising the number of comets visible in the X-rays
from 0.5~yr$^{-1}$ to $\sim$4~yr$^{-1}$, while also  permitting a comprehensive
study of scattered light at $E>1$~keV. 

\subsection{Variability of the Jovian Magnetosphere and Exosphere}

\begin{wrapfigure}{r}{0.68\textwidth}
\vspace*{-3mm}
    \centering
    \includegraphics[width=0.68\textwidth]{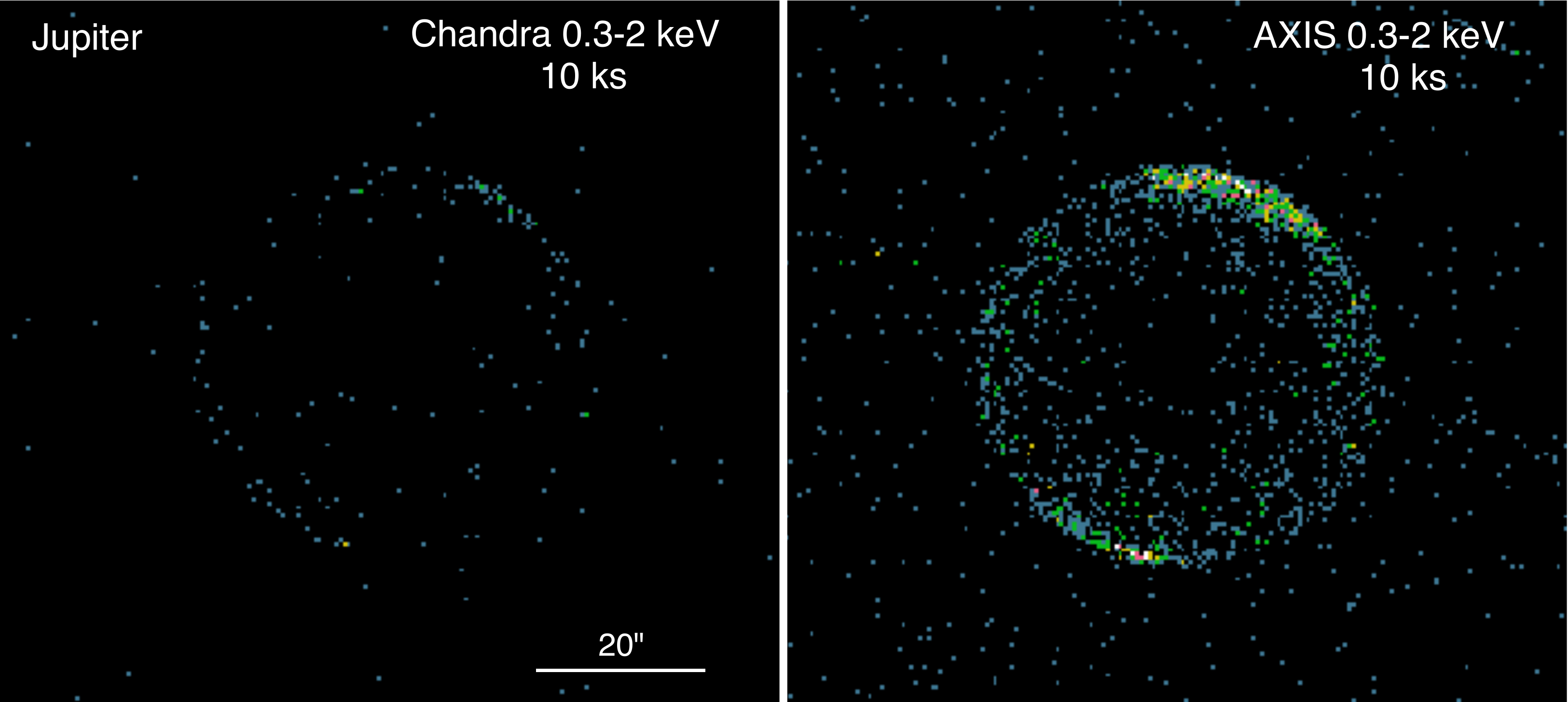}
    \caption{As shown by these simulated \Chandra{} ({\em left}) and \AXIS\ ({\em right}) observations 
      of Jupiter aurora, \AXIS' much larger soft X-ray effective area will resolve daily
      fluctuations in Jovian X-ray emissions and magnetic field reconnections on 1000-km scales, allowing 
      comparison of these changes to daily auroral and atmospheric weather events. The aurora is only in 
      view for 10-20 ks per 10-hour Jovian day.}
    \label{fig:PLANETS_jupiter}
\vspace*{-4mm}
\end{wrapfigure}

Jovian X-rays come from Jupiter's disk, poles, and flux
torus\cite{Waite1995}. The disk X-rays are scattered solar
photons\cite{Branduardi-raymont2008a,Branduardi-raymont2008b}, while the
polar and flux tube components are coupled to the particles, fields, and
``weather'' of the Jovian magnetosphere and
exosphere\cite{Crary1997,Connerney1998,Clarke2002}. Each source is strongly
variable, and \AXIS' sensitivity will allow, for the first time, monitoring
and mapping of each component with a cadence that can be correlated with the
\JUNO\ optical and UV maps (Fig.~\ref{fig:PLANETS_jupiter}). This will be instrumental in studying Jovian magnetic field properties known to
fluctuate on 1--3 day timescales, due to the loading/unloading rates of Io
plasma in the Jovian magnetosphere.

\subsection{Evolution of Planetary Atmospheres in the Solar System}

The role that X-rays play in the development and destruction of planetary
atmospheres is still not well understood. Solar UV light is known to cleave Earth's 
atmospheric ozone, and land-based life was not viable until terrestrial ozone 
was abundant enough to block $>$90\% of UV photons at the surface. Whereas extreme UV and X-ray
flares may damage atmospheres by removing bulk gas, depleting ozone (by
up to 90\% after X-ray flares\cite{Segura2010}), and dissociating water,
X-ray flares are also linked to higher abundances of various chemicals in
protoplanetary systems\cite{Glassgold2012,Cleeves2014,Cleeves2017}. \AXIS\
will provide robust spatial characterization of localized impacts of solar
X-rays/solar wind ions on planetary atmospheres, as well as probe the
dissipation of X-ray photons and excited secondary particles throughout the
atmosphere. Over larger time-scales, \AXIS\ will investigate the impact of
solar flares on the release of energetic particles from the upper
atmosphere, as well as on atmospheric size and density. Through studying the different 
atmospheres of our neighboring planets under extreme UV insolation, \AXIS\ will study the dependence
between atmospheric chemical abundance and high-energy solar activity at high spatial and temporal resolution.

\begin{figure}[t]
\vspace*{-1mm}
\colorbox{callout}{\color{white}\sfsm
\begin{minipage}{0.99\textwidth}\begin{minipage}{0.97\textwidth}
\vspace*{3mm}
\begin{itemize}[itemsep=5pt,labelwidth=0pt,labelindent=0pt]
\item \AXIS\ will investigate variations in comet emissions during perihelion
approach, including rapid outflow events and bow shock generation upon
crossing the snow line
\item \AXIS\ will observe high-energy particle transfer in Jupiter's upper
atmosphere and test results from \JUNO\ observations
\item \AXIS\ will study the impact of stellar X-ray activity on planetary
atmosphere evolution in Solar System and exoplanet environments
\end{itemize}
\vspace*{1pt}
\end{minipage}\end{minipage}}
\end{figure}

\subsection{High-Resolution Mapping of Elements on the Lunar Surface}

Lunar X-rays\cite{Adler1972} are dominated by fluorescence line emission
from O, Mg, Al, and Si\cite{Schmitt1991,Wargelin2004}. \AXIS\ can use these
lines to map abundances of the entire Moon with a resolution of 2~km in only 100~ks,
comparable to the 1-10~km resolution maps of only a small fraction of the lunar surface obtained by
lunar orbiters/landers (e.g., \textit{Clementine}, {\em LRO}), and an
order-of-magnitude improvement on the 20 km resolution of
{\em SELENE}\cite{Yokota2009}.

\subsection{Exoplanet Atmospheres}

Rocky planets have been discovered in the ``habitable zone'' around several
M-dwarfs, and \TESS\ will discover many more. However, such planets may lose
all their water due to stellar flares, which can dissociate water and then
evaporate the hydrogen. The extent to which this occurs is unknown,
especially among late M-dwarfs. \Chandra{} enabled a breakthrough study of
magnetic activity in young and evolved stellar
populations\cite{Feigelson2002,Nunez2015}, but \AXIS\ will perform deeper
stellar surveys and measure activity in smaller ``bins'' of spectral type
for many more stars, crucial to determining the typical activity of
planet-bearing stars.

\AXIS\ will directly study ``hot Jupiter'' planetary atmospheres through
transits. To date, the only X-ray transit observed was in HD~189733b, the first detected,
canonical hot Jupiter. The depth and length of the transit constrain the
atmospheric evaporation rate\cite{Poppenhaeger2013}, and \AXIS\ will enable
observations of at least an order of magnitude more transits than is
currently possible. There is also evidence that hot Jupiters affect the
magnetic fields of their host stars, as gauged by X-ray
flux\cite{Poppenhaeger2013,Poppenhaeger2014}. \AXIS\ will be able to confirm
or refute this picture with a much larger sample.

\vspace*{8mm}\noindent
\centerline{\textcolor{black}{\sf\LARGE MISSION IMPLEMENTATION}}
\vspace*{1mm}

\noindent New technology allowing a large, sharp, lightweight X-ray mirror is the major advance that makes the above scientific developments possible and sets \AXIS\ apart from all the existing and planned X-ray missions. The basis of the \AXIS\ mission design is simplicity --- one mirror, one detector with long heritage, no moving parts (beyond the post-launch focus adjustment), and
low-Earth orbit for low background, low-radiation environment and fast response.

\section{\AXIS\ MIRROR ASSEMBLY}
\label{section:mirrors}

The \AXIS\ mirror assembly represents a quantum leap from the state of art
represented by \Chandra{}. It is made possible by the recent conception and
development of the silicon meta-shell optics technology pursued by the Next
Generation X-ray Optics (NGXO) team at NASA Goddard Space Flight Center.
The design and implementation of the \AXIS\ mirror assembly drives the \AXIS\
science performance and programmatic requirements on mass, volume,
production cost, and schedule. It also incorporates knowledge and lessons
learned from designing and building mirror assemblies for past and current
observatories, including \Chandra{}, \XMM{}, \Suzaku{}, and \NuSTAR{}.
Table~\ref{table:MIRRORS_param} lists the characteristics of the \AXIS\ mirror
assembly for our baseline design.

{\rowcolors{3}{tablealt}{white}
\begin{table}[t]
    \centering
    \begin{tabularx}{\textwidth}{ L{4.6cm} c X}
        \rowcolor{callout}\textcolor{white}{\sfsm Parameter
          (unit)}\phantom{\raisebox{-3mm}{\rule{0cm}{8mm}}}& \textcolor{white}{\sfsm Value}
        & \textcolor{white}{\sfsm Comment} \\
        Focal length (mm)   & 9000 & Fits within a Falcon 9 fairing\\
        Outer diameter (mm) & 1700 & Fits within a Falcon 9 fairing; provides $A_{\text{eff}}$ at 1 keV\\
        Inner diameter (mm) & 300   & Provides $A_{\text{eff}}$  at 6 keV\\
        \rowcolor{tablealt} & 100 (axial) & Balances diffraction limits and off-axis response \\
        \rowcolor{tablealt} & 100 (azim) & Balances manufacturability and number of segments \\
            \multirow{-3}{4.6cm}{Dimensions of typical mirror segments (mm)} & 0.5 (thick) & Meets mass requirements\\
        Number of mirror segments & 16,568 & Not including stray light baffle segments\\
        Number of mirror modules & 188 & The mirror segments are assembled into 188 separate modules, each are separately and fully tested\\
        Number of meta-shells & 6 & The modules are assembled into 6 separate meta-shells.\\
        Mass of mirror assembly (kg) & 454 & Includes mass of mirror segments, stray light baffles and spider platform to which meta-shells are attached\\
        \rowcolor{white} &\phantom{1}1 keV:~ 7700 & \\
        \rowcolor{white} &\phantom{1}6 keV:~ 1600 & \\
        \rowcolor{white}\multirow{-4}{4.6cm}{Mirror-only \newline effective area
          (cm$^2$)}  & 12 keV:~ \phantom{1}180 & \multirow{-3}{3.5in}{Large effective areas are made possible by the small thickness of the mirror segments}\\
        \rowcolor{tablealt} Mirror on-axis PSF \newline 
         (HPD, arcsec) & 0.4 & 1.5$\times$ better
        than \Chandra's mirror; a combination of precision
        polishing technology and mono-crystalline silicon\\
        \rowcolor{white!100}Mirror 15$^{\prime}$ off-axis PSF \newline 
        (HPD, arcsec) & 1.0 &
        28$\times$ better than \Chandra{}; Wolter-Schwarzschild design and short axial length of mirror segments\\
        \hline
    \end{tabularx}
    \caption{Characteristics and key parameters of the \AXIS\ Mirror Assembly}
    \label{table:MIRRORS_param}
\end{table}
}

\subsection{The Silicon Metashell Approach}

\begin{wraptable}{r}{10cm}
    \centering
\vspace*{-4.5mm}
{\rowcolors{4}{tablealt}{white}
    \begin{tabularx}{9.52cm}{ C{1.7cm} C{1.cm} C{1.2cm} C{1.cm} C{1.cm} C{1.1cm} }
        \rowcolor{callout}\textcolor{white}{\sfsm Metashell}\phantom{\rule{0cm}{5mm}} & \textcolor{white}{\sfsm Mass} &
        \multicolumn{4}{c}{\cellcolor{callout}\textcolor{white}{\sfsm Effective Area (cm$^{\sf 2}$) at:}} \\
        \rowcolor{callout} & \textcolor{white}{\sfsm (kg)} & \textcolor{white}{\sfsm 1~keV} & \textcolor{white}{\sfsm 4~keV} & \textcolor{white}{\sfsm 6~keV} & \textcolor{white}{\sfsm 12~keV} \\
        1 (inner)& 37 & 480  & 390 & 410  & 180 \\
        2        & 49 & 840  & 610 & 620  & 3   \\
        3        & 58 & 1180 & 720 & 530  & 0   \\
        4        & 67 & 1500 & 670 & 60   & 0   \\
        5        & 73 & 1740 & 420 & 5    & 0   \\
        6        & 80 & 1960 & 120 & 1    & 0    \\
        Total   & 363 & 7700 & 2900 & 1600 & 180 \\
        \hline
    \end{tabularx}
    \caption{Contributions of each metashell to mass and area.}
    \label{table:MIRRORS_metashells}
    }
\vspace*{-2mm}
\end{wraptable}

The silicon meta-shell optics (SMO) technology has been in development since
2012. It combines the precision optical polishing technology that made
\Chandra{}'s exquisite PSF possible with the use of mono-crystalline
silicon material, fabricated into 0.5mm thick mirrors. Taking advantage of
the ready availability of mono-crystalline silicon, and the equipment and
processing knowledge accumulated by the semiconductor industry, this
technology meets the following three-fold requirement on the \AXIS\ X-ray
mirror assembly: (1) better PSF than the \Chandra{} mirror's 0.6\arcsec\ HPD, (2)
more than 10 times lighter per unit effective area, and (3) more than 10
times less expensive per unit effective area.

\AXIS\ uses a variant of the Type-I Wolter-Schwarzschild design, which meets the Abbe sine condition for imaging, giving better off-axis PSF than the Type-I Wolter design used by \Chandra{}\cite{Saha2019}. The off-axis PSF is further improved by a slight defocusing of the on-axis PSF from the geometrically perfect PSF to one that is somewhat larger (negligible compared to diffraction limits). Most importantly, since the off-axis PSF degradation is dominated by focal surface curvature, inversely proportional to the length of the mirror element in the optical axis direction and different for shells of different radius, \AXIS' use of shorter mirror segments (100~mm compared to \Chandra's 840~mm) improves the theoretical off-axis PSF by a large factor.

\begin{wrapfigure}{r}{0.52\textwidth}
\vspace*{-2mm}
\centering
 \includegraphics[width=0.5\textwidth,viewport=25 4 380 309,clip]{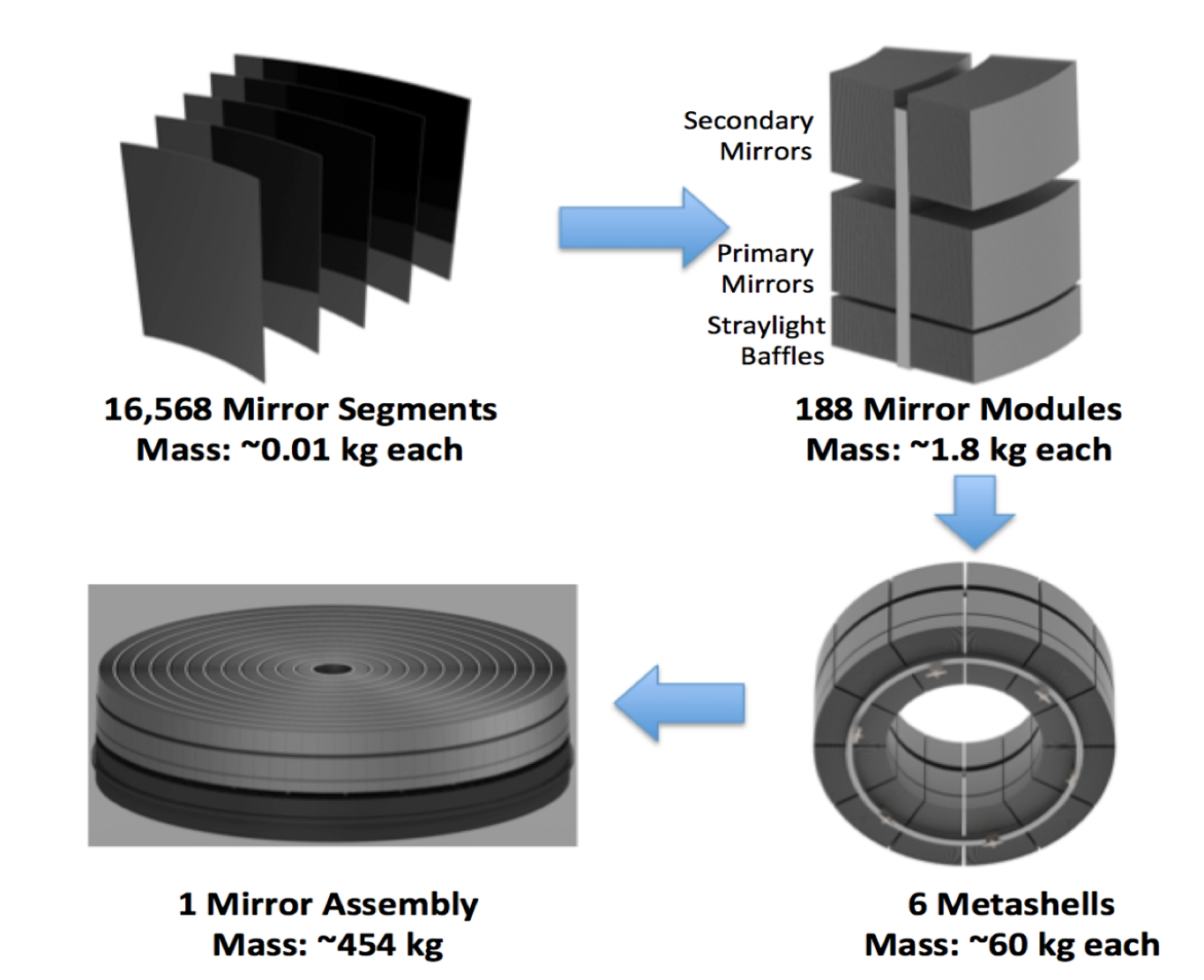}
\caption{A hierarchical approach to build the \AXIS\ mirror assembly that enables mass 
parallel production. It uses commercially available materials and equipment to minimize both
  production cost and schedule.}
    \label{fig:MIRROR_fig1}
\vspace*{-2mm}
\end{wrapfigure}

There are four major steps in building the \AXIS\ mirror assembly, as
illustrated in Fig.~\ref{fig:MIRROR_fig1}: (1) mirror segment fabrication,
(2) integration of mirror segments into mirror modules, (3) integration of
mirror modules into mirror meta-shells, and finally (4) integration of
mirror meta-shells into the final mirror assembly. The \AXIS\ mirror assembly
has several salient features that make it easy to accommodate on the
observatory. First, the entire mirror assembly is made of silicon, except
for trace amounts of iridium for enhancing X-ray reflectance, chromium for
binding the iridium to the silicon surface, silicon oxide for canceling
the thin film stress of the iridium, and epoxy for bonding mirror segments to
structures, all of which have no significant
thermal or structural impact. This effectively uniform material composition
enables the mirror assembly to operate at a temperature different from room
temperature ($\sim$20$^{\circ}$~C) at which it is built and tested. The lack
of a stringent operating temperature constraint significantly reduces
the need for precise thermal environmental control, reducing costs of building, testing, and operating the mirror assembly. Second, the modular approach makes it highly amenable to parallel production. The mirror segment's similarity in dimension to the semiconductor industry silicon wafer makes its production highly similar to the wafer production process, allowing use of commercially available equipment and knowledge to minimize cost and schedule. Third, the large numbers of mirror modules and mirror meta-shells make it easy to manage spares. The modest size of the mirror modules requires no special equipment to handle, test, or qualify.

Only the fabrication of mirror segments, and integrating (aligning and
bonding) them into mirror modules, require technology development. The other
two, integrating modules into meta-shells and integrating meta-shells in
turn into the mirror assembly at the required level of precision, are
routine engineering and I\&T work implemented in many past missions.

\subsection{Mirror Technology Development}

The NGXO team at NASA Goddard Space Flight Center
was established to develop X-ray mirror technology to meet the needs of
future X-ray missions such as \AXIS\ and  \Lynx{}. Their work started in
2001 with development of a precision glass slumping process to meet the
requirements of the \textit{Constellation-X} mission. Concurrent with that
development, the NGXO team successfully fabricated over 10,000 glass mirror
segments for the highly successful \NuSTAR{} mission. In 2012, the technique of combining precision polishing technology with mono-crystalline silicon was adopted. This allows us to achieve high angular resolution with a very light mass. Further, it has the added benefit that it uses readily-available semiconductor industry mono-crystalline silicon material, with corresponding processing equipment and technology.

\subsection{Mirror Segment Fabrication}

\begin{wrapfigure}{r}{0.67\textwidth}
\vspace*{-5mm}
\centering
  \includegraphics[width=0.65\textwidth,viewport=21 6 500 335,clip]{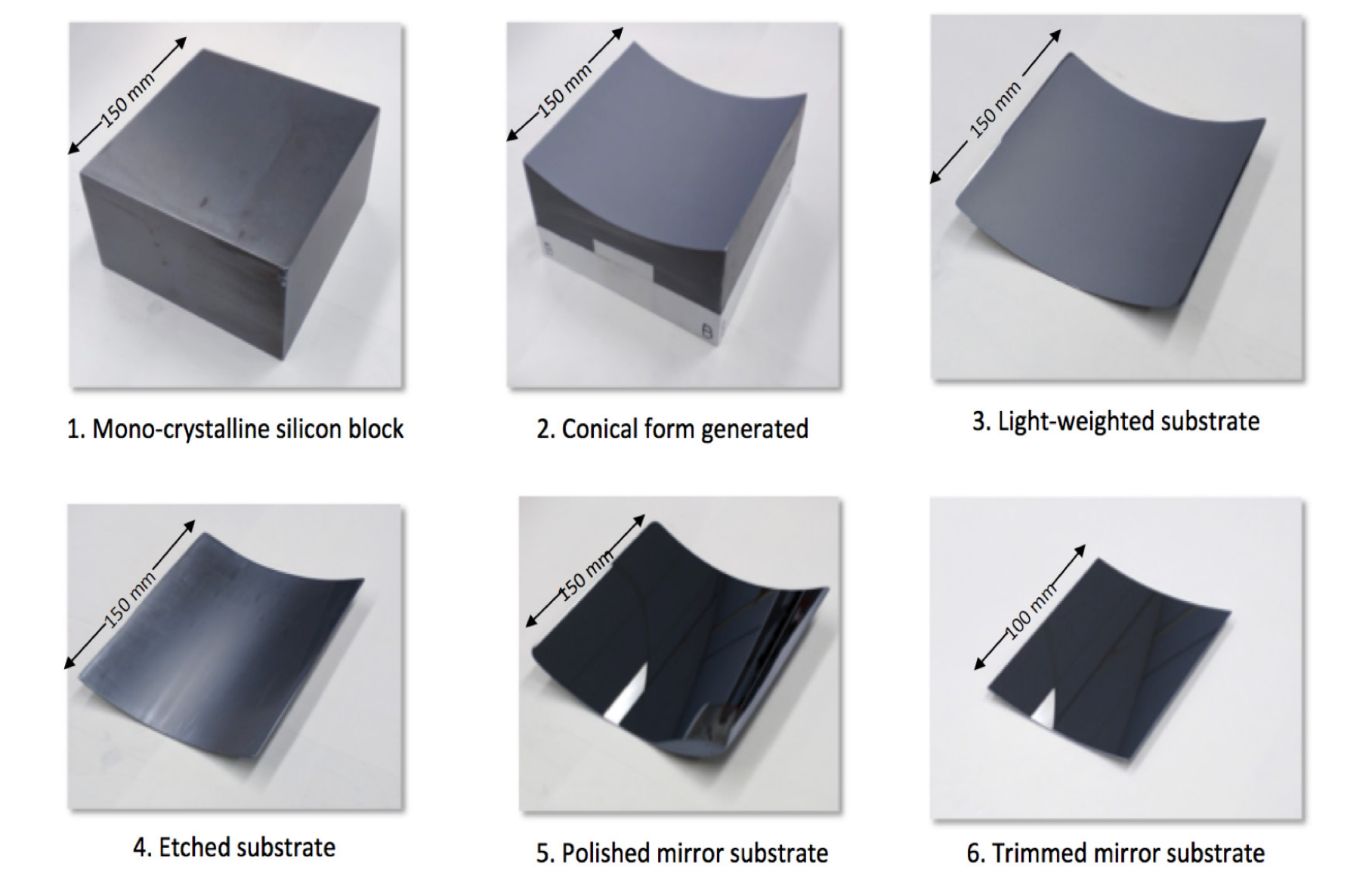}
\caption{Fabricating a mirror segment takes about four days of calendar time and 
10 hours of labor per segment.}
    \label{fig:MIRROR_fig2}
\end{wrapfigure}

The fabrication of a mirror segment (Fig.~\ref{fig:MIRROR_fig2}) starts
with a block of mono-crystalline silicon, measuring 150 $\times 150$       $\times75$ mm 
(Fig.~\ref{fig:MIRROR_fig2}.1). After an approximate conical contour is
cut into the block, it is ground on a computer numerical control
machine to improve the approximation to within 20~$\mu$m of the mathematical
prescription. The conical form is then lapped on a stainless steel cone that
is precisely machined with prescribed cone angle and radius. After
this step, the conical surface is an excellent first order approximation to
the prescribed mirror segment figure (Fig.~\ref{fig:MIRROR_fig2}.2). The block is next placed under a slicing saw to cut off the conical thin shell
(Fig.~\ref{fig:MIRROR_fig2}.3). Damage to the crystal structure from the
lapping and slicing operations results in unpredictable deformation. This
damage is removed, and therefore its conical shape restored, by etching the
thin shell in a standard HNA (hydrofluoric acid, nitric acid, and acetic
acid) solution that removes the atoms that were displaced from their lattice
locations (Fig.~\ref{fig:MIRROR_fig2}.4).  The conical shell is then
temporarily bent into a cylindrical shape, and polished on a cylindrical
surface padded with synthetic silk that has been developed by the
semiconductor industry for polishing silicon wafers
(Fig.~\ref{fig:MIRROR_fig2}.5). This is equivalent to the stress-polishing
invented by and successfully applied to mirror segments for the
Keck Telescopes.

At the end of the 20-hour polishing process, the mirror segment 
attains the required
micro-roughness. The mirror segment at this
stage, however, has severe figure errors near its four edges as a result of
the polishing process. In the next step, about 25mm from each of the four
sides of the mirror segment is trimmed with a standard abrasive saw,
resulting in a mirror segment that is approximately
100mm$\times$100mm$\times$0.5mm (Fig.~\ref{fig:MIRROR_fig2}.6). As an
integral part of each of these steps, any damage to the crystal structure is
carefully removed by either HNA etching or careful polishing, ensuring that
no surface or subsurface damage that stores energy and causes stress and
figure distortion remains. The last step in mirror fabrication uses a
state-of-the art ion-beam figuring machine to improve the figure of the
mirror segment to 0.2\arcsec\ HPD (two reflection equivalent), which meets
the \AXIS\ requirements. Typically the ion-beam figuring step removes about
300~nm of material from the mirror surface, improving the figure from about
3\arcsec\ to 0.3\arcsec\, while leaving the excellent micro-roughness
intact, at about 0.3~nm RMS as measured on 0.4mm$\times$0.4mm square. {\bs As of
February 2019, mirror segments with figure quality of 0.5\arcsec\ HPD have
been regularly fabricated.} Further improvement in the mirror fabrication process in
2019 is expected to result in mirror segments of 0.2\arcsec\ HPD or better.

\begin{table}[t]
    \centering
    \begin{tabularx}{\textwidth}{ L{0.77in} L{1.69in} L{1.69in} L{1.69in}}
    \rowcolor{callout} \phantom{\raisebox{-3mm}{x}\raisebox{3mm}{x}}&  
    \textcolor{white}{\centerline{\sfsm TRL-4}}
    & \textcolor{white}{\centerline{\sfsm TRL-5}} & \textcolor{white}{\centerline{\sfsm TRL-6}} \\
    \hline
    Illustration & \begin{minipage}{1.6in}
    \fcolorbox{white}{white}{
    \includegraphics[width=\linewidth, height=1in]{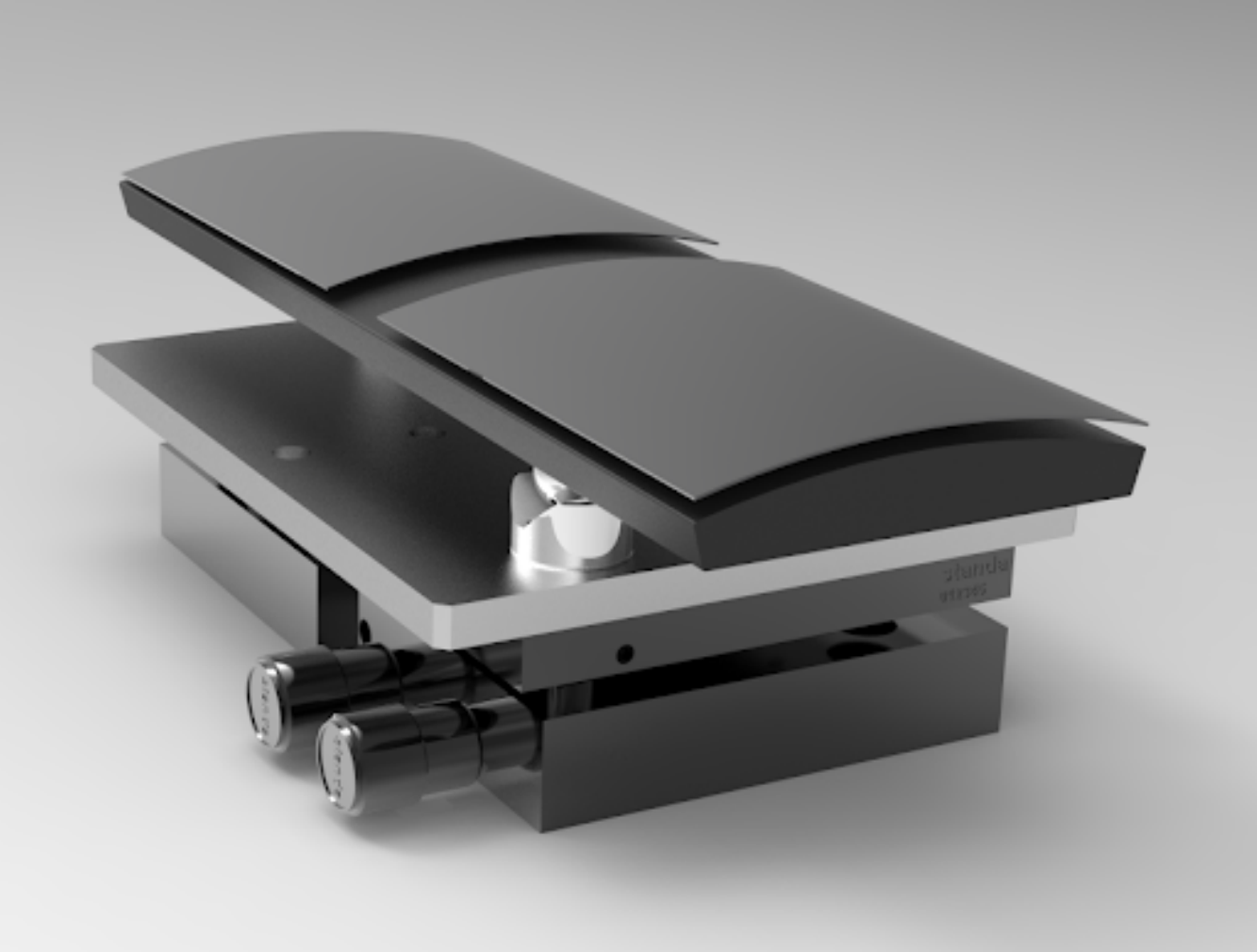}}
    \end{minipage} & \begin{minipage}{1.6in}
    \includegraphics[width=\linewidth, height=1in]{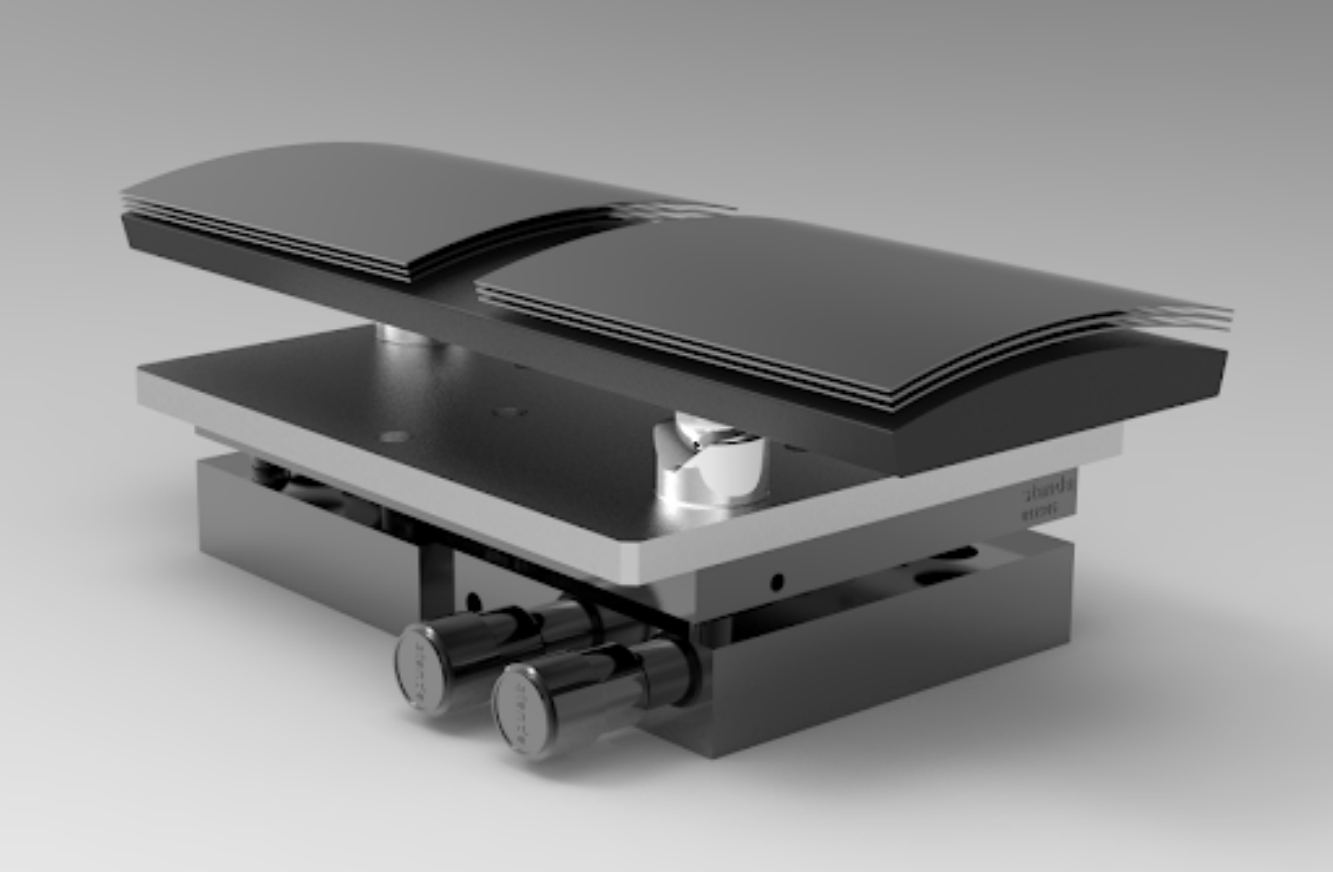}
    \end{minipage} & \begin{minipage}{1.6in}
    \includegraphics[width=\linewidth, height=1in]{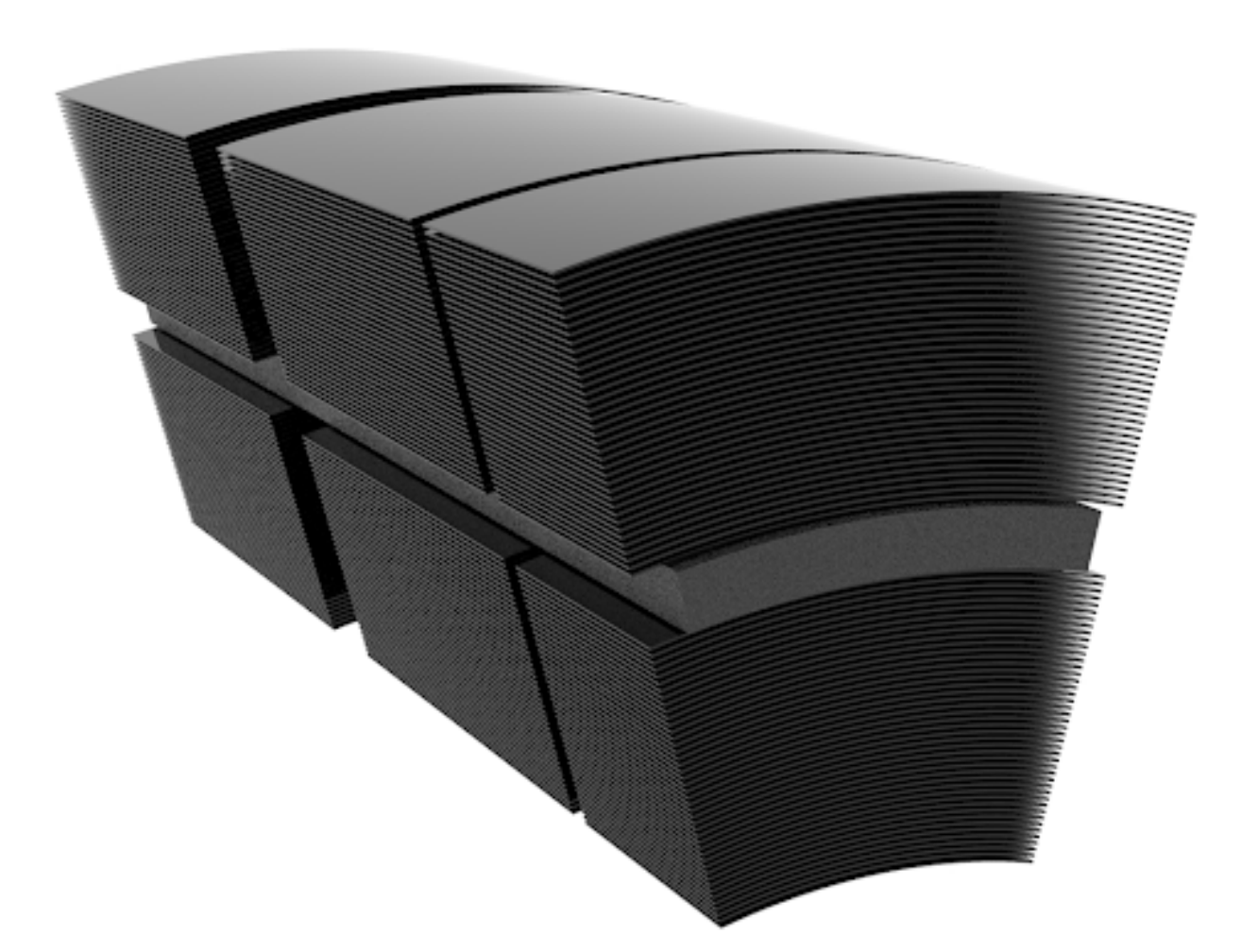}
    \end{minipage} \\
    \rowcolor{tablealt}
    Description &
    Fabrication, alignment, and
    bonding of single pairs of mirror segments to achieve progressively
    better PSF, culminating in 0.3\arcsec\ HPD. &
    Alignment and bonding of
    2-3 pairs of mirror segments to achieve 0.3\arcsec\ HPD. 
    Demonstrate the structural and other
    environmental integrity of the mirror bonds. &
    Alignment and bonding of many ($>$3) pairs of mirror segments to achieve
    0.3\arcsec\ HPD for a mirror module, passing all
    environmental tests. \\
    \rowcolor{white!100} Objectives &
    1. Develop and verify mirror
    fabrication and mirror coating processes. 2. Develop and verify the
    basic elements of alignment and bonding procedures for precision and
    accuracy. &
    1. Develop and verify mechanics and speed of co-alignment
    and bonding processes. 2. Conduct environmental tests: vibration,
    thermal vacuum, and acoustic to verify structural and performance
    robustness. &
    1. Develop and verify meta-shell production process:
    mirror fabrication, coating, alignment, and bonding. 2. Validate
    production schedule and cost estimates. 3. Develop plan for mass
    production.\\
    \rowcolor{tablealt} 2018 Status & Repeated building/testing,
    achieving $\sim$2\arcsec\ HPD. & Build and test one module, achieving
    $\sim$5\arcsec\ HPD. & In progress.\\
    \rowcolor{white!100} Completion & December 2020 & December 2022 & December 2024\\
    \hline
    \end{tabularx}
    \caption{The status and maturation plan for the mirror technology.}
    \label{table:MIRRORS_plan}
\end{table}

\subsection{Mirror Segment Coating}

\begin{wrapfigure}{r}{0.6\textwidth}
    \centering
    \includegraphics[width=0.58\textwidth,viewport=13 14 484 120,clip]{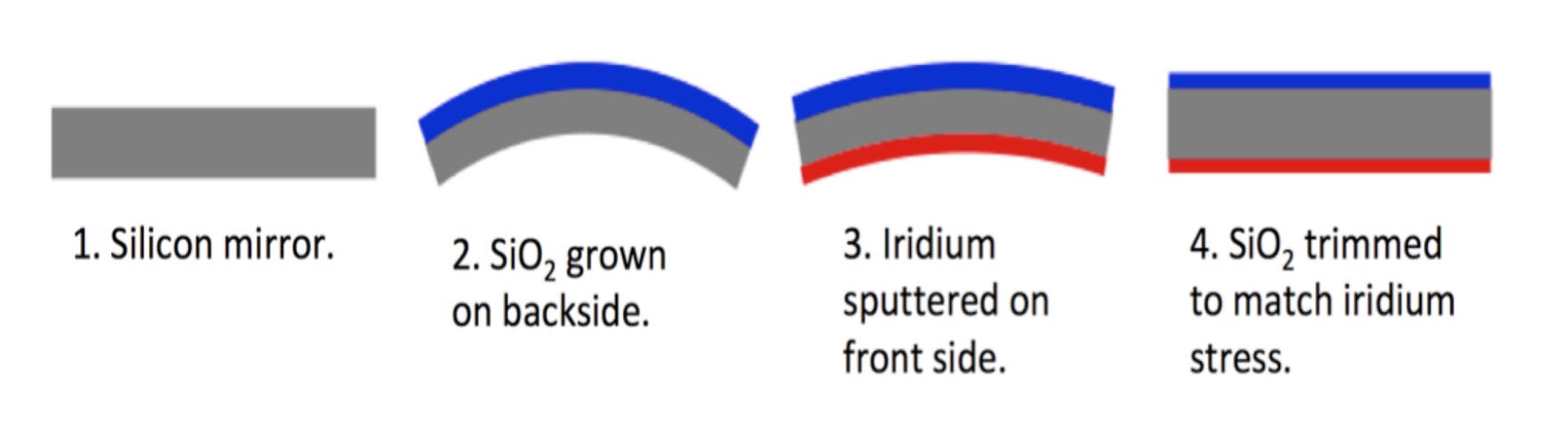}

    \caption{Coating a mirror segment with
      reflectance-enhancing iridium layer without degrading its figure has been demonstrated in the laboratory.}
    \label{fig:MIRROR_fig3}
\vspace*{-1mm}
\end{wrapfigure}

After it is qualified by extensive measurements, the mirror segment is coated with
30~nm of reflectance-enhancing iridium. The compressive stress of the
iridium coating, which can severely distort the figure, is compensated for
by approximately 300~nm of silicon oxide on the backside. The coating is
accomplished in steps illustrated in Fig.~\ref{fig:MIRROR_fig3}. After the mirror
segment is fabricated, it is heated to 1,050$^{\circ}$~C to grow a
$\sim$300-nm layer of silicon oxide on its backside, a standard, proven
industrial process. As with the iridium thin film, the silicon oxide film
has compressive stress. Next, $\sim$30~nm of iridium is
sputtered on the mirror segment. The stresses of the oxide film and the
iridium film largely cancel each other out. However, the cancellation is not always 
complete, leaving behind a residual figure distortion. 

In the final step, the residual figure distortion is measured on an 
interferometer, from which a residual stress map
is calculated. It is then converted into a thickness map of the silicon oxide
film. This thickness map is finally fed into an ion-beam figuring machine
for residual stress removal, achieving precise cancellation of figure
distortion. The validity of this process was fully demonstrated in a
collaborative effort between the NGXO group and scientists at MIT's Kavli
Institute using a wet chemical process to trim the silicon oxide thickness\cite{Yao2019}. 
Several mirror segments have been
coated and treated in this manner and display a net PSF degradation of less
than 0.2\arcsec\ HPD (two reflections equivalent). 

As of February 2019, the NGXO group had installed a state-of-the-art 
ion beam figuring machine. The machine is fully operational and the 
NGXO group is making 0.5\arcsec\ mirror segments on a routine basis. It is 
anticipated that in 2019 the entire process of coating a mirror 
segment shown in Fig.~\ref{fig:MIRROR_fig3} will be fully demonstrated 
to meet \AXIS\ and \Lynx\ requirements.

\subsection{Mirror Segment Alignment}

\begin{wrapfigure}{r}{0.53\textwidth}
\vspace*{-10mm}
    \centering
    \includegraphics[width=0.5\textwidth,viewport=29 10 420 323,clip]{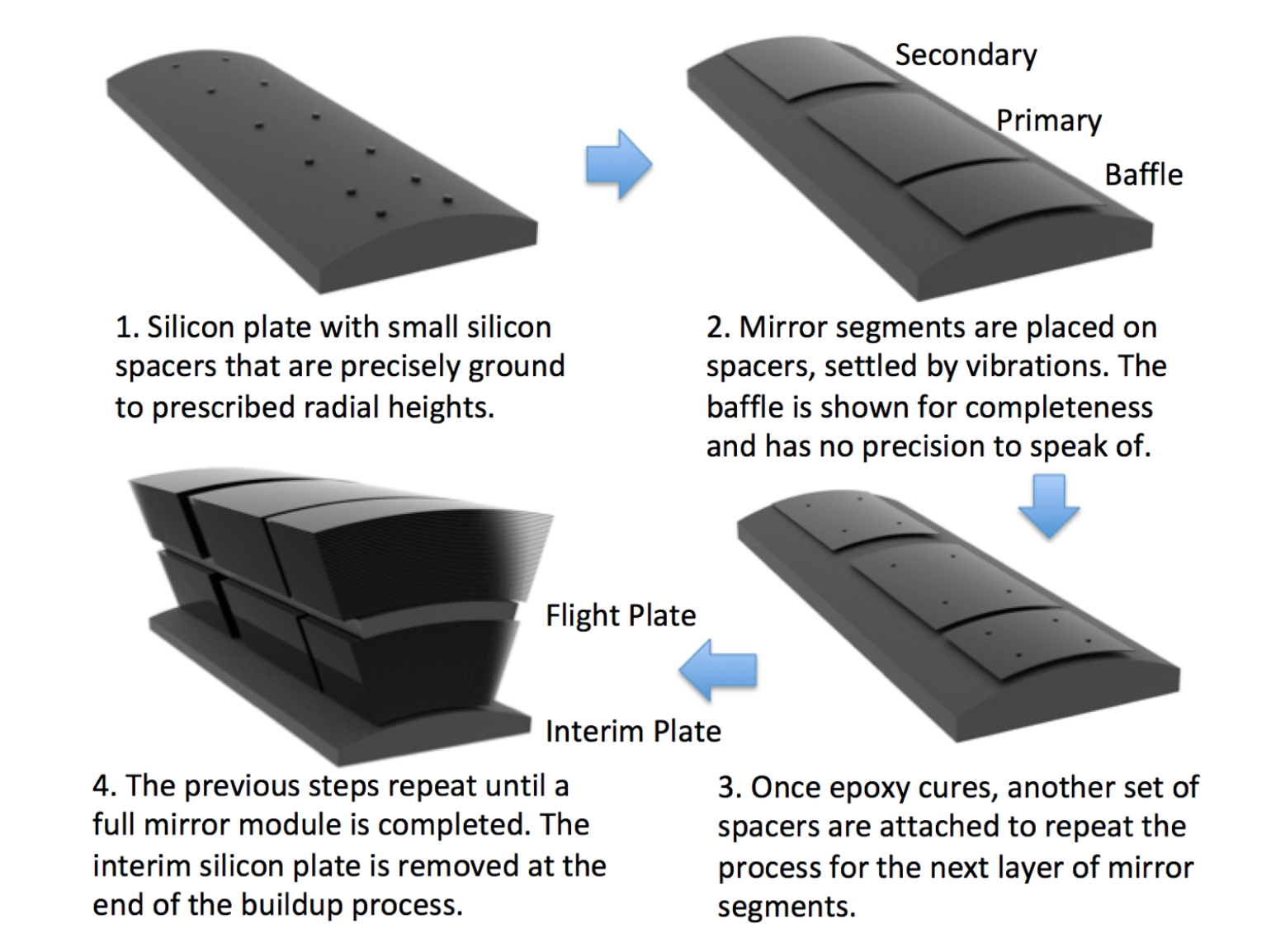}
    \caption{The process of aligning and bonding mirror segments to make a
      mirror module is straightforward. Note that, except for the epoxy and trace amounts of
      iridium and silicon oxide on each mirror segment, all components are
      made of silicon, including mirror segment, spacers, the interim plate,
      and the flight plate.}
    \label{fig:MIRROR_fig4}
\vspace*{-5mm}
\end{wrapfigure}

Fig.~\ref{fig:MIRROR_fig4} shows schematically the integration of mirror segments into a
mirror module. One important point to note is that, while three points
uniquely determine the location and orientation of a flat mirror, it takes
four points to uniquely determine the location and orientation of a curved
mirror. Therefore, each mirror segment is aligned and bonded using 4 spacers
(Fig.~\ref{fig:MIRROR_fig4}). The process starts by
attaching four spacers on a silicon plate for each mirror segment. The four
spacers are then ground to precise heights under the guidance of a radius
gauge. The mirror segment is next placed on the spacers, and its alignment
is checked by a set of precise Hartmann measurements using a laser beam. The
alignment errors determined by the Hartmann measurements are then used for
further grinding of the heights of the spacers. This iteration of
measure-and-grind continues until the mirror segment achieves its prescribed
alignment. Once alignment is achieved, a small amount of epoxy is applied on
top of each of the four spacers and the mirror segment is placed on them.
When the epoxy cures, the mirror segment is permanently bonded. This process 
is done in parallel with the segment 
fabrication and takes 4 hours of clock time and 10 hours of labor per segment.

\begin{wrapfigure}{r}{0.58\textwidth}
\vspace*{2.7mm}
    \centering
   \includegraphics[width=0.53\textwidth,viewport=13 11 517 257,clip]{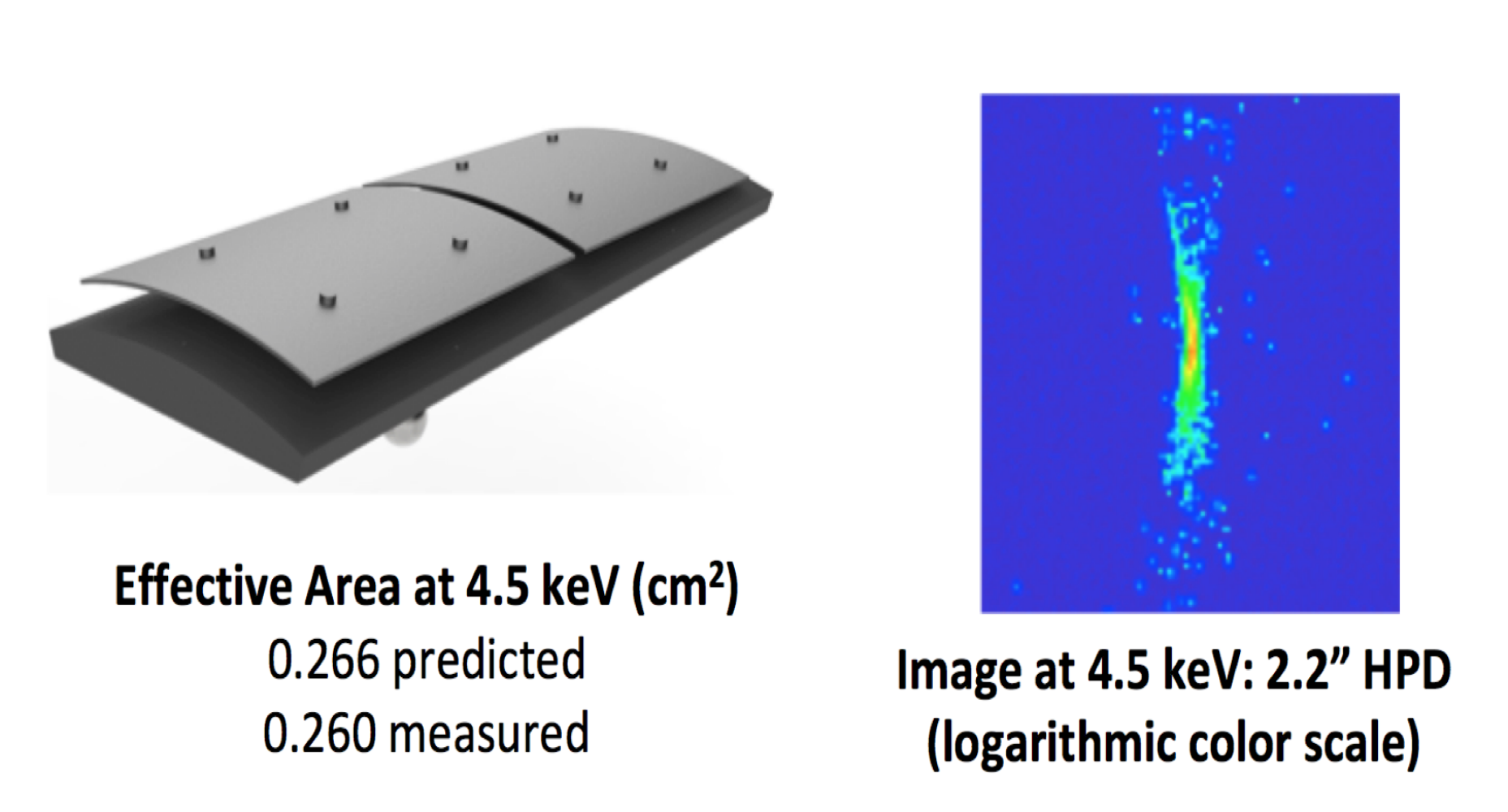}
    \caption{Laboratory measurements have validated the mirror development approach. 
    A pair of mirror segments have been fabricated, aligned,
      bonded, and tested in an X-ray beam, achieving a PSF of 2.2\arcsec\
      HPD (uncorrected for gravity distortion) with 
      expected effective areas. Further optimization and improvement in
      alignment and bonding are needed to realize the full potential of the
      mirror segments of 0.3\arcsec\ HPD.}
    \label{fig:MIRROR_fig5}
\vspace*{-8mm}
\end{wrapfigure}

The validity of the alignment and bonding process has been demonstrated by 
testing of several mirror modules containing one or
two pairs of mirror segments, achieving 2.2\arcsec\ HPD under full
illumination with 4.5 keV photons in an X-ray beam line. A test image 
shown in Fig.~\ref{fig:MIRROR_fig5} was obtained with the mirror 
module in horizontal position (due to limitations of the beam line) and
includes the effect of gravity, which is analyzed and determined to be 
about 1.5\arcsec, leaving about 1.6\arcsec\ for the intrinsic image error for the module. The intrinsic error further includes the effect of 
energy-dependent scattering. After correcting the image from 4.5 keV 
to 1 keV X-rays, the image quality becomes 1.3\arcsec\ HPD (bottom of 
Table\ \ref{table:MIRRORS_error}).

\begin{table}[b]
    \centering
    \begin{tabularx}{\textwidth}{L{1in} L{0.85in} C{0.58in} C{0.76in} L{2.48in}}
        \rowcolor{callout}
        \multicolumn{2}{L{2.02in}}{\cellcolor{callout}\centerline{\textcolor{white}{\sfsm Source of error} \phantom{\rule{0cm}{5mm}}  } }
            & \textcolor{white}{\sfsm Req.} & \textcolor{white}{\sfsm 2018 status} & \centerline{\textcolor{white}{\sfsm Notes}} \\
        \rowcolor{callout} \multicolumn{2}{L{2.02in}}{\cellcolor{callout}} &
        \textcolor{white}{\sfsm (\arcsec\ HPD)} & \textcolor{white}{\sfsm (\arcsec\ HPD)} & \\            
        \rowcolor{white!100} Optical prescription & Diffraction \newline Geometric PSF (on-axis) & 0.1 \newline 0.05 & 0.1 \newline 0.05 & Weighted mean of all shells at 1~keV. \newline The on-axis design PSF is degraded to achieve best possible off-axis PSF\\
        \rowcolor{tablealt}Mirror segment fabrication & Mirror substrate & 0.2 & 0.5 & With one iteration on an ion-beam machine. With two iterations substrates should meet requirements.\\
        \rowcolor{tablealt} & Coating & 0.1 & 0.2 & 20nm iridium and 5nm of chrome as a bonding layer on the frontside. 300 nm of silicon oxide on the backside. Better stress cancellation achievable by finer trimming of the oxide.\\
        \rowcolor{white!100}Mirror module & Alignment & 0.1 & 1.1 & Dominated by Hartmann measurement precision (to be improved by using shorter $\lambda$ and less coherent light).\\
        \rowcolor{white!100} & Bonding & 0.2 & 0.4 & 0.4\arcsec\ is an upper limit. Additional measurements/testing are needed.\\
        \rowcolor{tablealt} Meta-shell construction & Alignment & 0.1 & 0.1 & Each module's image to be within 0.1\arcsec\ of the total image (de-center error $=3$~$\mu$m, de-space error $=10$~$\mu$m, and roll angle error $<$0.1\arcsec). Pitch and yaw requirements $\approx$10\arcsec.\\
        \rowcolor{tablealt}& Bonding & \cellcolor{tablealt!80!gray}0.1 & \cellcolor{tablealt!80!gray}0.1 & \cellcolor{tablealt!80!gray} \\
        \rowcolor{white!100} Integration of meta-shells & Alignment & \cellcolor{gray!20!white}0.1 & \cellcolor{gray!20!white}0.1 & \cellcolor{gray!20!white} \\
        \rowcolor{white!100} & Attachment & \cellcolor{gray!20!white}0.1 & \cellcolor{gray!20!white}0.1 & \cellcolor{gray!20!white} These numbers are based on\\
        \rowcolor{tablealt} Ground-to-orbit effects & Launch shift &
        \cellcolor{tablealt!80!gray}0.1 &
        \cellcolor{tablealt!80!gray}0.1 &
        \cellcolor{tablealt!80!gray}  preliminary analyses, but show that \newline \hspace{0.0cm} requirements can be met. \\ 
        \rowcolor{tablealt}& Gravity Release & \cellcolor{tablealt!80!gray}0.1 & \cellcolor{tablealt!80!gray}0.15 & \cellcolor{tablealt!80!gray} \\
       \rowcolor{tablealt}& Thermal & \cellcolor{tablealt!80!gray}0.1 & \cellcolor{tablealt!80!gray}0.15 & \cellcolor{tablealt!80!gray} \\
       \rowcolor{white!100}In-orbit mirror performance &  & 0.4 & 1.3 & On-axis mirror assembly (not including jitter/detector pixellation).\\
        \hline
    \end{tabularx}
    \caption{The mirror error budget.}
    \label{table:MIRRORS_error}
\end{table}

Further refinement of the alignment and bonding process will be done in 2019
to fully realize the 0.3\arcsec\ potential of the mirror segments, including
better control of the absolute radial placement of each mirror segment, and
fine adjustment of the mirror segment in the axial direction. We expect
that, by sometime in 2019, X-ray images of 0.4\arcsec\ HPD will be obtained
with a fully aligned and bonded mirror pair with full illumination.

\subsection{Testing and Qualification of the Mirror Module}

Successful technology development requires constructing mirror modules
demonstrated to meet all requirements, including those involving science
performance, environment, production budget and schedule. The mirror module,
typically composed of dozens of mirror segments, once completed, undergoes a
battery of science performance and spaceflight environmental tests. For
science performance, it is subject to measurement in an X-ray beam for both
PSF and effective area at several energies, both before and after
environmental tests that include vibration, thermal vacuum, acoustic, and
shock components.
 
\subsection{Technology Development Schedule}

X-ray mirror technology was recognized by the astronomical community and
NASA to be of strategic importance. The NGXO
team is funded at \$2.4M a year to advance the silicon meta-shell optics
technology. We expect that this level of funding is adequate to achieve
TRL-5 for \AXIS\ by 2022 (Table~\ref{table:MIRRORS_plan}). Additional
project-specific funding is needed to achieve TRL-6.
Table~\ref{table:MIRRORS_error} shows a comparison of major error terms to
requirements. Most of the error terms are at present within
a factor of $\sim 2$ of meeting \AXIS\ requirements.
We are in the process of procuring necessary tooling to build and test a mirror module with 12 pairs of mirror segments, but designed to have a angular resolution of 5".  This mirror module will be subject a set of X-ray performance tests before and after a battery of environmental tests, including vibrations, acoustics, and thermal vacuum, a necessary step towards better understanding of the mirror construction process.

\section{\AXIS\ DETECTOR ASSEMBLY}
\label{section:detector}

The design of the \AXIS\ focal plane is driven by the need to take 
advantage of the \AXIS\ mirror assembly. The key technical challenges for \AXIS\
are: small pixel size; high readout rate with low noise
to ensure good low energy response; and fast, low-noise onboard processing
electronics. The \AXIS\ design exploits ongoing technical advances toward
``fast, low-noise, megapixel X-ray imaging arrays with moderate spectral
resolution,'' identified by the 2017 Physics of the Cosmos Program Annual
Technology Report as a top-priority technology
development\footnote{https://ntrs.nasa.gov/search.jsp?R=20170009472}. Our
plan utilizes both fast parallel-readout CCDs, capitalizing on decades of
heritage provided by X-ray CCD detectors used to great success aboard
\Chandra{}, \Suzaku{}, \Swift{}, and \XMM{}, and a fast, low-power CMOS active
pixel sensor with less heritage. At present, considering the use of both
technologies minimizes technical risk in the design phase and allows us to take advantage of
detector technology developments of the last 25 years and those anticipated
over the next few years based on ongoing development work. The final design 
will likely incorporate a single detector technology and greatly simplify
many aspects of the focal plane reducing risk and cost in the construction phase.  A summary of the \AXIS\ detector assembly 
design and drivers is given in Table~\ref{table:DETECTOR_params}.

{\rowcolors{3}{tablealt}{white!100}
\begin{table}[t]
    \centering
    \begin{tabularx}{\textwidth}{L{2.1in} L{1.9in} L{2.003in}}
        \rowcolor{callout}
        \textcolor{white}{\sfsm Parameter}{\rule{0cm}{5mm}} & \textcolor{white}{\sfsm Value} & \textcolor{white}{\sfsm Comment}\\
        Focal plane layout & 4 CCD, 1 CMOS & See Fig.~\ref{fig:DETECT_fpa_schem} \\
        Field of view      & $24^{\prime}\times 24^{\prime}$, ~6.4$\times$6.4~cm, 4000$\times$4000~pixels & Set by chip size and science\\
        Pixel size         & 16~$\mu$m, ~0.37\arcsec\ & Same pixel size for
        both detector types; sub-pixel positioning localizes photons
        to 0.15\arcsec\ HPD (samples mirror PSF)\\
        CCD format (pixels) & 2500$\times$1500 pixels, ~4$\times$2.4cm 32 output nodes per CCD & Backside-illuminated frame-transfer devices\\
        CCD serial / parallel transfer rates & 2.5 MHz / 0.6 MHz & \\
        CCD frame rate & 20 fps & \\
        CCD radiation damage mitigation & Charge injection, trough & \\
        CMOS format & 1000$\times$1000~pixels, 1.6$\times$1.6~cm & \\
        CMOS frame rate & $>$20~fps (goal $>$100~fps) & Optional 100$\times$100~pixel sub-array  (bright sources)\\
        Depletion depth & 100~$\mu$m & \\
        Energy band & 0.2--12~keV & Telescope and depletion depth \\
        Readout noise & $<$4e$^{-}$ & Energy resolution requirement \\
        Spectral resolution (FWHM) & 60\,eV@1\,keV, 150\,eV@6\,keV & \\
        Optical and \newline contamination blocking filters & 40~nm Al\newline 30~nm Al + 45~nm polyimide & On-chip \newline Warm offset filter at $+20$~$^{\circ}$C\\
        Focal plane temperature & $-90$~$^{\circ}$C & Reduces radiation-induced CTI in CCDs. CMOS can be warmer.\\
        Data rate: FPA $\rightarrow$ FEE \newline Data rate: FEE $\rightarrow$ MEB & 3840 Mbps \newline 1 Mbps & All 5 sensors \newline Assume 200 bit/evt, 1000 evt/s\\
        \hline
    \end{tabularx}
    \caption{Characteristics and key parameters of the baseline \AXIS\ Focal Plane Array}
    \label{table:DETECTOR_params}
\end{table}
}

\subsection{Focal Plane Detectors}
\label{sec:detectors}

The baseline \AXIS\ Focal Plane Array (FPA) incorporates a hybrid approach that
exploits two technologies currently in advanced development: CCDs with low
clock power and fast, massively parallel readout; and, CMOS active pixel
sensors, which are fast, low-power, radiation-hard devices. The \AXIS\ FPA
uses four 1.5k$\times$2.5k CCDs to tile the majority of the focal plane
outside of the center, and a single, smaller 1k$\times$1k CMOS in the center
(see Fig.~\ref{fig:DETECT_fpa_schem}) to minimize pile-up of bright
targets. The CCDs are tilted to match the curved focal plane, minimizing
image distortion. Both detector types are back-illuminated and fully
depleted to 100~$\mu$m to ensure high QE across the
\AXIS\ band of 0.2--12~keV.  The 16$\mu$m (0.37\arcsec) pixel size is
sufficient to sample the 0.4\arcsec\ mirror PSF, because charge from 
a single photon is spread across multiple pixels and can be centroided
through sub-pixel positioning\cite{Li2004,Bray2018}
(and thus each photon's position can be determined) to 0.15\arcsec\ (HPD).
This accuracy will be significantly better than that for \Chandra\ ACIS, even though it
has a comparable pixel size. Even with the necessary multi-pixel event
reconstruction, readout noise of less than 4~$e^{-}$ ensures good soft
response\cite{Miller2018}.

Both types of detectors are baselined to read out at 20 frames/s
(fps) --- 64$\times$ that of \Chandra\ ACIS --- with the CMOS projected
to read out faster than 100 fps for pointed observations of bright sources.
With this design (and taking into account the \AXIS\ greater collecting area),
sources up to $6\times$ brighter than the
the \Chandra{} pile-up limit will be observed free of pile-up. Faster readout
improves time resolution and allows timing studies to take better advantage of
the large collecting area. It also results in fewer optical photons
contaminating each frame, thus allowing for thinner filters and much higher
soft X-ray sensitivity.

\subsubsection{CCD Technology Development.}

Si-based CCD detectors offer excellent QE, and near-theoretical spectral
resolution across the \AXIS\ energy band. The primary challenges for \AXIS\ are
(1) fast, low-noise readout at low power; and (2) mitigating the
charge-transfer effects of on-orbit radiation damage.  Both of these
obstacles have been overcome in recent years.

The MIT Lincoln Laboratory Digital CCD (DCCD) development effort combines
fast, low-noise readout amplifiers with low-voltage charge 
transfer\cite{Bautz2018}. Devices with high-speed p-channel JFET
outputs incorporating clock swings of 3V were tested at speeds of 2.5~MHz,
producing responsivity over 20~$\mu$V/$e^{-}$ and only 5.5~$e^{-}$ of
amplifier readout noise, close to the \AXIS\ requirement of $<4$~$e^{-}$ total
readout noise. Achievement of 4~$e^{-}$ in the next few years is likely.
The effects of radiation damage on \AXIS\ are greatly ameliorated by the
benign radiation environment of low inclination LEO and use of charge injection whereby
sacrificial charge is periodically introduced during frame transfer to fill
traps; this successfully mitigated radiation damage to the \Suzaku{}/XIS, 
with a loss of only 5\% of the field of view\cite{Koyama2007}. SPENVIS 
simulations show that \AXIS\ is subject to $\sim$100x less non-ionizing 
radiation damage than \Suzaku{}, since in its $\leq 8^{\circ}$ inclination
LEO, \AXIS\ does not traverse as deeply into the South Atlantic 
Anomaly as \Suzaku{} in its $31^{\circ}$ inclination orbit.

\subsubsection{CMOS Technology Development.}

Hybrid CMOS X-ray detectors have been developed and successfully
flight-proven over the past decade. These devices are active pixel sensors,
made by hybridizing a Silicon detection layer to a Silicon
readout-integrated-circuit (ROIC) layer through an indium bump bond.  Each
individual pixel has its own readout circuitry, and there is no transfer of
charge from pixel to pixel.  As a result, the device is inherently radiation
hard since any radiation damage is limited to a narrow site in the silicon
lattice with no effect on other pixels, and the device is inherently low
power since there is no need to drive large capacitive loads to transfer
charge throughout the device.  These active pixel sensors have the
ability to readout the signal through multiple output lines and to readout
designated windows at higher clock rates.  The Teledyne H1RG$^{\text{TM}}$ HyVisI and H2RG$^{\text{TM}}$ HyVisI
devices\footnotemark\footnotetext{http://www.teledyne-si.com/products/Documents/H2RG\%20Brochure\%20-\%20September\%202017.pdf} are high TRL, as is the SIDECAR$^{\text{TM}}$ ASIC\footnotemark\footnotetext{http://www.teledyne-si.com/products-and-services/imaging-sensors/sidecar-asic} that typically runs them, with
flight missions such as OCO-2.  A more recent device, the X-ray version of a
1024$\times$1024 pixel HyVisI array bonded to an H2RG$^{\text{TM}}$ ROIC with 32 output lines,
was recently flown successfully on the \textit{WRX-R} rocket flight\cite{Chattopadhyay2018}.  
A standard H1RG$^{\text{TM}}$ has 1024$\times$1024 pixels and can
readout through 16 parallel output lines at pixel rates ranging from
100~kHz to 5~MHz for each of the individual lines.  These
devices can also read out an individual window through a dedicated line in
order to achieve very rapid frame rates in a small window for bright source
observations.

\begin{wrapfigure}{R}{0.58\textwidth}
\vspace*{-2mm}
    \centering
    \includegraphics[width=0.57\textwidth]{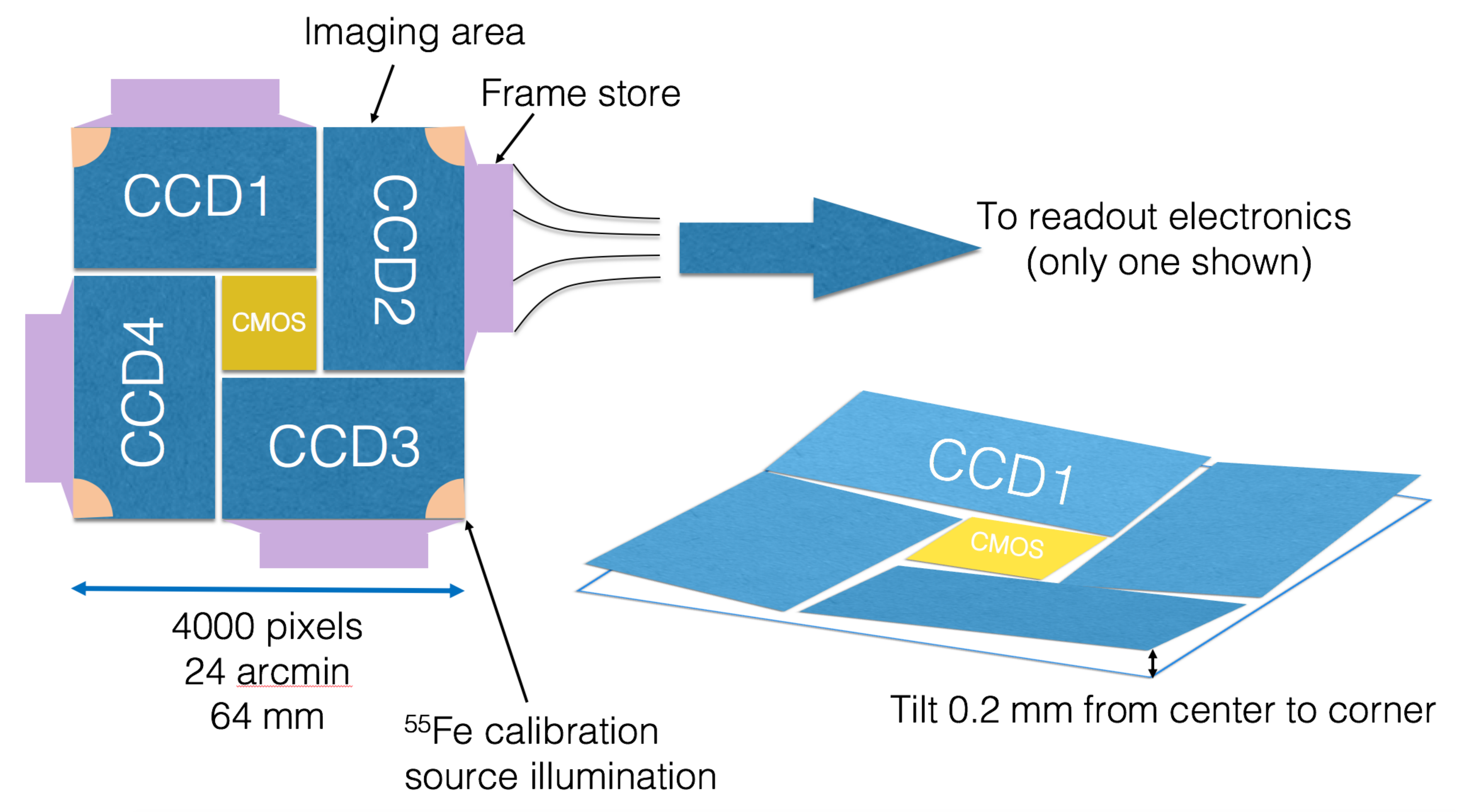}
    \caption{The baseline \AXIS\ Focal Plane Array utilizes a CCD/CMOS hybrid design.
    Both detector technologies are included to capitalize on parallel 
    technological development; the final design will likely use 
    a single detector technology.}
    \label{fig:DETECT_fpa_schem}
\vspace*{-1mm}
\end{wrapfigure}

Read noise of these devices has improved, reaching 5.5~$e^{-}$ (RMS) for the
most recent test devices and 6.5~$e^{-}$ (RMS) for the technologically
mature devices\cite{Chattopadhyay2018}.  The energy resolution of the best
devices is presently 148~eV (FWHM) at 5.9~keV and 78~eV (FWHM) at 0.53~keV\cite{Hull2018},
and we expect to reach a read noise of less than 4~$e^{-}$.
The detector power is $\sim 200$~mW for a typical
1k$\times$1k device. Currently, the best read noise is achieved when
operating the devices with many parallel readout lines at individual line
rates of $\sim$200~kHz. Future developments are expected to improve the
read noise at higher rates and/or to multiplex more parallel output lines.
There are also less mature devices that can achieve several orders of
magnitude faster effective frame rates through readout of only the pixels
with a valid X-ray event above a set threshold; small versions of these
devices have been tested, and larger versions are currently being
fabricated.

\subsection{Sensor Housing}

The FPA is housed in a vacuum chamber mounted to a baseplate, and features a
commandable vacuum door and $^{55}$Fe calibration sources that illuminate
the corners of the focal plane for monitoring and calibrating sensor
performance as in \Suzaku.  The FPA itself is mounted on a plate that is thermally
isolated from the housing and that regulates the temperature of the FPA
through a heat pipe to an external radiator
(Fig.~\ref{fig:DETECT_fpa_block}).  The FPA is maintained at
$-90.0\pm0.5$~$^{\circ}$C during normal operation with this radiator and a
trim heater thermally coupled to the mounting plate. A louvered radiator fin
on the dark side of the telescope barrel shielded from the sun can support
passive cooling of the FPA to $-65$~$^{\circ}$C, assuming 9W of power
dissipation from the FPA.  We expect the power dissipation to be
significantly less for the flight focal plane given the lower power use of
the CCDs under technology development as described above, and in CMOS
detectors. Higher operating temperature (up to $-60$~$^{\circ}$C) is also
possible given the fast readout that reduces the effects of dark current,
and charge injection that mitigates CCD radiation damage.

\subsection{Optical and Contamination Blocking Filters}

\AXIS\ incorporates a hybrid approach to block optical and UV photons from
reaching the light-sensitive detectors, and prevent build-up of molecular
contamination on the cold surfaces in the light path.  The detectors have
40~nm of Al directly deposited on the sensor surface to eliminate
light-leak, a smaller amount than previous instruments due to the faster
readout and resulting looser light-blocking requirements of the baseline \AXIS\ FPA. A
contamination blocking filter composed of an additional 30~nm Al and 45~nm
polyimide is located 4~cm above the focal plane beneath the door of the
sensor housing, where it is held at $+20$~$^{\circ}$C to provide additional
light blocking and prevent the kind of molecular contamination that has
built up on the cold filters of previous instruments (e.g. \Chandra\ ACIS,
\Suzaku\ XIS).  The FPA may be heated to $+20$~$^{\circ}$C in a
decontamination mode.

\subsection{Focus Mechanism}

\begin{wrapfigure}{r}{0.67\textwidth}
\vspace*{-13mm}
    \centering
    \includegraphics[width=0.70\textwidth,viewport=4 1 686 406,clip]{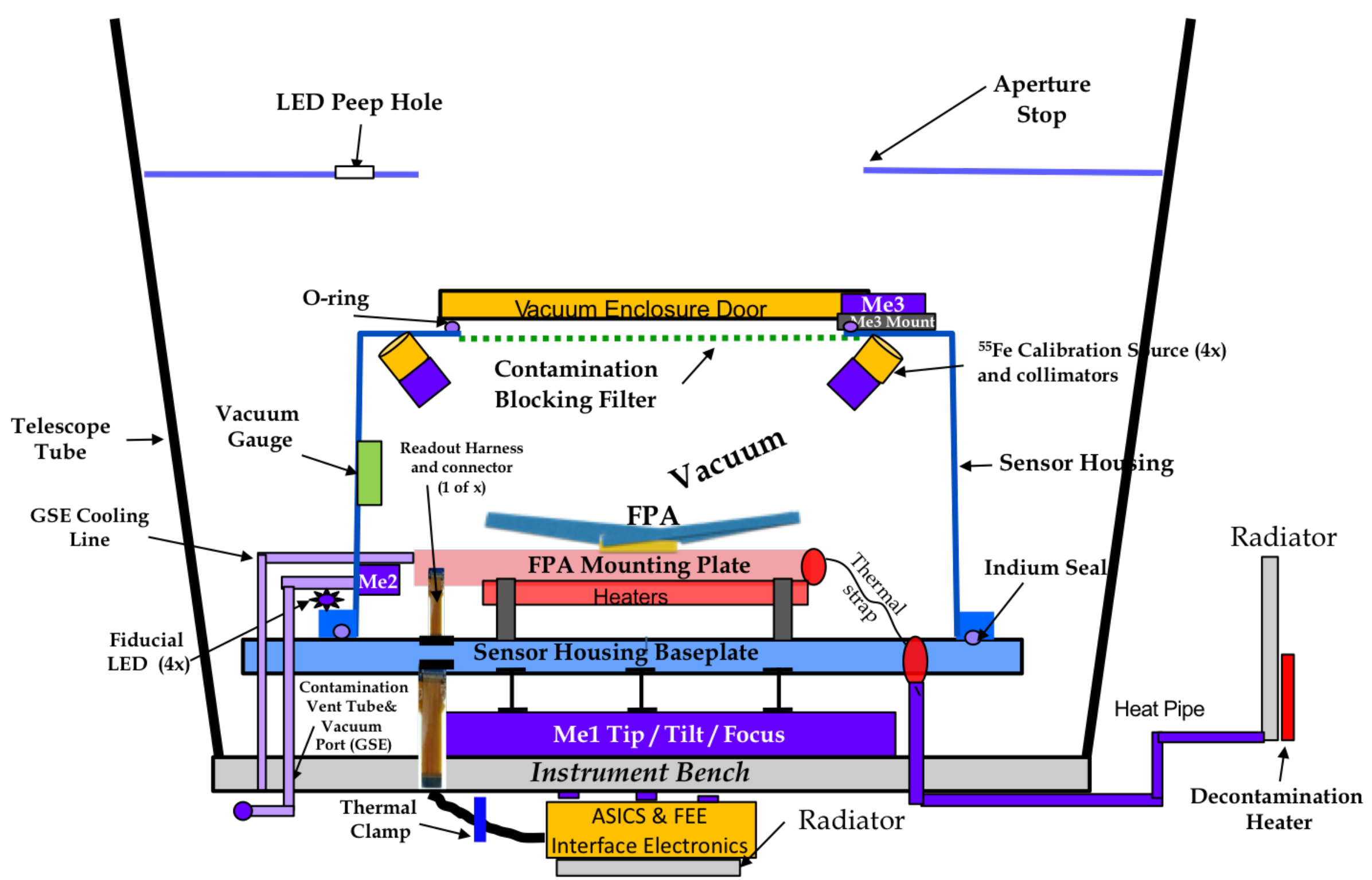}
    \caption{\AXIS\ baseline Focal Plane Assembly block diagram strongly resembles 
    those of successful X-ray CCD instruments.}
    \label{fig:DETECT_fpa_block}
\vspace*{-2mm}
\end{wrapfigure}

The sensor housing is directly mounted on a focus mechanism
(Fig.~\ref{fig:DETECT_fpa_block}). The design suspends the Detector
Assembly on three tangential flexures and controls tip, tilt, and piston
with three linear actuators. The focus mechanism moves the estimated 20~kg
of suspended mass with a resolution of $\pm$1 arcmin in tip and tilt and
$\pm$12~$\mu$m in piston, over a range of $\pm$1$^{\circ}$ in tip and tilt
and $\pm$2~mm in piston---sufficient to sample the expected on-orbit focal
length uncertainty. Once focus is achieved during commissioning, the
actuators will be locked in place, though later adjustments may be made if
needed.

\subsection{Front-End Electronics Design and Technology Development}

The Front-End Electronics (FEE) digitize the output signal from the
detectors. This may be accomplished using a dedicated SIDECAR$^{\text{TM}}$ ASIC
for each CCD and CMOS detector, located in the FEE box on the opposite side
of the instrument bench from the sensor housing.  In this configuration, the
FEE ASICs can run warm.

The SIDECAR$^{\text{TM}}$ ASICs provide bias voltages and clock control to the detectors and amplify and convert the analog output to digital signal. This signal,
totaling 900 (240) Mbps for each CCD (CMOS) running at 20~fps and 12~bits pixel$^{-1}$ is transferred to a set of FPGAs, one per detector,
which perform event processing to reduce the large data stream to $\sim 1$~Mbps total X-ray plus background events.  These are transferred by digital
line to the Master Electronics Box (MEB) on the spacecraft end of the
observatory for packaging, filtering, and telemetry.

The FEE design requires placing it on the warm side of the instrument bench
with a dedicated radiator to maintain the $-10$ to $+40$~$^{\circ}$C
operating temperature range.  To eliminate parasitic heat loads between this
stage and the $-90$~$^{\circ}$C FPA, the cable harness between the FPA and
FEE box is run through a thermal clamp maintained at $-40$~$^{\circ}$C.

While current technology is sufficient to achieve the necessary frame time,
\AXIS\ is expected to capitalize on ongoing technology development. For
example, the VERITAS family of ASIC chips\cite{Herrmann2018} currently under
development provides fast, multichannel, low-noise readout for CCDs and
DEPFET active pixel sensors, with the latter planned as the detector
technology for the \Athena\ Wide-Field Imager\cite{Meidinger2017}.  This
technology can be adopted with some modifications for the \AXIS\ detector FEE.

\section{SPACECRAFT AND MISSION OPERATIONS}
\label{sec:ops}

\begin{wrapfigure}{r}{0.6\textwidth}
\vspace*{-4mm}
\centering
 \includegraphics[width=0.4\textwidth,angle=-90,viewport=25 78 534 790,clip]{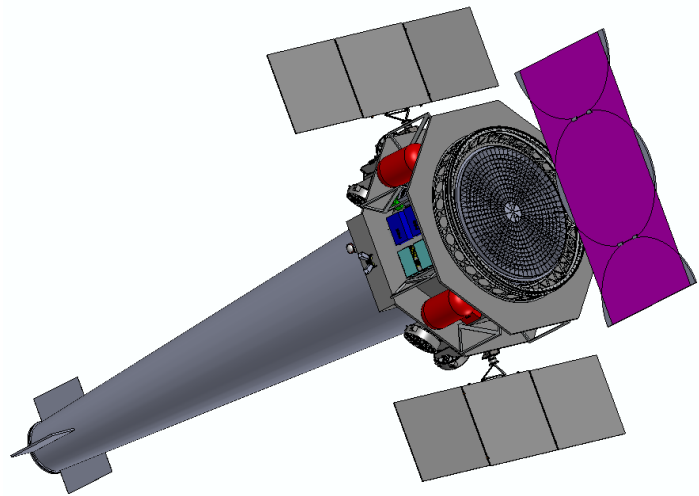}
\caption{Mission Design Lab model of the \AXIS\ observatory.}
\vspace*{-2mm}
\label{fig:OPS_spacecraft}
\end{wrapfigure}

The GSFC Mission Design Lab (MDL) studied the \AXIS\ mission concept using the
instrument (X-ray telescope and detector) point design output from the IDL
described above. The estimated instrument mass and power are 750\,kg and
300\,W, respectively. The total wet mass, including the de-orbit systems, is
2300\,kg (including 20\% margin). The estimated average \AXIS\ power
consumption is 720\,W for the entire spacecraft and instrument, and peak power is 1200\,W, provided by 8.2\,m$^2$ of solar panels
producing 2600\,W at launch and 1200\,W after 10 years. A
145~A\,hr battery provides power during eclipses. The low-inclination
LEO minimizes the particle background and allows for rapid
communication and response times.

The resulting spacecraft design, shown in Fig.~\ref{fig:OPS_spacecraft},
meets all of the \AXIS\ science requirements in a class~B mission with a
nominal five year lifetime (with consumables sized for at least ten years).
The spacecraft meets the mass, length, and diameter specifications for
launch into this orbit on a \textit{SpaceX} Falcon~9.

{\rowcolors{3}{white!100}{tablealt}
\begin{table}[t]
    \centering
    \begin{tabularx}{\textwidth}{L{1.75in} C{0.55in} L{3.703in}}
        \rowcolor{callout}
        {\centerline{\textcolor{white}{\sfsm Error }}}\phantom{\rule{0cm}{4mm}} & \textcolor{white}{\sfsm Req. \newline (\arcsec\ HPD)} &{\centerline{\textcolor{white}{\sfsm Note}}}\\
        \rowcolor{white}
        \multicolumn{3}{c}{\bs Pointing errors not 
        affecting angular resolution:\phantom{\raisebox{-2.5mm}{\rule{0cm}{7mm}}}} \\ 
        Pointing maneuver accuracy & 15 & Target should fall near detector center\\
        Absolute celestial location from startrackers
        & 1 & Can be significantly reduced in post-processing using
        IDs of hundreds of X-ray sources in every observation \\
    \hline
        \rowcolor{white}
        \multicolumn{3}{c}{\bs Contributions
        to on-axis angular resolution:\phantom{\raisebox{-2.5mm}{\rule{0cm}{7mm}}}} \\ 
        Mirror PSF & 0.4 & from Table \ref{table:MIRRORS_error} \\
        Detector photon positioning & 0.15 & from Table \ref{table:DETECTOR_params}, 
        prediction based on CCD simulations \\
        Startracker accuracy \newline for relative pitch, yaw & 0.1 & Knowledge of relative
        tilt of telescope w.r.t.\ sky during observation. Roll error has negligible
        effect on on-axis PSF. \\
        Telescope rigid-body\newline pitch, yaw in 50\,ms & 0.1 & Uncorrectable telescope 
        motion over the 50\,ms detector integration time. Motions on longer
        timescales are corrected
        in ground processing using startracker data. \\
        Telescope flex in 50\,ms & 0.1 & Uncorrectable motion of detector in 
        mirror ref.\ frame over detector integration time. 
        Motions on longer timescales are corrected in ground processing
        using metrology system data. \\
        Detector metrology system accuracy & 0.1 & Corresponds to 5$\mu$m detector translation
        in plane perpendicular to optical axis; roll, pitch, yaw are negligible.\\
        Telescope length stability & 0.1 & Corresponds to detector defocusing by 10$\mu$m \\
        Total angular resolution & 0.5 & All contributions added in quadrature\\
        \hline
    \end{tabularx}
    \caption{Angular resolution and pointing error budget. (For a 2D Gaussian, HPD$=2.35\sigma$)}
    \label{table:OPS_pointing_error}
\end{table}
}

\subsection{Pointing accuracy and in-orbit angular resolution}
\label{sec:aspect}

Achieving a final on-axis angular resolution of 0.5\arcsec\ HPD
requires keeping all non-mirror contributions to $\lax0.1$\arcsec. 
The in-orbit PSF budget is given in Table \ref{table:OPS_pointing_error}.
The startrackers mounted at the mirror track the motion of the 
mirror axis with a 0.1\arcsec\ accuracy. These offsets are applied in
ground processing to the positions of the X-ray photons as registered 
by the detector (\S\ref{sec:detectors}). Because the photon arrival
time is quantized by 50\,ms\ (the detector readout period), 
any motions within that interval cannot be corrected and will
result in broadening of the PSF. In addition to the rigid-body 
motions of the telescope, 
it will flex due to thermal distortions, etc., which will mostly 
result in translation of the detector with respect to the mirror axis. These
motions will be tracked using a metrology system employing fiducial
lights mounted at the detector assembly and observed by a startracker
mounted at the mirror (using either the \Chandra\ scheme\cite{Aldcroft2000}
or a separate startracker pointed toward the detector). These offsets are 
applied to the X-ray photon positions to map them to the mirror reference frame.
Again, motions within 50\,ms\ cannot be corrected (but will be very small).
Preliminary metrology system design work has been initiated with our
industry partner, Northrop Grumman. GSFC Attitude Control System 
engineers consulted by the \AXIS\ team expect the requirements in 
Table \ref{table:OPS_pointing_error} to be achievable 
using current motion damping technology and reaction wheels and three
startracker camera heads with a 4~Hz readout rate and standard 
interpolation algorithms.

We note that, while ambitious, the above requirements
are certainly technically feasible. \Hubble, a telescope of similar 
size and shape, working in a similar LEO, launched 3 decades ago, 
achieved a 20 times better in-orbit angular resolution than our
goal for \AXIS\ --- without the luxury of fast detector readout and
photon-by-photon image reconstruction used in the X-ray. \HST\ 
is held {\em stable}\/ to 0.01\arcsec\ over the span of $>$1000~s
exposures. This compares to our requirement of 0.1\arcsec\ 
telescope stability over the 50~ms detector integration time and
0.1\arcsec\ attitude {\em knowledge}\/ for longer timescales, 
which is applied to each X-ray photon to map it onto the sky.

All the above requirements are for the {\em relative}\/ attitude motions within an observation. The startrackers will also provide $\sim$1\arcsec\ knowledge of the absolute
celestial position. In each observation, \AXIS\ will have hundreds of serendipitous X-ray point
sources. If needed, their identification with optical/IR/radio counterparts from available surveys will
allow a much more accurate determination of the position in post-processing. 

\subsection{Rapid response to transient sources}
\label{sec:slew}

Optimizing observing efficiency and allowing for rapid response to
transients requires a slew rate of 120$^{\circ}$ in $<6$~minutes. With 
this slew rate and our nominal observing program, the efficiency was 
estimated by the MDL to be at least 70\%, giving a net observing time of 
$\sim 2.2\times 10^7$~s per year. 
Six Honeywell HR-16 reaction wheels, each with 100~N\,m\,s
capacity, were baselined by the MDL. Three magnetic torquers dissipate
accumulated angular momentum using the Earth's magnetic field. The field of
regard is set primarily by requiring a 45$^{\circ}$ Sun exclusion angle to
avoid stray light and maintain thermal control and avoidance of the bright
Earth and Moon.

\AXIS\ has a straightforward operational concept. After launch and commissioning, the instrument will undergo a $\sim$1-month
standard checkout and calibration phase to determine the optical axis,
confirm the effective area and angular/spectral resolution, and optimize the
focus. The only instrument mechanism is the focus adjuster which is expected to be used
only during commissioning, although it will be available during the mission
if necessary.

\begin{wrapfigure}{r}{0.69\textwidth}
    \centering
    \vspace*{-7mm}
    \includegraphics[width=0.69\textwidth]{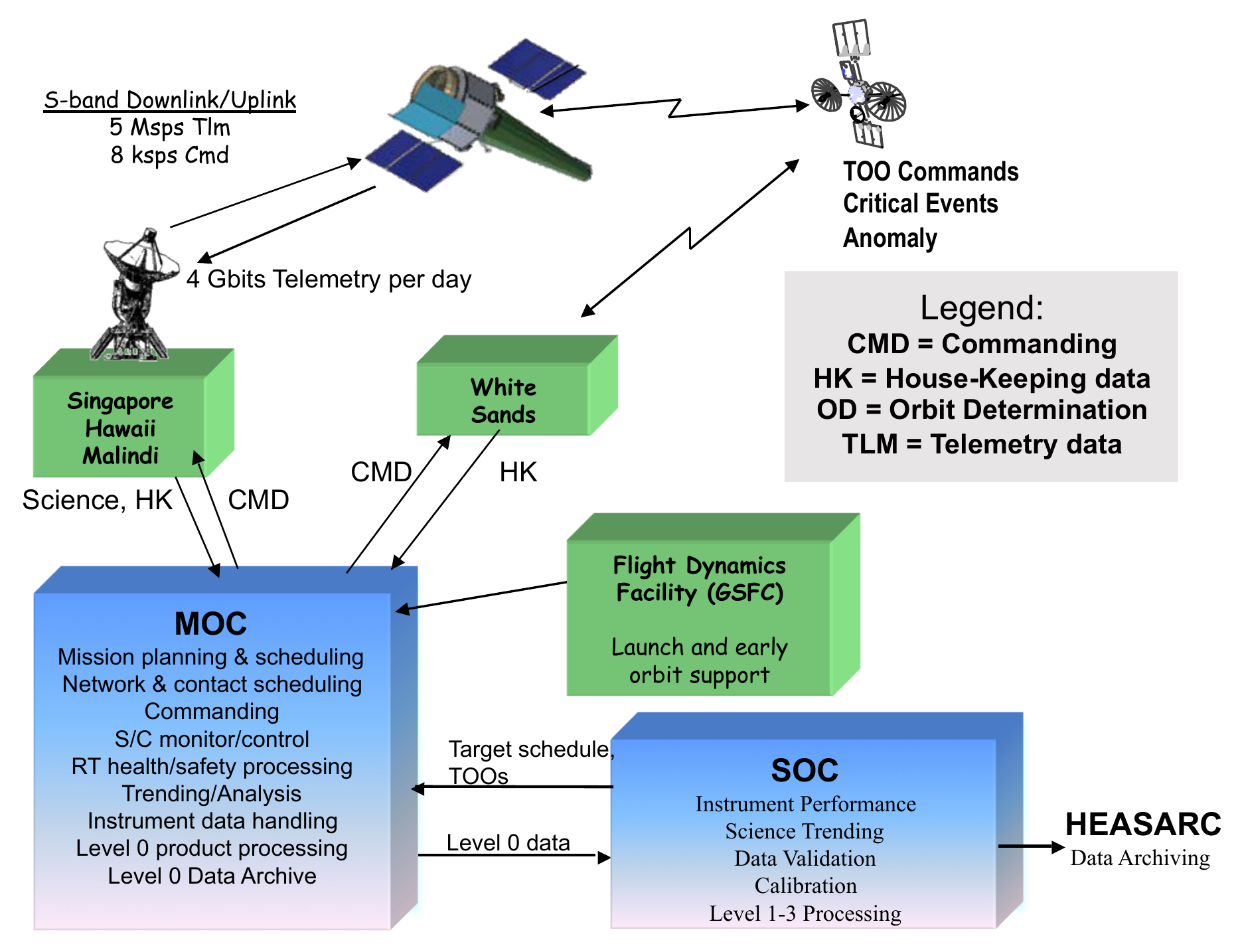}
    \caption{The Ground Operations plan for \AXIS\ is simple and utilizes proven elements.}
    \label{fig:OPS_ground_ops}
    \vspace*{-2mm}
\end{wrapfigure}

During normal operations, target sequences will be generated on a weekly
basis by the Science Operations Center (SOC) and transmitted to the Mission
Operations Center (MOC) for uploading. Most fields will be observed for
total integration times of 50-100~ks and will not require continuous
observation to build up the required exposure. This flexibility enables a
smooth and automatic restart after interruptions caused by ToOs, described
below. The detectors primarily operate in full-readout mode, with a
secondary partial readout mode used to reduce pile-up for bright sources.
Despite the large format detectors, the total data rate is modest as only
X-ray photon events identified by on-board processing are telemetered to the
ground, except for daily bias frames for calibrating read noise, and full
frame dumps during low telemetry (typically, three every few months) for
dark current calibration. The onboard storage (128~Gbit) and telemetry
system were designed to support 4~Gbit/day downlink using two 10-minute
S-band ground station passes, and allows for the handling of ground system
outages. Data volumes as high as 40~Gbit/day may be accommodated with
additional downloads. There are no requirements for rapid downlink.

\begin{figure}[t]
\centering
  \includegraphics[width=0.9\textwidth,viewport=34 70 678 434,clip]{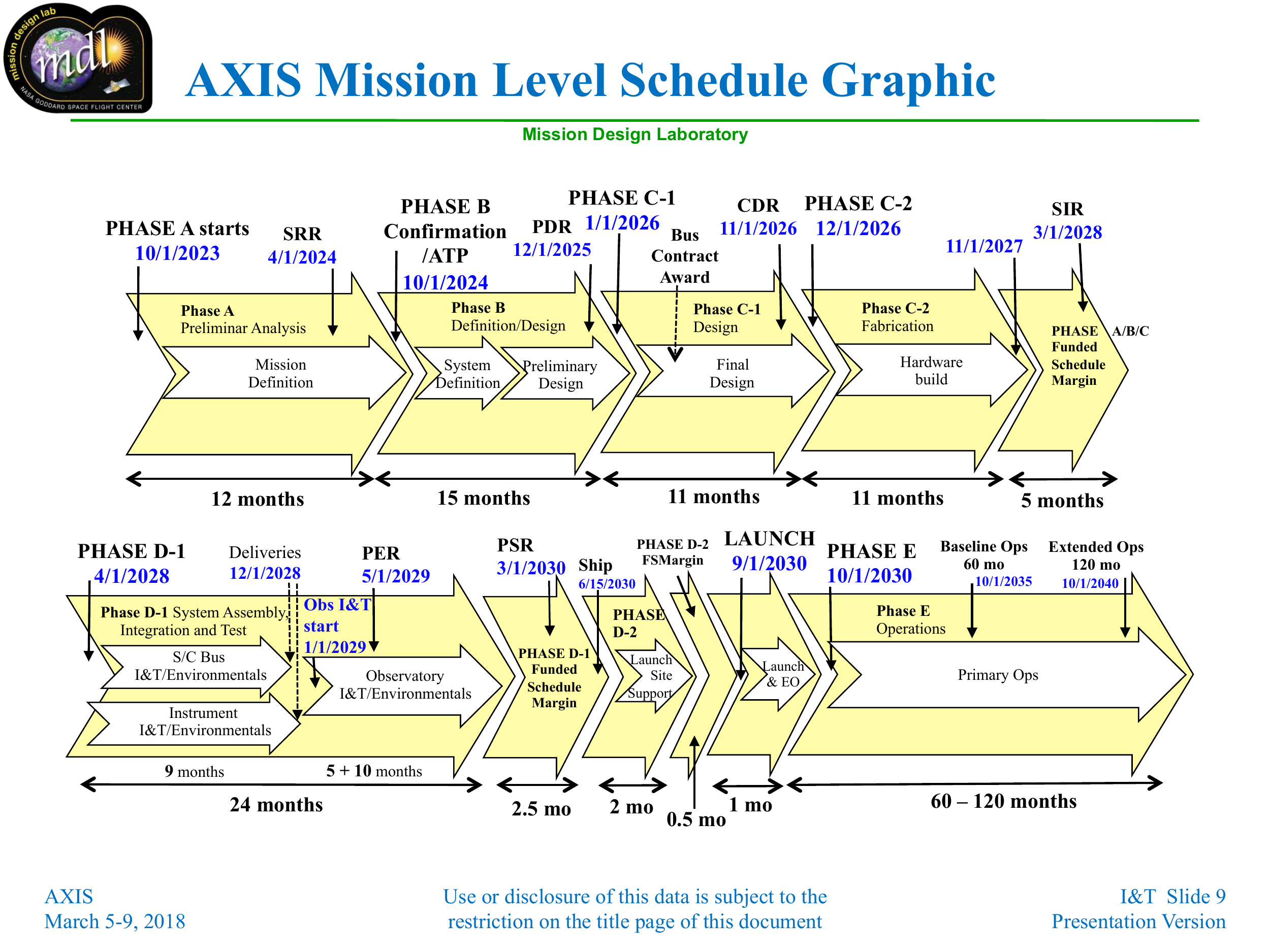}
  \caption{The \AXIS\ mission schedule used for estimating cost assumes an experience-based 
  seven-year development period (phases A-D).}
  \label{fig:OPS_schedule}
\end{figure}

\AXIS\ will support approximately five ToOs per week.  Most of these are
expected to be responses to the community. After approval by
the SOC, a ToO will be sent to the MOC for upload to \AXIS. The spacecraft
will be designed to accept a ToO interrupt to observe a new target and then
automatically return to the preset target sequence when complete, similar to
\Swift{} operations. The \AXIS\ requirement for ToO response, based on the
extensive \Swift{} legacy, is four hours. Normal operations are designed
assuming two downlink passes per day using the Near-Earth Network (NEN),
although there are many more opportunities for ToO uplinks. In rare cases,
response times as short as one hour are possible by taking advantage of
TDRSS. TDRSS would be used during launch and for critical command and
control communication.

The ground system architecture is shown in Fig.~\ref{fig:OPS_ground_ops}.
Science and housekeeping (HK) data will be downloaded on a daily basis. HK
will be checked for anomalies and trended at the MOC. After Level~0
processing, both datastreams will be transferred to the SOC for pipeline
processing. The pipeline software will be based on heritage from existing
missions including \Chandra{}, \Swift{}, and \Suzaku{}. In normal
operations, data will undergo validation and verification at the SOC within
one week of observation and then be submitted to the HEASARC for
distribution and archiving.

The mission schedule assumed for costing is shown in
Fig.~\ref{fig:OPS_schedule}. \AXIS\ launch commences $\sim$7 years after the
start of Phase-A assuming that TRL~5 is achieved before early 2023.

\section{COST ESTIMATE}
\label{section:cost}

Redacted. {\em Note: Our in-house cost estimates show that the AXIS cost is consistent with
the Probe guideline. Once the costs have been validated by NASA, we will
update this document to include a cost breakdown.}

\subsection{Descopes}

Almost no primary \AXIS\ science objective requires a given \textit{instantaneous} collecting area.  Hence, the mirror size could be reduced to fit \AXIS\ within a particular cost envelope.  The scientific trade is prioritizing the science objectives against the longer exposures that would be needed to achieve them.  Since the \AXIS\ sensitivity is photon limited for all the observations described in this report, the scaling of exposure time with mirror area is essentially linear, and performing these science trade studies is straightforward. The obvious approach to reducing mirror size, which has the least cross coupling to other systems and could be applied nearly any time during the mission development, is removal of one or more metashells. Table~\ref{table:MIRRORS_metashells} shows the relative reduction of mass and effective area at several energies provided by the removal of each metashell.  

If a mirror descope is required, taking it early in the program offers substantial benefits.  The mirror size drives the entire observatory configuration.  This means that a mirror size reduction prior to Preliminary Design Review (PDR) could lead to substantial savings throughout the system (e.g., less massive optical bench, smaller wheels, less massive propellant system).  A scaling estimate performed during the X-ray Probe studies earlier in the decade suggests that an integrated design without the outer metashell could reduce the overall mission cost by as much as 10 percent.

Our baseline cost for the detector is conservative, because we plan on selecting a single type of chips from the
two included in the current hybrid design. This simplifying ``descope'' will occur very early in the program. 
The cost saving here is nominal, but the technical risk reduction could be substantial.

\subsection{Enhancements}

\AXIS\ was designed expressly to provide the most sensitive, and thus the largest, possible X-ray observatory within the \$1B cost guideline.  The availability of such a powerful capability might induce NASA to explore broadening the scientific reach.  One incremental enhancement to our current concept would be to enlarge the detector array to cover the
off-axis angles where the mirror PSF broadens above 1\arcsec\ but is still under
5\arcsec\ HPD  (\Athena's resolution). This can be achieved by adding an outer 
ring of four more chips (Fig.\ \ref{fig:DETECT_fpa_schem}) to extend the
field of view to $36^{\prime}\times 36^{\prime}$ (double our nominal FOV by solid angle) and
take full advantage of the mirror's superb imaging capability.

Another possible enhancement would be to add a second focal plane instrument.  
While the team did not study adding instruments to \AXIS, some plausible enhancements were studied for similar-sized missions during the X-ray Probes Concepts study earlier in the decade\footnotemark\footnotetext{https://pcos.gsfc.nasa.gov/studies/completed/x-ray-probe-2013-2014.php}. The incremental costs for enhancements identified there are likely valid estimates for \AXIS.  As an example,one of the Probe concepts studied  (the Notional Calorimeter mission) could have been augmented  with a grating spectrometer at  an estimated increase in cost of  \$150M. (The cost of the Notional X-ray Grating Spectrometer probe, a mission with similar capability to that of adding a grating to \AXIS, was  estimated  to cost \$750M.)  Other modest augmentations might include a side-mounted hard X-ray telescope (for which arcsecond angular resolution would not be required), a polarimeter, or a redundant imaging detector. 

\clearpage
\section{ACRONYMS AND ABBREVIATIONS}

\begin{table}[b]
    \centering
    \begin{tabularx}{\textwidth}{ L{4cm} X }
        \hline
        {\bs Missions and Facilities} & \\
        \hline
        \ALMA\     & Atacama Large Millimeter Array \\
        \Athena\   & Advanced Telescope for High ENergy Astrophysics (future)\\
        \Chandra\  & Chandra X-ray Observatory \\
        ACIS    & Advanced CCD Imaging Spectrometer (\Chandra\ detector) \\
        \CTA\      & Cherenkov Telescope Array (future)\\
        \ELT\      & European Extremely Large Telescope (future)\\
        \eROSITA\  & Extended Roentgen Survey with an Imaging Telescope Array (future)\\ 
        ESA        & European Space Agency \\
        \Euclid\   & European space cosmology mission (future) \\
        \GMT\      & Giant Magellan Telescope (future) \\
        \GMRT\     & Giant Meter-wave Radio Telescope\\
        GSFC       & NASA's Goddard Space Flight Center\\
        HEASARC    & High Energy Astrophysics Science Archive Research Center\\    
        \Herschel\ & Herschel Space Observatory \\
        \HST\      & Hubble Space Telescope \\
        \JVLA\     & Karl G. Jansky Very Large Array \\
        \JWST\     & James Webb Space Telescope (future)\\
        \LIGO\     & Laser Interferometer Gravitational-wave Observatory\\
        \LISA\     & Laser Interferometer Space Antenna (future)\\
        \LOFAR\    & Low-Frequency Array\\
        \LSST\     & Large Synoptic Survey Telescope \\
        \Lynx\     & Lynx X-ray Surveyor (NASA Flagship study)\\
        \MWA\      & Murchison Wide-Field Array\\
        NASA       & National Aeronautics and Space Administration \\
        NGXO       & Next-generation X-ray Optics lab at GSFC\\
        \NuSTAR\   & Nuclear Spectroscopic Telescope Array \\
        \SKA\      & Square Kilometer Array (future)\\
        \Spitzer   & Spitzer Space Telescope \\
        \Suzaku\   & Japan-US X-ray imaging spectroscopy mission (past)\\
        \Swift\    & Neil Gehrels Swift Observatory\\
        \TESS\     & Transiting Exoplanet Survey Satellite\\
        \TMT\      & Thirty Meter Telescope (future)\\
        \WFIRST\   & Wide Field Infrared Survey Telescope (future)\\
        \textit{WRX-R}  & Water Recovery X-ray Rocket\\
        XIS        & X-ray Imaging Spectrometer (\Suzaku\ detector)\\
        \XMM\      & X-ray Multi-Mirror Mission (Newton) \\ 
        \XRISM\    & X-ray Imaging and Spectroscopy Mission (future)\\
        \hline
    \end{tabularx}
\end{table}

\begin{table}
    \centering
    \begin{tabularx}{\textwidth}{ L{4cm} X }
        \hline
        {\bs Astronomical Terms} & \\
        \hline
        AGN     & Active galactic nucleus (powered by an accreting massive black hole) \\
        BCG     & Brightest cluster galaxy (in a galaxy cluster)\\
        BH      & Black hole \\
        CMB     & Cosmic microwave background \\
        CGM     & Circumgalactic medium (gas around, but bound to, galaxies) \\
        CX      & Charge exchange (between an ion and neutral atom)\\
        CXB     & Cosmic X-ray background (can be resolved into AGNs)\\
        FUV     & Far ultraviolet (typically $\lambda < 2000$\AA) \\
        GC      & Galactic Center\\
        GW      & Gravitational wave \\
        HMXB    & High-mass X-ray binary (compact object with a massive star)\\
        ICM     & Intracluster medium (hot gas in a galaxy cluster)\\
        IGM     & Intergalactic medium (diffuse gas between galaxies) \\
        IMF     & Initial mass function (of stars)\\
        ISCO    & Innermost stable circular orbit ($3 GM/c^2$ in the Schwarzschild metric)\\
        ISM     & Interstellar medium \\
        LMC     & Large Magellanic Cloud\\
        LMXB    & Low-mass X-ray binary (compact object with a dwarf star) \\
        LSS     & Large-scale structure (of the Universe)\\
        MHD     & Magnetohydrodynamic \\
        NIR     & Near-infrared\\
        NS      & Neutron star\\
        Pop~III & Population III stars (first generation of stars)\\
        PWN(e)  & Pulsar wind nebula(e)\\
        RMS     & Root-mean square \\
        SED     & Spectral energy distribution\\
        SFR     & Star-formation rate\\
        SMBH    & Supermassive black hole \\
        SMC     & Small Magellanic Cloud \\
        SN(e)   & Supernova(e) \\
        SNR     & Supernova remnant \\
        TDE     & Tidal disruption event (a star shredded by a black hole)\\
        ULX     & Ultraluminous X-ray source ($L_X > L_{\text{Edd}}$ for $10 M_{\odot}$) \\
        WHIM    & Warm-hot intergalactic medium \\
        XLF     & X-ray luminosity function \\
        XRB     & X-ray binary (neutron star or black hole with a companion star) \\
        \hline   
    \end{tabularx}
\end{table}

\begin{table}
    \centering
    \begin{tabularx}{\textwidth}{ L{4cm} X }
        \hline
        {\bs Technical Terms} & \\
        \hline
        ASIC    & Application-specific integrated circuit\\
        CCD     & Charge-coupled device (detector) \\
        CMOS    & Complementary metal-oxide-semiconductor (active pixel detector) \\
        CTI     & Charge-transfer inefficiency (in a detector)\\
        DEPFET  & DEpleted P-channel Field Effect Transistor (active pixel detector)\\
        E/PO    & Education and public outreach\\
        ETU     & Engineering test unit \\
        FEE     & Front-end electronics \\
        FOV     & Field of view \\
        FPA     & Focal plane array (\AXIS\ detector array)\\
        FPGA    & Field-programmable gate array \\
        fps     & Frames per second \\
        FWHM    & Full width at half-maximum \\
        GSE     & Ground systems engineering \\
        H(1,2)RG$^{\text{TM}}$    & Teledyne focal plane array detector \\
        HK      & Housekeeping (data) \\
        HPD     & Half-power diameter (image region containing 50\% of flux)\\
        IDC     & Integrated Design Center \\
        IDL     & Instrument design lab \\
        I\&T    & Integration and Testing \\
        JFET    & Junction gate field-effect transistor \\
        LEO     & Low-earth orbit (like \ROSAT, \Swift, \Suzaku; unlike \Chandra, \Athena, \Lynx) \\
        MDL     & Mission design lab \\
        MEB     & Master electronics box\\
        MEL     & Master equipment list\\
        MOC     & Mission operations center\\
        NEN     & Near-Earth Network \\
        PDR     & Preliminary design review\\
        PSF     & Point-spread function (image of a point source)\\
        QE      & Quantum efficiency (probability for a detector to register an X-ray photon)\\
        ROIC    & Readout integrated circuit \\
        SIDECAR$^{\text{TM}}$ & Teledyne System Image, Digitizing, Enhancing, Controlling, and Retrieving\\
        SMO     & Silicon metashell optics \\
        SOC     & Science operations center\\
        STM     & Science traceability matrix \\
        TDRSS   & Tracking and Data Relay Satellite System \\
        ToO     & Target of opportunity \\
        TRL     & Technology readiness level \\
        WBS     & Work breakdown structure (for costing)\\
        \hline
    \end{tabularx}
\end{table}

\clearpage


\section{REFERENCES}

\small
\parskip=0mm
\vspace{-6mm}

\bibliographystyle{naturemag}
\bibliography{references}

\end{document}